\documentclass[a4paper, 12pt]{article}
\usepackage{geometry}
\pdfoutput=1
\usepackage{setspace}
\usepackage[utf8]{inputenc}
\usepackage[T1]{fontenc}
\usepackage[english]{babel}
\usepackage{lmodern}
\usepackage{amsmath}
\usepackage{amssymb}
\usepackage{bm}

\usepackage{xcolor}

\usepackage{graphicx}
\usepackage{float}
\usepackage{subfig}

\usepackage[hyphens]{url}
\usepackage{hyperref}
\usepackage{csquotes}
\usepackage[    sorting=none,
                style=phys,
                articletitle=false,
                biblabel=brackets,
                chaptertitle=false,
                pageranges=false,
                citestyle=numeric-comp] {biblatex}
\DeclareMathOperator{\Y}{Y}
\DeclareMathOperator{\bL}{\mathbf{\hat{L}}}
\DeclareMathOperator{\Lsquared}{\hat{L}^2}
\DeclareMathOperator{\Laplace}{\Delta}

\title{ Perturbation Theory of Optical Resonances of Randomly Deformed Dielectric Spheres}
\author{    Julius Gohsrich \\
\href{mailto:julius.gohsrich@mpl.mpg.de}{julius.gohsrich@mpl.mpg.de}\\
\href{mailto:julius.gohsrich@fau.de}{julius.gohsrich@fau.de}}

\date{\today}

\begin{document}

\newcommand{\ETE}{\mathbf{E}^\text{TE}}
\newcommand{\BTE}{\mathbf{B}^\text{TE}}
\newcommand{\ETM}{\mathbf{E}^\text{TM}}
\newcommand{\BTM}{\mathbf{B}^\text{TM}}
\newcommand{\bD}{\mathbf{D}}
\newcommand{\bB}{\mathbf{B}}
\newcommand{\bH}{\mathbf{H}}
\newcommand{\bE}{\mathbf{E}}
\newcommand{\bF}{\mathbf{F}}
\newcommand{\bX}{\mathbf{X}}
\newcommand{\bY}[2]{\mathbf{Y}_{{#1} \, {#2}}}
\newcommand{\bPsi}[2]{\mathbf{\Psi}_{{#1}\,{#2}}}
\newcommand{\bPhi}[2]{\mathbf{\Phi}_{{#1}\,{#2}}}
\newcommand{\sY}[2]{\Y_{{#1} \, {#2}}}
\newcommand{\sPsi}[2]{\mathrm{\Psi}^{{#1} \, {#2}}}
\newcommand{\sPhi}[2]{\mathrm{\Phi}^{{#1} \, {#2}}}
\newcommand{\sF}[2]{\F^{{#1}\,{#2}}}
\newcommand{\sX}[2]{X^{{#1}\,{#2}}}
\newcommand{\br}{\mathbf{r}}
\newcommand{\bn}{\mathbf{n}}
\newcommand{\Kdelta}[2]{\delta_{{#1} \, {#2}}}
\newcommand{\intdOmega}{\int \! \mathrm{d}\Omega \ }
\newcommand{\und}{&&\text{and}&&}
\newcommand{\komma}{, &&}

\newcommand{\bnabla}{\bm{\nabla}}
\newcommand{\mH}{\mathscr{H}}
\newcommand{\bM}{\mathbf{M}}
\newcommand{\bP}{\mathbf{P}}
\newcommand{\bV}{\mathbf{V}}
\newcommand{\bI}{\mathbf{I}}
\newcommand{\hId}{\hat{I}}
\newcommand{\hP}{\hat{P}}
\newcommand{\hQ}{\hat{Q}}
\newcommand{\hV}{\hat{V}}
\newcommand{\hW}{\hat{W}}
\newcommand{\hH}{\hat{H}}
\newcommand{\hM}{\hat{M}}
\newcommand{\hN}{\hat{N}}
\newcommand{\ta}{\tilde{\alpha}}
\newcommand{\tb}{\tilde{\beta}}
\newcommand{\hmA}{\hat{\mathcal{A}}}
\newcommand{\hmB}{\hat{\mathcal{B}}}
\newcommand{\hmD}{\hat{\mathcal{D}}}
\newcommand{\hmI}{\hat{\mathcal{I}}}
\newcommand{\hmP}{\hat{\mathcal{P}}}
\newcommand{\hmQ}{\hat{\mathcal{Q}}}
\newcommand{\hmV}{\hat{\mathcal{V}}}
\newcommand{\hmW}{\hat{\mathcal{W}}}
\newcommand{\hmL}{\hat{\mathcal{L}}}
\newcommand{\hmM}{\hat{\mathcal{M}}}
\newcommand{\hmN}{\hat{\mathcal{N}}}
\newcommand{\er}{\hat{\mathbf{e}}_r}
\newcommand{\etheta}{\hat{\mathbf{e}}_\theta}
\newcommand{\ephi}{\hat{\mathbf{e}}_\phi}
\newcommand{\z}{{(0)}}
\newcommand{\one}{{(1)}}
\newcommand{\two}{{(2)}}
\newcommand{\lz}{{l_0}}
\newcommand{\abs}[1]{\ensuremath{\left| #1 \right|}}
\newcommand{\norm}[1]{\lVert#1\rVert}
\newcommand{\ave}[1]{\langle#1\rangle}
\newcommand{\ket}[1]{|#1\rangle}
\newcommand{\bra}[1]{\langle#1|}
\newcommand{\brak}[2]{\langle#1|#2\rangle}
\newcommand{\proj}[2]{|#1\rangle \! \langle#2|}
\newcommand{\mean}[3]{\langle#1|#2|#3\rangle}

\renewcommand{\epsilon}{\varepsilon}
\newcommand{\order}[1]{\mathcal{O}(#1)}

\newcommand{\pdv}[2]{\frac{\partial^{#1}}{\partial {#2}^{#1}}}
\newcommand{\dv}[2]{\frac{\mathrm{d}^{#1}}{\mathrm{d} {#2}^{#1} }}

\newcommand{\be}{\mathbf{e}}
\newcommand{\bA}{\mathbf{A}}

\newcommand{\bpsi}{\boldsymbol{\psi}}
\newcommand{\bvarphi}{\boldsymbol{\varphi}}
\newcommand{\bolda}{\boldsymbol{a}}
\newcommand{\boldta}{\boldsymbol{\tilde{a}}}

\newcommand{\TE}{\text{TE}}
\newcommand{\TM}{\text{TM}}

\renewcommand*{\multicitedelim}{\addcomma}
\newcommand*\EmptyPage{\newpage\null\thispagestyle{empty}\newpage}
\numberwithin{equation}{section}
\numberwithin{figure}{section}

\thispagestyle{empty}
\begin{titlepage}
\centering
{\LARGE \bf{Perturbation Theory of Optical Resonances}}\par
\vspace{0.4cm}
{\LARGE \bf{of Deformed Dielectric Spheres}}\par
\vspace{2.5cm}
{\Large \bf{Master's Thesis in Physics}}\par
\vspace{2.5cm}
Presented by \par
\vspace{0.4cm}
{\Large \bf{Julius Gohsrich}}\par
\vspace{0.4cm}
September 1, 2020 \par
\vspace{2.5cm}
Friedrich–Alexander–Universität Erlangen–Nürnberg
\vspace{0.4cm}
\begin{figure}[H]
    \centering
    \includegraphics[width=0.5\linewidth]{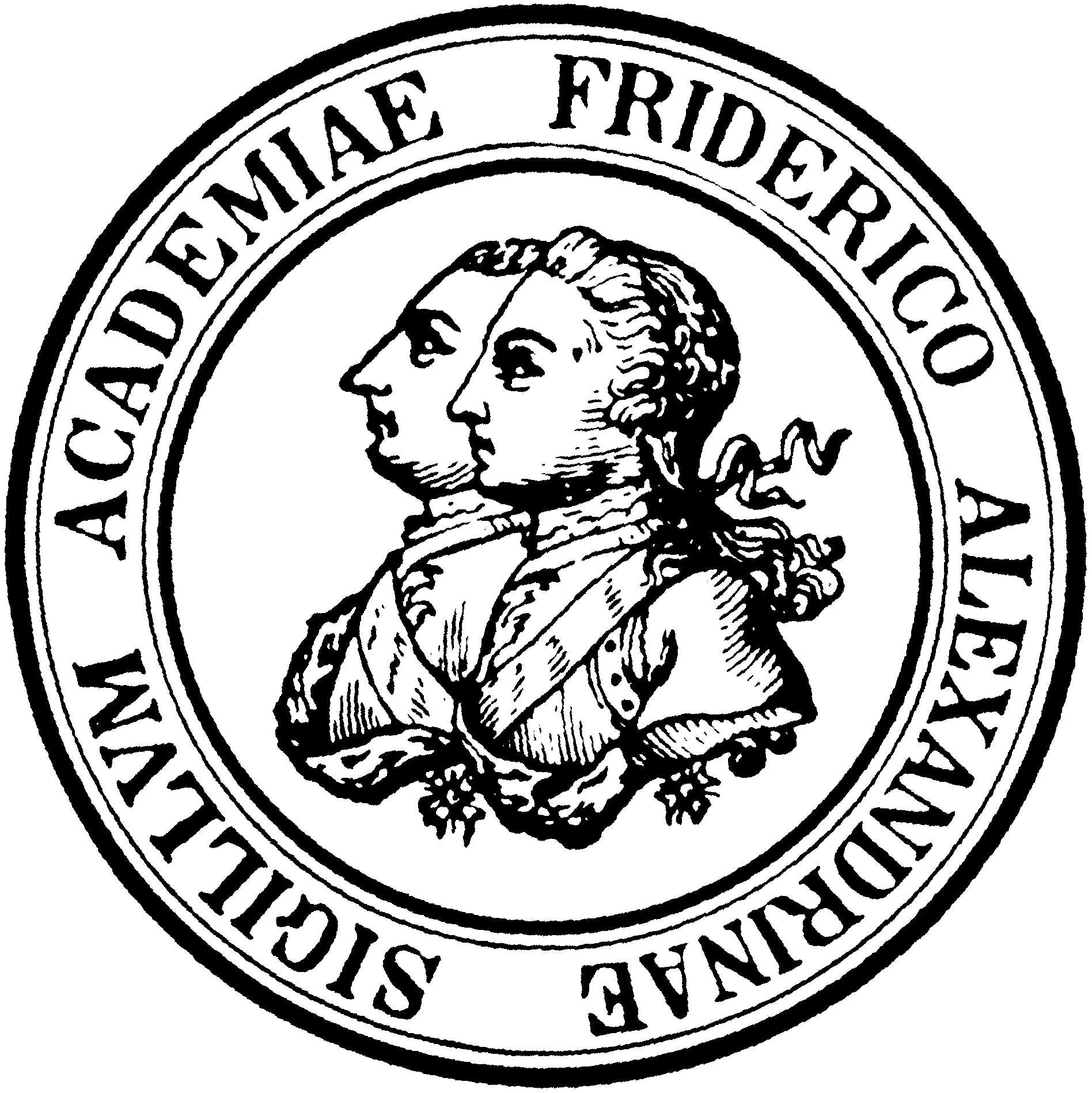}
\end{figure}
\vspace{0.4cm}
\begin{tabular}{rl}
Supervisor: & Prof. Dr. Florian Marquardt \\
Co-Supervisor: & Dr. habil. Andrea Aiello
\end{tabular}\par
\vspace{1.5cm}
\end{titlepage}
\newpage

\thispagestyle{empty}
\begin{abstract}
\noindent
Light injected into a spherical dielectric body may be confined very efficiently via the mechanism of total internal reflection. The frequencies that are most confined are called resonances. If the shape of the body deviates from the perfect spherical form the resonances change accordingly. In this thesis, a perturbation theory for the optical resonances of such a deformed sphere is developed. The optical resonances of such an open system are characterized by complex eigenvalues, where the real part relates to the frequency of the resonant light and the imaginary part to the energy leakage out of the system. As unperturbed and analytically solvable problem serves the homogeneous dielectric sphere, and the corrections to its eigenvalues are determined up to and including second order for any polarization of light. For each order, the corrections of the optical resonances are determined by a finite-dimensional linear eigenvalue equation, similar to degenerate time-independent perturbation theory in quantum mechanics. Furthermore, geometrically intuitive applicability criteria are derived. To check the validity of the presented method, it is applied and compared to an analytically solvable problem.
\end{abstract}
\newpage

\thispagestyle{empty}
\tableofcontents
\thispagestyle{empty}
\newpage

\setcounter{page}{1}
\section{Introduction}
\label{sec:intro}
The description of light scattering from dielectric bodies like, e.g., raindrops, atmospheric dust, crystals, et cetera, is an active topic for over a century. The first exact mathematical solution was derived by Gustav Mie in his seminal 1908 work \cite{mie08} concerning dielectric spheres. Until today, Mie theory remains an active topic with applications ranging from microscopic resonators to astrophysics \cite{hergert12}. \par
One of the most remarkable phenomenon described by this theory is the occurrence of optical resonances. A resonance manifests as a sharp peak in the total scattering cross-section of the scatterer and can be explained by resonances of the underlying structure. For a dielectric sphere, such resonant behavior can be explained by whispering gallery modes (WGMs), which propagate close to the inner surface due to near-total internal reflection \cite{oraevsky02,lai90,lai90-2}. As some light leaks out of the dielectric sphere, such a system can be described as an open system, characterized by complex eigenvalues of an associated non-Hermitian operator \cite{sternheim72,rotter09,brody13}.\par
\begin{figure}[htb]
    \centering
    \vspace{-0.5cm}
    \includegraphics[width=\textwidth]{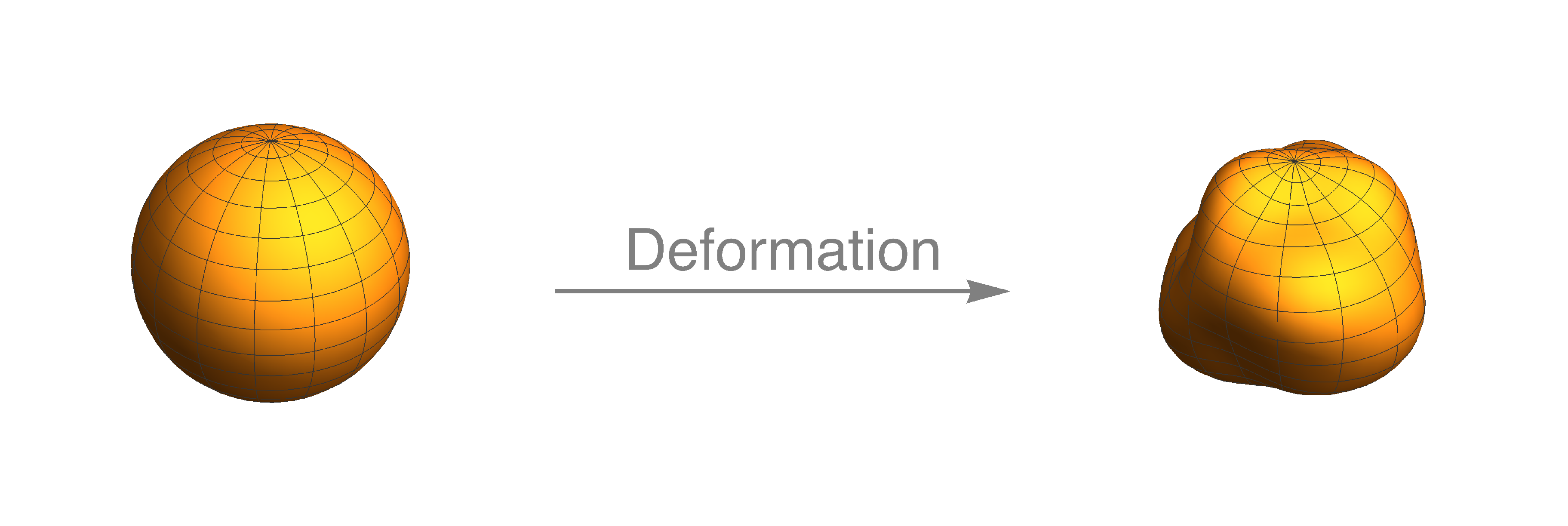}
    \vspace{-1cm}
    \caption{Representation of an artificially huge arbitrary deformation.}
    \label{fig:scheme3d}
\end{figure}

The WGMs are used in a plethora of optical and optomechanical applications. Microscopic glass spheres are used to realize biological, chemical and physical sensors \cite{foreman15,lin17}. In such systems, the WGMs probe the surface of the dielectric body as well as objects close to the surface, and small deviations from the spherical form drastically change the frequencies and the losses of the WGMs. Such deviations are the result of the manufacturing process as well as due to surface roughness and are in general unwanted. For optomechanical systems using levitated drops of liquids on the other hand, the surface of the droplet acts as mechanical resonator and therefore is needed to change its shape. Furthermore, flattening due to rotation, thermally excited capillary waves and other surface waves deforms the surface of such droplets \cite{childress17}. Figure \ref{fig:scheme3d} illustrates an arbitrarily deformed dielectric body. \par
Due to the increasing interest in such optical and optomechanical systems, a wealth of numerical \cite{greengard15,yan20} and perturbative methods have been developed to determine the resonances of such slightly deformed dielectric spheres. The most sophisticated perturbative methods are the resonant state expansion (RSE) \cite{lind93,muljarov10,doost14} and the Kapur-Peierls (KP) formalism \cite{aiello19}. Both of these methods originate in the quantum theory of scattering \cite{more73,johnson93}. The RSE uses an optical analogue of Gamow \cite{gamow28} or Siegert \cite{siegert39} states and employs Brillouin-Wigner perturbation theory. While this approach is well-suited for numerical considerations, an order of perturbation, and thus the notion of error estimation, does not exist. On the other hand, the KP formulation employs Rayleigh-Schr\"odinger perturbation theory which does not suffer from this weakness, however it makes use of some strong assumptions and is not able to predict corrections of transverse magnetic modes. Furthermore, both approaches do not deliver applicability criteria. \par
In order to overcome those shortcomings, we want to employ another perturbative method called \emph{\emph{B}oundary \emph{C}ondition \emph{P}erturbation \emph{T}heory} (BCPT), the progenitor of the popular Rayleigh-Schr\"odinger perturbation theory. It was originally developed by Lord Rayleigh in his book \emph{The Theory of Sound} \cite{rayleigh94}, where he investigated how the mechanical resonances of a circular membrane change if it gets slightly deformed. Recently, this method has successfully been applied to effective two-dimensional cavities, which slightly deviate from a circular form \cite{dubertrand08,ge13,badel19}. \par
The general idea of BCPT is quickly outlined: First one needs to find the general solution of the electromagnetic fields, solely depending on a set of field coefficients and an eigenvalue. The boundary condition encodes the geometry and thus some small parameter. By inserting the general solution into the boundary condition and expanding the boundary condition around the small parameter, one finds a chain of equations relating the field coefficients, the resonances, and their respective corrections. Solving this chain of equations order by order determines the optical resonances. This approach differs significantly from the ones in the current literature, where the boundary condition is brought into a particular simple form allowing to shift the problem to an expansion of the electromagnetic fields. \par
Employing BCPT we are able to determine the optical resonances up to and including second-order corrections for any polarization of light including geometrically intuitive applicability criteria. We remark that, at the best of our knowledge, the second-order perturbative solutions for both $\TE$- and $\TM$-polarizations of the electromagnetic fields, were never derived, in a correct form, before. Thus, the results presented in this thesis are perfectly original and solve a long-standing open problem. \par
This thesis is structured as follows. In Section \ref{sec:perfect} we determine the electromagnetic fields solving Maxwell's equations for a dielectric body embedded in another dielectric material. Considering the boundary condition appropriate for the dielectric sphere we determine its optical resonances. Having solved the unperturbed problem, we generalize our approach in Section \ref{sec:deformed} by considering a more general boundary condition, encoding the geometry of a more general dielectric body, and bring it to a form suitable to apply BCPT. In Sections \ref{sec:solutionTE} and \ref{sec:solutionTM} we finally apply BCPT and determine the optical resonances of such a dielectric body up to and including second-order corrections for both polarizations of light. In Section \ref{sec:firstExplicit} we rewrite the first-order equations in a more explicit way and reason the validity of our method by applying and compare it to an analytically solvable problem in Section \ref{sec:shrink}. We conclude our work in Section \ref{sec:summary}.
\newpage

\section{Resonances of a Dielectric Sphere}
\label{sec:perfect}
Before we tackle the hard problem of finding the optical resonances of a deformed dielectric sphere, we need to understand the general physical framework and the \emph{unpertubed} problem, the dielectric sphere. \par
We start by adapting Maxwell's equations to our problem in Section \ref{sec:maxwell}. Following that, we introduce the Debye potentials to derive a general solution for Maxwell's equations in Section \ref{sec:debye}. Afterwards we refine the Debye potentials so that they describe physical fields in Section \ref{sec:phfields} and finally find the optical resonances for a perfectly spherical body in Section \ref{sec:perfectsolution}.
\subsection{Maxwell's Equations for a Dielectric Body}
\label{sec:maxwell}
In classical electrodynamics, Maxwell's equations describe the evolution of electromagnetic fields. Considering the fields inside a dielectric medium, Maxwell's equations read
\begin{align}
    \begin{split}
        \bnabla \cdot \bD &= 0 , \\
        \bnabla \cdot \bB &= 0 , \\
        \bnabla \times \bE &= - \dot{\bB} , \\
        \bnabla \times \bH &= \dot{\bD} ,
    \end{split}
\end{align}
where the dot denotes the time derivative and all vector fields depend on $(\br,t)$. Let us consider a dielectric body, the interior of which defines a region $A$, made of a homogeneous isotropic dielectric material (medium~$1$) surrounded by another one (medium $2$). Both media are separated by the interface denoted $\partial A$. As we have two regions, we label them by $\alpha$, so $\alpha = 1$ corresponds to medium $1$ inside $A$, and $\alpha = 2$ corresponds to medium $2$ outside $A$. \par
Maxwell's equations in this form are incomplete and need to be completed by the 
constitutive equations, which relate the electromagnetic fields $\bE_\alpha$ and~$\bB_\alpha$ with $\bD_\alpha$ and $\bH_\alpha$. In this work we assume a linear response described by
\begin{align}
    \bD_\alpha = \epsilon_0 \epsilon_\alpha \bE_\alpha, \und \bB_\alpha = \mu_0 \mu_\alpha \bH_\alpha .
\end{align}
Here, $\epsilon_0$ is the vacuum permittivity, $\epsilon_\alpha$ the relative permittivity, $\mu_0$ the vacuum permeability and $\mu_\alpha$ the relative permeability. As the speed of light in vacuum is $c_0 = (\epsilon_0 \, \mu_0)^{-1/2}$ and the speed of light in matter is $c_\alpha = c_0 \, (\epsilon_\alpha \, \mu_\alpha)^{-1/2}$, we introduce the refractive index
\begin{align}
    n_\alpha = c_0/c_\alpha = \sqrt{\epsilon_\alpha \, \mu_\alpha} .
\end{align}
Throughout this work, we will assume that $n_1>n_2$. Collecting the previous statements and assuming time-harmonic fields, i.e., all fields vary as $\exp(-i \omega t)$, we can rewrite Maxwell's equations as
\begin{subequations}
    \label{eq:TimeIndepMaxwellEB}
    \begin{align}
        \bnabla \cdot \bE_\alpha &= 0 \label{eq:Maxa} , \\
        \bnabla \cdot \bB_\alpha &= 0 \label{eq:Maxb} , \\
        \bnabla \times \bE_\alpha &= i \omega \, \bB_\alpha \label{eq:Maxc} , \\
        \bnabla \times \bB_\alpha &= - (i \omega / c_\alpha^2) \, \bE_\alpha \label{eq:Maxd} ,
    \end{align}
\end{subequations}
where the fields lost their time dependence and only depend on $\br$. In this form it is apparent that Maxwell's equations are invariant under the discrete transformation
\begin{align}
    \label{eq:dualSymmetry}
    \bE_\alpha \rightarrow c_\alpha \, \bB_\alpha, \und \bB_\alpha \rightarrow -\bE_\alpha/c_\alpha ,
\end{align}
which is called dual symmetry and will help us finding new results out of old ones. \par
Sometimes it is useful to decouple Maxwell's equations. By taking the $\text{curl}$ of \eqref{eq:Maxc}, using the vector identity $\bnabla \times (\bnabla \times \bE) = - \Laplace\bE + \bnabla(\bnabla \cdot \bE)$ on the left hand side of the equation and inserting \eqref{eq:Maxd} on the right side of the equation, one finds the vector Helmholtz equation
\begin{align}
    \label{eq:vectorHelmholtzEquation}
    \left( \Laplace + \frac{\omega^2}{c_\alpha^2} \right) \bE_\alpha = 0 ,
\end{align}
which can be physically interpreted as time-independent wave equation for the electric field. Employing the dual symmetry \eqref{eq:dualSymmetry}, one finds that the magnetic field also satisfies the Helmholtz equation.\par
If one assumes plane wave solutions of \eqref{eq:vectorHelmholtzEquation}, i.e., the fields are proportional to \linebreak $\exp(i \, \mathbf{k}_\alpha \cdot \br)$, Helmholtz's equation gives us the well-known dispersion relation
\begin{align}
    \omega = c_\alpha \, k_\alpha \equiv c_0 \, k_0 ,
\end{align}
where the wave number in matter $k_\alpha$ is given by $k_\alpha = \abs{\mathbf{k}_\alpha}$ and $k_0$ denotes the vacuum wave number. \par
As partial differential equations, Maxwell's equations need to be accompanied by boundary conditions. In the context of electromagnetic interfaces, these are also called matching conditions and can be determined to be
\begin{align}
    \label{eq:generalBoundaryCondition}
    \bn \times \left. \left( \bE_2 - \bE_1 \right) \right|_{\partial A} = 0, \und \bn \times \left. \left( \bB_2 - \bB_1 \right) \right|_{\partial A} = 0 ,
\end{align}
where $\bn$ is the normal of $\partial A$.

\subsection{Debye Potentials}
\label{sec:debye}
Now we want to find a general solution for Maxwell's equations \eqref{eq:TimeIndepMaxwellEB}. Every divergenceless electromagnetic field can be split into \cite{gray78}
\begin{subequations}
    \label{eq:EBL}
    \begin{align}
        \bE = i \bL u^\TE(\br) -\frac{1}{k} (\bnabla \times \bL) \, u^\TM(\br) , \label{eq:EL} \\[4pt]
        c \bB = i \bL u^\TM(\br) + \frac{1}{k} (\bnabla \times \bL) \, u^\TE(\br) , \label{eq:BL}
    \end{align}
\end{subequations}
where $\bL$ is the orbital angular momentum operator\footnote{A list of all properties of the angular momentum operator used in the work can be found in Appendix~B of \cite{gray78}.}
\begin{align}
    \label{eq:bL}
    \bL = -i \br \times \bnabla.
\end{align}
The scalar fields $u^\TE$ and $u^\TM$ are the so-called (scalar) \emph{Debye potentials} \cite{debye09} and respectively encode two different polarized fields, namely the \emph{transverse electric} ($\TE$)
\begin{align}
    \label{eq:EBTE}
    \bE^\TE = i \bL u^\TE(\br), && c\bB^\TE = \frac{1}{k} (\bnabla \times \bL) \, u^\TE(\br),
\end{align}
and \emph{transverse magnetic} ($\TM$) ones
\begin{align}
    \label{eq:EBTM}
    c\bB^\TM = i \bL u^\TM(\br), && \bE^\TM = -\frac{1}{k} (\bnabla \times \bL) \, u^\TM(\br).
\end{align}
The names come from the fact that the $\TE$ polarized part of the electric as well as the $\TM$ polarized part of the magnetic field are transverse to the radial direction $\er$, i.e.
\begin{align}
    \er \cdot \bE^\TE = 0 , && \er \cdot \bB^\TM = 0.
\end{align}
This is also called a toroidal field. In addition to that, the respective remaining part of the electromagnetic field are poloidal fields satisfying
\begin{align}
    \er \cdot( \bnabla \times \bE^\TM) = 0, && \er \cdot( \bnabla \times \bB^\TE) = 0.
\end{align}
Furthermore, the fields of the same polarization are orthogonal to each other
\begin{align}
    \bE^\TE \cdot \bB^\TE = 0, && \bE^\TM \cdot \bB^\TM = 0.
\end{align}
\par
The fields described by \eqref{eq:EBL} satisfy the first two Maxwell equations (\ref{eq:Maxa}-\ref{eq:Maxb}) by definition, as
\begin{align}
    \nabla \cdot \bL = 0, && \bnabla \cdot (\bnabla \times \bL) = 0.
\end{align}
The remaining equations (\ref{eq:Maxc}-\ref{eq:Maxd}) are fulfilled if the Debye potentials satisfy the scalar Helmholtz equation
\begin{align}
    \label{uScalarHelmholtz}
    (\Laplace + k^2) \, u^\sigma(\br) = 0,
\end{align}
where $\sigma =\TE,\TM$ denotes the polarization. Thus we found that by introducing the Debye potentials, we reduced the two vector Helmholtz equations \eqref{eq:vectorHelmholtzEquation} for the electromagnetic fields to two scalar Helmholtz equations for the Debye potentials. \par
Before we actually solve this equation, let us write the electromagnetic fields in terms of the Debye potentials in a more suitable way. First we notice from the form of \eqref{eq:EBL} that the dual symmetry of Maxwell's equations \eqref{eq:dualSymmetry} can be expressed using the Debye potentials as
\begin{align}
    \label{eq:dualSymmetry2}
    c\bB[u^\TM, u^\TE] = \bE[u^\TE, -u^\TM] ,
\end{align}
where the square brackets denote functional dependence. With this property we can restrict our discussions to one of the two fields. As the next step, we expand the Debye potentials in their Laplace series \eqref{eq:LaplaceSeries} as
\begin{align}
    \label{eq:uSigmaLaplace}
    u^\sigma(\br) = \sum_{l=0}^\infty \sum_{m=-l}^l u_{l \, m}^\sigma(r) \sY{l}{m}(\theta,\phi),
\end{align}
where $\sY{l}{m}(\theta,\phi)$ are the scalar spherical harmonics and we denote with $u_{l \, m}^\sigma(r)$ the reduced Debye potentials. Inserting Laplace's series into $\bE^\TE$ from \eqref{eq:EBTE} we find
\begin{align}
    \bE^\TE &= \sum_{l=0}^\infty \sum_{m=-l}^l i \bL \left(u_{l \, m}^\TE(r) \sY{l}{m}(\theta,\phi) \right)\notag \\
    &= \sum_{l=0}^\infty \sum_{m=-l}^l u_{l \, m}^\TE(r) \, \left(i \bL \sY{l}{m}(\theta,\phi) \right) \notag \\
    &\equiv \sum_{l=0}^\infty \sum_{m=-l}^l u_{l \, m}^\TE(r) \, \bPhi{l}{m}(\theta,\phi) , \label{eq:ETEVSH}
\end{align}
where in the second line we used the fact that $\bL$ commutes with any functions solely depending on $r$ and in the third line we introduced the vector quantity $\bPhi{l}{m}$, which will be defined soon. Using
\begin{align}
    \bnabla \times \bL =  \er \frac{i}{r} \Lsquared + \left(\er \times \bL\right) \, \frac{1}{r} \frac{\mathrm{d}}{\mathrm{d}r} r ,
\end{align}
where $\Lsquared = \bL \cdot \bL$, we can rewrite the transverse magnetic part of the electric field in \eqref{eq:EBTM} as
\begin{align}
    \bE^\TM &= \sum_{l=0}^\infty \sum_{m=-l}^l \bigg\{ -\frac{1}{k} \left(\bnabla \times \bL\right) \left(u_{l \, m}^\TM(r) \, \sY{l}{m}(\theta,\phi)\right)\bigg\} \notag \\[6pt]
    &= \sum_{l=0}^\infty \sum_{m=-l}^l \bigg\{ -\frac{i}{kr} \bigg[ l(l+1) \, u_{l \, m}^\TM(r) \left(\er \sY{l}{m}(\theta,\phi)\right) \notag \\[4pt]
    &\hphantom{\mathrel{=} \sum_{l=0}^\infty \sum_{m=-l}^l \bigg\{ -\frac{i}{kr} \big[}+ \frac{\mathrm{d}}{\mathrm{d}r}\left(r \, u_{l \, m}^\TM(r)\right) \left(-i \er \times \bL \sY{l}{m}(\theta,\phi)\right) \bigg] \bigg\} \notag \\[6pt]
    &\equiv \sum_{l=0}^\infty \sum_{m=-l}^l \left\{ -\frac{i}{kr} \left[ l(l+1) u_{l \, m}^\TM(r) \bY{l}{m}(\theta,\phi) + \frac{\mathrm{d}}{\mathrm{d}r}\left(r \, u_{l \, m}^\TM(r)\right) \bPsi{l}{m}(\theta,\phi) \right] \right\},
\end{align}
where in the second line, we used $\Lsquared\sY{l}{m}=l(l+1)\sY{l}{m}$ from \eqref{eq:LsquaredY}. In the last line and in \eqref{eq:ETEVSH} we introduced the three vector quantities
\begin{align}
    \bY{l}{m}(\theta,\phi) &= \er \sY{l}{m}, &&
    \bPsi{l}{m} = r \, \bnabla \sY{l}{m} \und
    \bPhi{l}{m} = (\br \times \bnabla) \sY{l}{m},
\end{align}
called vector spherical harmonics. It can be shown that the vector spherical harmonics are orthogonal \cite{barrera85} and complete \cite{kristensson16} and therefore permit a \emph{multipole expansion} of the electromagnetic fields as
\begin{subequations}
    \label{eq:EBDebye}
    \begin{align}
        \bE_\alpha &= \sum_{l=1}^\infty \sum_{m=-l}^l \left\{ u_{\alpha \, l \, m}^\TE \bPhi{l}{m} - \frac{i}{k_\alpha r} \left[ l(l+1) u_{\alpha \, l \, m}^\TM \bY{l}{m} + \frac{\mathrm{d}}{\mathrm{d}r} \left( r \, u_{\alpha \, l \, m}^\TM \right) \bPsi{l}{m} \right] \right\} , \label{eq:EDebye} \\[6pt]
        c_\alpha \,\bB_\alpha &= \sum_{l=1}^\infty \sum_{m=-l}^l \left\{u_{\alpha \, l \, m}^\TM \bPhi{l}{m} + \frac{i}{k_\alpha r} \left[ l(l+1) u_{\alpha \, l \, m}^\TE \bY{l}{m} + \frac{\mathrm{d}}{\mathrm{d}r} \left( r \, u_{\alpha \,l \, m}^\TE \right) \bPsi{l}{m} \right] \right\} , \label{eq:BDebye}
    \end{align}
\end{subequations}
where we reintroduced the label $\alpha$ and used \eqref{eq:dualSymmetry2} to get the magnetic field. Furthermore we want to draw attention to the fact that the sum over $l$ actually starts at $l=1$, since $\bPhi{0}{0}$, $\bPsi{0}{0}$ as well as the prefactor of $\bY{0}{0}$ vanish. Up to this point, the reduced Debye potentials are not yet specified.
\subsection{Physical Fields}
\label{sec:phfields}
In the previous section, we brought the electromagnetic fields in the form \eqref{eq:EBDebye} and showed that they satisfy Maxwell's equations if the Debye potentials $u^\sigma(\br)$ satisfy the scalar Helmholtz equation \eqref{uScalarHelmholtz}. Multiplying \eqref{uScalarHelmholtz} by $r^2$ and rewriting the Laplacian in terms of the angular momentum operator using \eqref{eq:laplacianL}, we find
\begin{align}
    \left[\frac{\partial}{\partial r} \left(r^2 \frac{\partial}{\partial r}\right)
    + (kr)^2 - \Lsquared \right]\, u^\sigma(\br) = 0.
\end{align}
Inserting Laplace's series for $u^\sigma$ from \eqref{eq:uSigmaLaplace} and using the product rule results in
\begin{align}
    r^2 \frac{\partial^2}{\partial r^2} u_{l \, m}^\sigma(r) + 2r \frac{\partial}{\partial r} u_{l \, m}^\sigma(r)
    + \left[(kr)^2 - l(l+1) \right] u_{l \, m}^\sigma(r)  = 0,
\end{align}
and finally substituting $r \rightarrow kr \equiv z$ one finds
\begin{align}
    z^2 \frac{\partial^2}{\partial z^2} u_{l \, m}^\sigma(z) + 2z \frac{\partial}{\partial z} u_{l \, m}^\sigma(z)
    + \left[z^2 - l(l+1) \right] u_{l \, m}^\sigma(z) = 0,
\end{align}
which is the \emph{spherical Bessel (differential) equation} (cf. App. \ref{app:bessel}). Its fundamental solutions are the spherical Bessel functions $j_l(z)$ and $y_l(z)$ and the general solution is the superposition of both
\begin{align}
    \label{eq:ujy}
    u_{l \, m}^\sigma(z) &= A_{l \, m}^\sigma \, j_l(z) + B_{l \, m}^\sigma \, y_l(z).
\end{align}
Another way to express this solution is by employing the spherical Hankel functions $h_l^{(1)}(z)$ and $h_l^{(2)} (z)$ from \eqref{eq:sphHankelDef} to write
\begin{align}
    \label{eq:uhh}
    u_{l \, m}^\sigma(z) &= S_{l \, m}^\sigma \, h_l^{(1)}(z) + I_{l \, m}^\sigma \, h_l^{(2)}(z).
\end{align}
The physical solutions however are more restrictive. Let us first consider the reduced Debye potentials inside the dielectric body $A$ ($\alpha = 1$). Here we have to require that the Debye potentials are everywhere regular. As the spherical Bessel function of second kind $y_l(z)$ diverges at the origin ($y_l(z) \propto 1/z^{l+1}$ for $z \rightarrow 0$), the everywhere regular solution is given by \eqref{eq:ujy} with $B_{l \, m}^\sigma = 0$. \par
For the Debye potential outside $A$ ($\alpha = 2$), the linear combination of the spherical Hankel functions \eqref{eq:uhh} is suited best, as in the far field, they describe outgoing spherical waves ($h_l^{(1)}(z) \propto e^{iz}/z$ for $z \to \infty$) and incoming spherical waves ($h_l^{(2)} \propto e^{-iz}/z$ for $z \to \infty$). Thus, the physical solutions need to be of the form
\begin{subequations}
    \label{eq:uReducedAlpha}
    \begin{align}
        u_{1, l \, m}^\sigma(k_1 r) &= A_{l \, m}^\sigma \, j_l(k_1 r) , & \alpha &= 1,\\[4pt]
        u_{2, l \, m}^\sigma(k_2 r) &= S_{l \, m}^\sigma \, h_l^{(1)}(k_2 r) + I_{l \, m}^\sigma \, h_l^{(2)}(k_2 r) , & \alpha &= 2.
    \end{align}
\end{subequations}
These reduced Debye potentials, and therefore the corresponding electromagnetic fields, describe a scattering process: The incident field is encoded in $I_{l \, m}^\sigma$, the scattered field in $S_{l \, m}^\sigma$ and the internal field in $A_{l \, m}^\sigma$. Such a scattering problem can be symbolically solved by introducing the transition matrix $T$ connecting the incident field coefficients with the scattered field coefficients via
\begin{align}
    S_{l \, m}^\sigma = \sum_{\sigma' \, l' \, m'} T_{l \, l' \, m \, m'}^{\sigma \, \sigma'} \ I_{l' \, m'}^{\sigma'},
\end{align}
and the boundary condition \eqref{eq:generalBoundaryCondition} encoding the geometry of the problem fully determines the transition matrix \cite{kristensson16}. This $T$-matrix encodes all physical relevant information, however, only the $T$-matrix of the dielectric sphere is analytically known in our setup. \par
The resonances, characterized by sharp peaks in the total scattering cross-section, are associated to the analytic continuation of the $T$-matrix \cite{taylor72,kristensson16}. As the $T$-matrix, and therefore its poles, is in general not analytically known, one has to find a different approach to determine the resonances. \par
One possibility to determine the resonances was introduced by Gamow in the context of quantum mechanical scattering theory \cite{gamow28}. By imposing outgoing waves only, in our case setting $I_{l \, m}^\sigma$ in \eqref{eq:uReducedAlpha} to zero, one can determine the resonances without determining the $T$-matrix. For electromagnetic scattering, this idea was already used in \cite{muljarov10,doost14} to construct the resonant-state expansion. Using Gamow's approach has two important implications for us. The first one is that the resonances are characterized by complex eigenvalues, where the imaginary part relates to the energy leakage out of the dielectric body $A$. This is in contrast to the scattering solutions described in the previous paragraphs, where the wave numbers are real. The second implication is that the associated electromagnetic fields are not physical as they are not normalizable in the standard sense\cite{bohm89,delamadrid12}. Usually, this problem needs to be addressed by constructing some non-standard normalization in order to determine the resonances. In this work however, we find the resonances without introducing such a normalization. We demonstrate this fact for the dielectric sphere in the next section. \par
Before we do so, let us write down the electromagnetic fields for $I_{l \, m}^\sigma=0$ explicitly. Therefore we insert \eqref{eq:uReducedAlpha} into \eqref{eq:EBDebye}. To distinguish the Gamow approach from the scattering approach, we rename $A_{l \, m}^\sigma \rightarrow a_{l \, m}^\sigma$ and $S_{l \, m}^\sigma \rightarrow b_{l \, m}^\sigma$ and find for the fields inside the dielectric body
\begin{align}
    \label{eq:Einside}
    \hphantom{c_1 \,} \bE_1 = \sum_{l=1}^\infty \sum_{m=-l}^l \bigg\{ &a_{l\,m}^\text{TE} \bigg[ \, j_l(k_1 r) \bPhi{l}{m} \bigg] \notag \\[4pt]
    &- i \, a_{l \, m}^\text{TM} \bigg[ l(l+1) \frac{j_l(k_1 r)}{k_1 r} \bY{l}{m} + \frac{[(k_1 r) \, j_l(k_1 r)]'}{k_1 r} \bPsi{l}{m} \bigg] \bigg\},
\end{align}
and
\begin{align}
    \label{eq:cBinside}
    c_1 \,\bB_1 = \sum_{l=1}^\infty \sum_{m=-l}^l \bigg\{& a_{l \, m}^\text{TM} \bigg[ j_l(k_1 r) \bPhi{l}{m} \bigg] \notag \\[4pt]
    &+ i \, a_{l \, m}^\text{TE} \bigg[ l(l+1) \frac{j_l(k_1 r)}{k_1 r} \bY{l}{m} + \frac{[(k_1 r) \, j_l(k_1 r)]'}{k_1 r} \bPsi{l}{m} \bigg] \bigg\},
\end{align}
where we rewrote the $\Psi$-component using
\begin{align}
    \frac{\mathrm{d}}{\mathrm{d}r} (r f(k r))
    &= \frac{\mathrm{d}}{\mathrm{d}(k r)} \, [(k r) f(k r)] \notag \\
    &\equiv [(k r) f(k r)]' .
\end{align}
It can be shown that the vector quantities in the square brackets in \eqref{eq:Einside} and \eqref{eq:cBinside} are part of a complete set called \emph{regular spherical vector waves} \cite{kristensson16}. \par
Similarly one has for the electromagnetic fields outside the dielectric body
\begin{align}
    \label{eq:Eoutside}
    \hphantom{c_2 \, } \bE_2 = \sum_{l=1}^\infty \sum_{m=-l}^l \bigg\{ & b_{l\,m}^\text{TE} \bigg[ h_l(k_2 r) \bPhi{l}{m} \bigg] \notag \\
    &- i \, b_{l \, m}^\text{TM} \bigg[ l(l+1) \frac{h_l(k_2 r)}{k_2 r} \bY{l}{m} + \frac{[(k_2 r) \, h_l(k_2 r)]'}{k_2 r} \bPsi{l}{m} \bigg] \bigg\},
\end{align}
and
\begin{align}
    \label{eq:cBoutside}
        c_2 \,\bB_2 = \sum_{l=1}^\infty \sum_{m=-l}^l \bigg\{ &b_{l \, m}^\text{TM} \bigg[ h_l(k_2 r) \bPhi{l}{m} \bigg] \notag \\
        &+ i \, b_{l \, m}^\text{TE} \bigg[ l(l+1) \frac{h_l(k_2 r)}{k_2 r} \bY{l}{m} + \frac{[(k_2 r) \, h_l(k_2 r)]'}{k_2 r} \bPsi{l}{m} \bigg] \bigg\},
\end{align}
where we dropped the superscript $(1)$ in $h_l^{(1)}$ as the spherical Hankel function of second kind $h_l^{(2)}$ does not occur anymore. Likewise to the regular spherical vector waves, the quantities in square bracket are denoted \emph{out-going} or \emph{radiating} spherical vector waves. \par
To conclude this section let us remark that at no point in the derivation of the electromagnetic fields we made any assumption on the form of the dielectric body. Thus, all information needs to be encoded in the field coefficients $a_{l \, m}^\sigma$ and $b_{l \, m}^\sigma$, which will be determined by imposing the electromagnetic boundary conditions \eqref{eq:generalBoundaryCondition} at the interface between the two dielectric media.
\subsection{Resonances of the Dielectric Sphere}
\label{sec:perfectsolution}
Let us start to develop our perturbative approach by finding the resonances of the dielectric sphere with radius $r=r_0$. To do so we need to rewrite the general boundary conditions \eqref{eq:generalBoundaryCondition}. Using the fact that the normal is given by $\bn = \er$, the boundary conditions can evidently be rewritten as
\begin{align}
    \label{eq:sphericalBoundaryCondition}
    \er \times \left. \left( \bE_2 - \bE_1 \right) \right|_{r=r_0} = 0, \und \er \times \left. \left( \bB_2 - \bB_1 \right) \right|_{r=r_0} = 0 .
\end{align}
Inserting the multipole expansions (\ref{eq:Einside},\ref{eq:cBinside},\ref{eq:Eoutside},\ref{eq:cBoutside}) into these boundary conditions, we find
\begin{align}
    \label{eq:sphericalBoundaryE}
    0 = \sum_{l=1}^\infty \sum_{m=-l}^l \bigg\{ &\bigg[ a_{l\,m}^\text{TE} \, j_l(k_1 r_0) - b_{l\,m}^\text{TE} \, h_l(k_2 r_0)  \bigg] \bPsi{l}{m} \notag \\
    &+ i\bigg[ a_{l \, m}^\text{TM} \frac{[(k_1 r_0) \, j_l(k_1 r_0)]'}{k_1 r_0} - b_{l \, m}^\text{TM} \frac{[(k_2 r_0) \, h_l(k_2 r_0)]'}{k_2 r_0} \bigg] \bPhi{l}{m} \bigg\},
\end{align}
and
\begin{align}
    \label{eq:sphericalBoundaryB}
    0 = \sum_{l=1}^\infty \sum_{m=-l}^l \bigg\{ &\bigg[ \frac{a_{l\,m}^\text{TM}}{c_1} \, j_l(k_1 r_0) - \frac{b_{l\,m}^\text{TM}}{c_2} \, h_l(k_2 r_0)  \bigg] \bPsi{l}{m} \notag \\
    &- i\bigg[ \frac{a_{l \, m}^\text{TE}}{c_1} \, \frac{[(k_1 r_0) \, j_l(k_1 r_0)]'}{k_1 r_0} - \frac{b_{l \, m}^\text{TE}}{c_2} \, \frac{[(k_2 r_0) \, h_l(k_2 r_0)]'}{k_2 r_0} \bigg] \bPhi{l}{m} \bigg\} ,
\end{align}
where we used the cross products of $\er$ with the vector spherical harmonics given by \eqref{eq:ercrossvsh}
\begin{align}
    \er \times \bY{l}{m} = 0 \komma  \er \times \bPsi{l}{m} = \bPhi{l}{m}, \und \er \times \bPhi{l}{m} = - \bPsi{l}{m} .
\end{align}
From the orthogonality of the vector spherical harmonics \eqref{eq:VSHorthogonality} it follows that each square bracket in \eqref{eq:sphericalBoundaryE} and \eqref{eq:sphericalBoundaryB} that multiplies a vector spherical harmonics must vanish. Due to the linear independence of $j_l$ and $h_l$, the first line of \eqref{eq:sphericalBoundaryE} results in
\begin{align}
    \label{eq:TE-E}
    a_{l\,m}^\text{TE} = a_{l\,m}^\text{E} \frac{1}{j_l(k_1 r_0)},
    \und
    b_{l\,m}^\text{TE} = a_{l\,m}^\text{E} \frac{1}{h_l(k_2 r_0)} ,
\end{align}
and the first line of \eqref{eq:sphericalBoundaryB} gives
\begin{align}
    \label{eq:TM-M}
    a_{l\,m}^\text{TM} = a_{l\,m}^\text{M} \frac{1}{n_1 \, j_l(k_1 r_0)},
    \und
    b_{l\,m}^\text{TM} = a_{l\,m}^\text{M} \frac{1}{n_2 \, h_l(k_2 r_0)},
\end{align}
where $a_{l\,m}^\text{E}$ and $a_{l\,m}^\text{M}$ are arbitrary complex constants depending on $l$ and $m$. Substituting \eqref{eq:TM-M} into the second line of \eqref{eq:sphericalBoundaryE} we obtain the condition for the TE modes
\begin{align}
    \label{eq:TETranscendental}
    \frac{[(k_1 r_0) j_l(k_1 r_0)]'}{ j_l(k_1 r_0)} - \frac{[(k_2 r_0) h_l(k_2 r_0)]'}{ h_l(k_2 r_0)} = 0 ,
\end{align}
and by inserting \eqref{eq:TE-E} into the second line of \eqref{eq:sphericalBoundaryB}, we likewise find the condition for the TM modes
\begin{align}
    \label{eq:TMTranscendental}
    \frac{[(k_1 r_0) \, j_l(k_1 r_0)]'}{n_1^2 \, j_l(k_1 r_0)} - \frac{[(k_2 r_0) \, h_l(k_2 r_0)]'}{n_2^2 \, h_l(k_2 r_0)} = 0 .
\end{align}
These equations precisely describe the values for $k_0=k_\alpha/n_\alpha$, for which the $T$-matrix introduced in the previous section becomes singular \cite{kristensson16}. Thus by assuming outgoing waves only, we found the resonances of the dielectric sphere without the need to determine the $T$-matrix first. Furthermore, an overall normalization of the electromagnetic field is not needed. \par
\begin{figure}[p]
    \centering
    \subfloat[Contour plot for TE-modes]{
        \includegraphics[width=\textwidth]{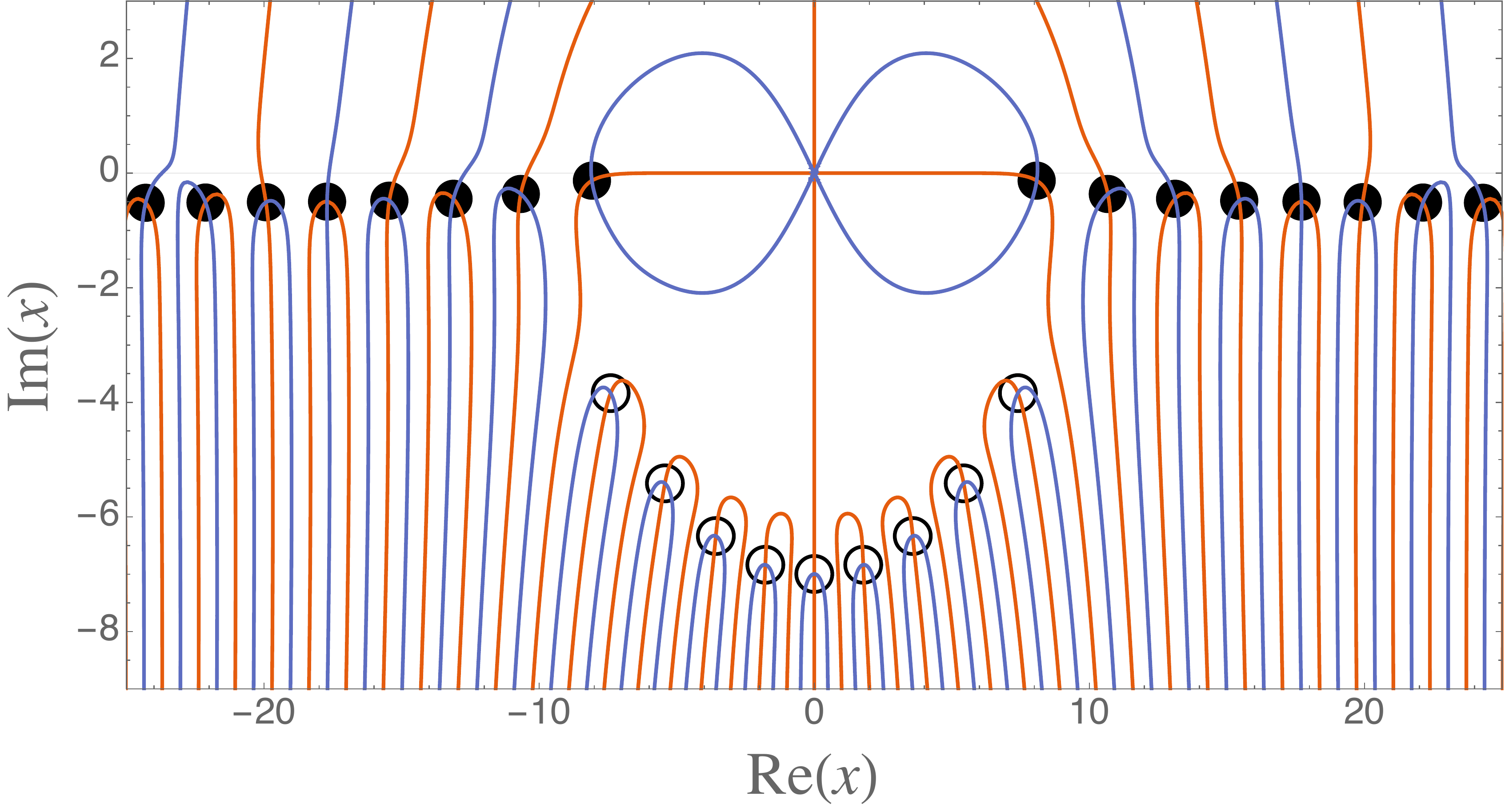}
        \label{fig:ContourTE}
    }
    \hspace{0.7cm}
    \hfill
    \subfloat[Contour plot for TM-modes]{
        \includegraphics[width=\textwidth]{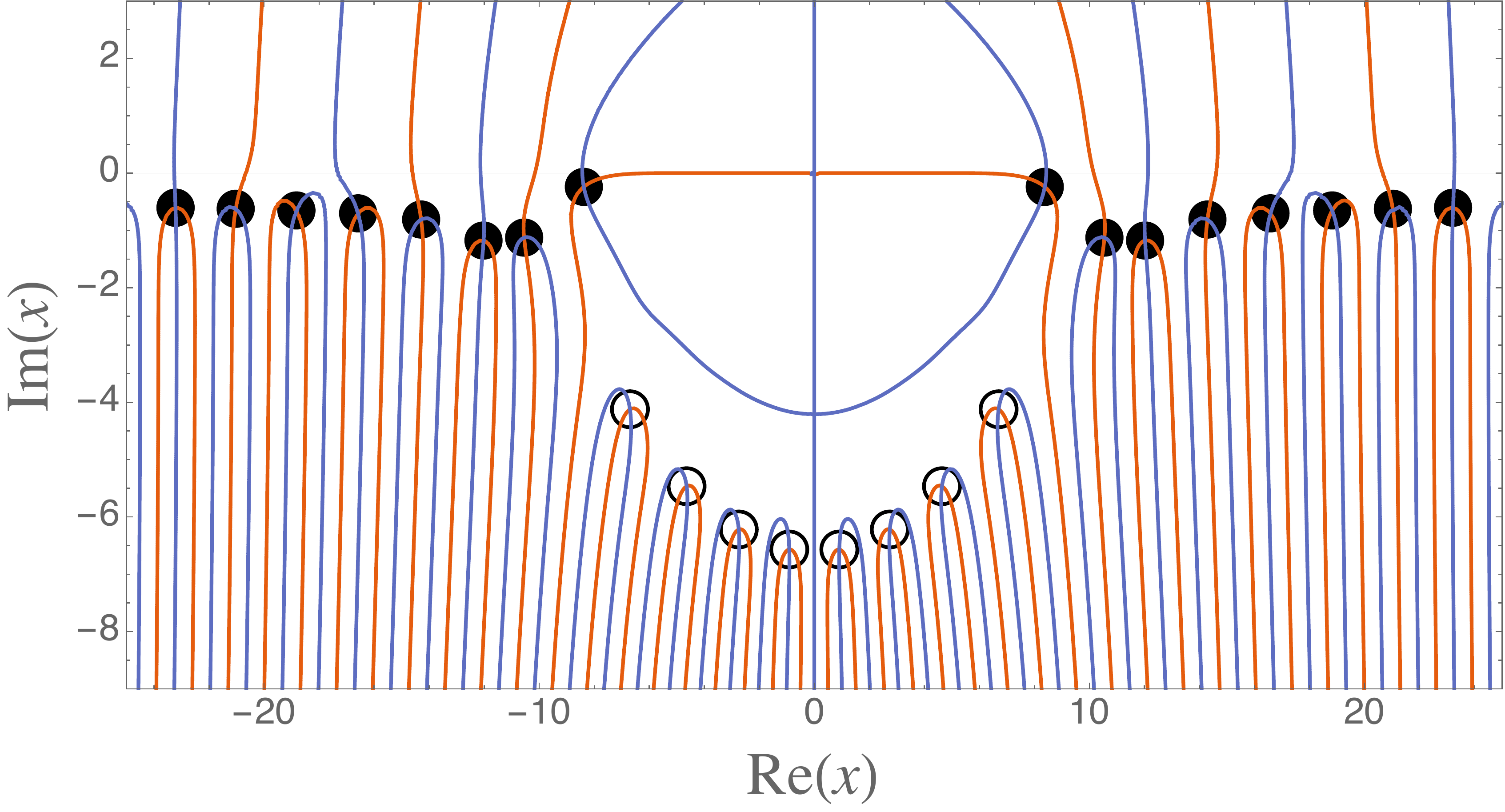}
        \label{fig:ContourTM}
    }
    \caption{Contour plots in the complex plane of the zeros of an equivalent formulation of \eqref{eq:fTEfTM} from \cite{aiello19} for the respective polarization, $l=9$, $n_1=1.5$ and $n_2=1$. The red lines correspond to solutions of $\text{Re}\,f_l^\sigma(x)=0$ and the blue ones to solutions of $\text{Im}\,f_l^\sigma(x)=0$. The intersections of those lines correspond to the eigenvalues $x_{l \,n}^\sigma$. As discussed in \cite{kukulin13}, not all solutions of \eqref{eq:fTEfTM} correspond to resonances: Open circles correspond to non-resonant eigenvalues, filled circles to resonances.}
    \label{fig:TETM}
\end{figure}
\begin{figure}[htbp]
    \centering
    \includegraphics[width=\textwidth]{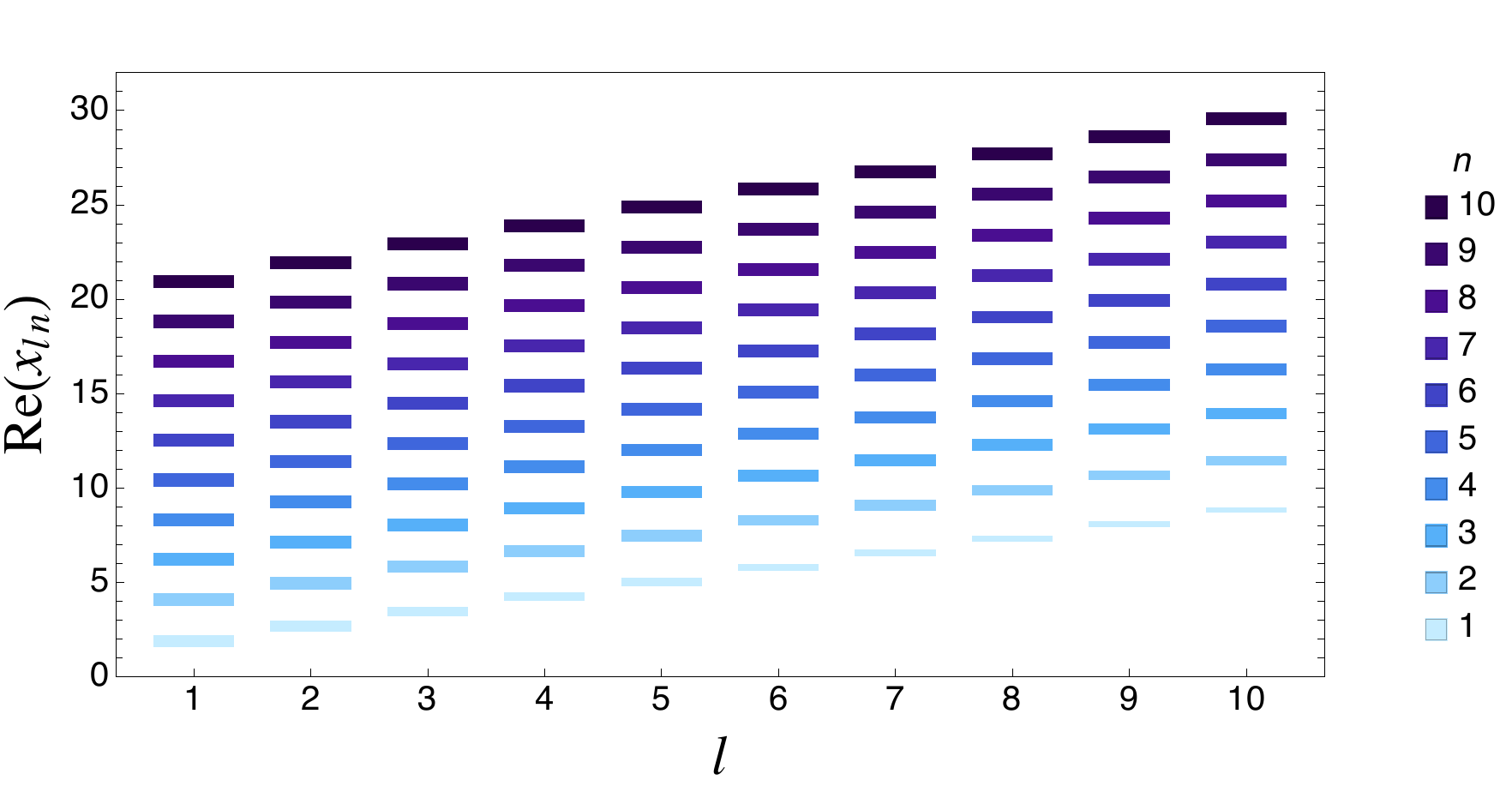}
    \caption{Spectrum of the $\TE$-modes of the dielectric unit sphere for $n_1=1.5$ and $n_2=1$. The vertical position of the spectral lines correspond to the real part of the corresponding resonance, the thickness of each line to the imaginary part. For $n=1$, the imaginary part decreases quickly in $l$, whereas for $n>1$, the imaginary part decreases slowly.}
    \label{fig:spectrumTE}
\end{figure}

To proceed further, let us introduce the dimensionless wave number
\begin{align}
    \label{eq:xdef}
    x = k_0 \, r_0,
\end{align}
which we denote in anticipation of further calculations as eigenvalue of the system. With this, we have $k_\alpha \, r_0 = n_\alpha \, x$ and we furthermore introduce the functions
\begin{subequations}
    \label{eq:fTEfTM}
    \begin{align}
        f_l^\text{TE}(x) &= \frac{[(n_1 x) j_l(n_1 x) ]'}{ j_l(n_1 x)} - \frac{[(n_2 x) h_l(n_2 x)]'}{ h_l(n_2 x)} \label{eq:fTE}, \\[4pt]
        f_l^\text{TM}(x) &= \frac{[(n_1 x) j_l(n_1 x)]'}{(n_1 x)^2 \, j_l(n_1 x)} - \frac{[(n_2 x) h_l(n_2 x)]'}{(n_2 x)^2 \, h_l (n_2 x)}, \label{eq:fTM}
    \end{align}
\end{subequations}
so that we can compactly rewrite \eqref{eq:TETranscendental} and \eqref{eq:TMTranscendental} as
\begin{align}
    \label{eq:fTEfTMvanish}
    f_l^\text{TE}(x) = 0, \und f_l^\text{TM}(x) = 0,
\end{align}
respectively. These transcendental equations can be solved numerically and a detailed discussion is presented in \cite{aiello19}, Appendix~B. We briefly discuss the approach in Figure \ref{fig:TETM}. The most important properties of the resonances $x$ are that they are complex numbers with negative imaginary part, and we express this fact writing $x = x_r + i\,x_i$, where $x_i<0$. Furthermore, one finds for each $l$ and $\sigma$ a countably infinite set of resonances labeled by $n$ as $x \equiv x_{l \, n}^\sigma$. From the analytic continuation of $j_l$ and $h_l$ in \eqref{eq:besselAnalyticContinuation} it follows that $[f_l^\sigma(x)]^*=f_l^\sigma(-x^*)$, and thus the resonances are symmetrically distributed around the imaginary axis. Therefore, one can label the resonances with positive real part with positive $n$ as
\begin{align}
    x = x_{l \,n}^\sigma, && n=1,2,\ldots,
\end{align}
and the resonances with negative real part satisfying $x_{l,-n}^\sigma = -(x_{l,n}^\sigma)^*$ where $n>0$. The spectrum for $\TE$-modes is shown in Figure \ref{fig:spectrumTE}. All the resonances $x_{l \,n}^\sigma$ are intrinsically non-degenerate with respect to $l$, $n$ and $\sigma$, but quasi-degeneracies occur \cite{dubertrand08}. In the context of quantum mechanical scattering theory, the resonances with negative real part are denoted as anti-resonances \cite{kukulin13}. 
\par
\begin{figure}[htb]
    \vspace{2.5cm}
    \centering
    \includegraphics[width=0.85\textwidth]{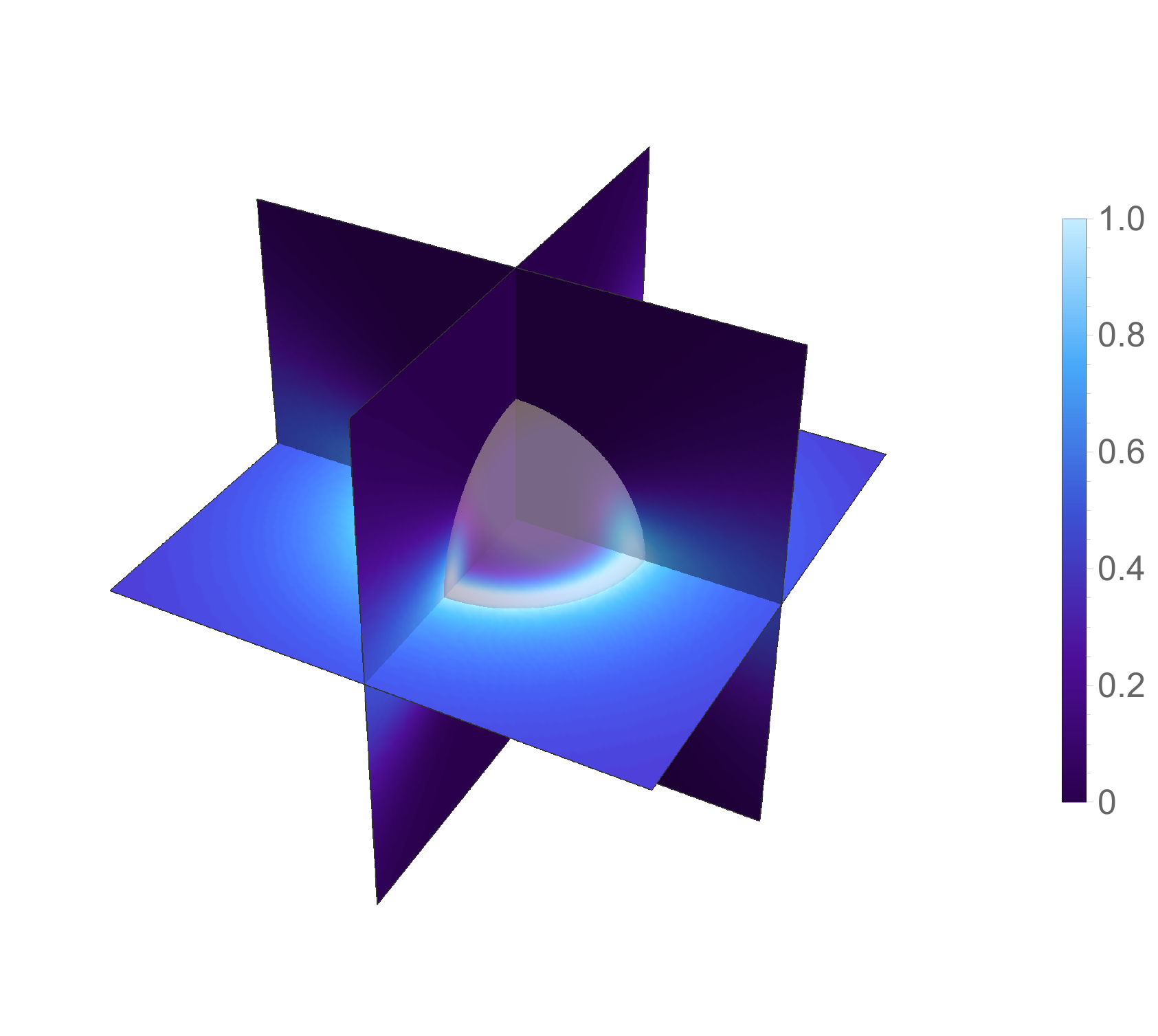}
    \caption{Electric energy density ($\sim |E(r,\theta,\phi)|^2$) of the $\TE$-mode of the opaque dielectric sphere characterized by $(l,m,n)=(9,9,1)$. The color scale is normalized such that the highest intensity equals one. On can clearly see that such a WGM is localized at the equator. The non-zero intensity outside the dielectic sphere indicates the losses to the external environment.}
    \label{fig:plot3d}
\end{figure}
\begin{figure}[htbp]
    \centering
    \includegraphics[width=\textwidth]{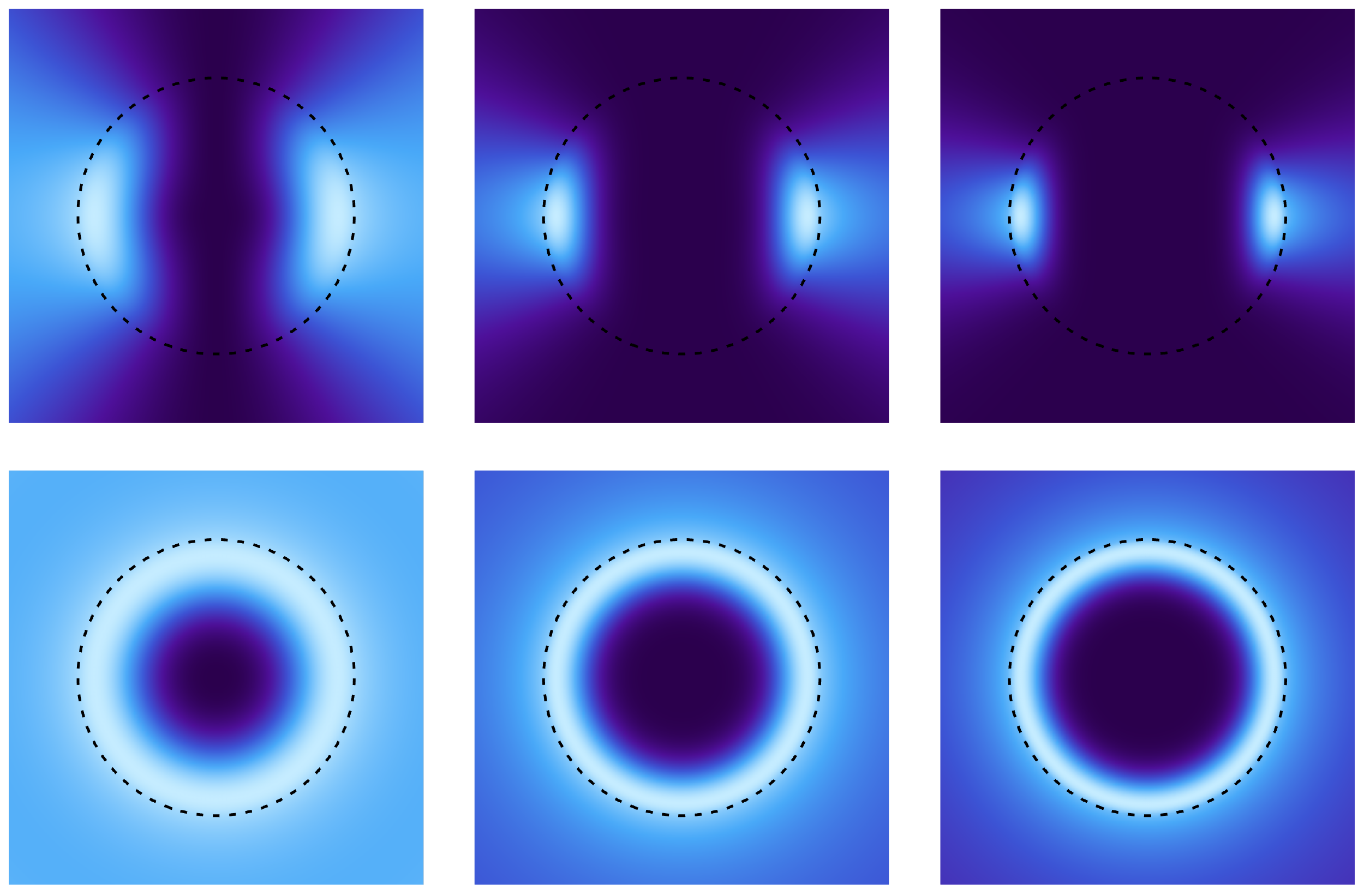}
    \caption{Electric energy densities for the $\TE$-modes characterized by, from left to right, $(l,m,n)=(3,3,1),(7,7,1),(11,11,1)$. Color scale as in Figure \ref{fig:plot3d}. \\
    Top row: Side view of the WGMs. For increasing $l$, the WGMs are more localized at the equator. Furthermore the losses occur more and more only in the plane of the equator.
    Bottom row: Corresponding top views at the equator. This again visualizes the increasing localization at the equator.}
    \label{fig:fieldsl}
\end{figure}
\begin{figure}[htbp]
    \centering
    \includegraphics[width=\textwidth]{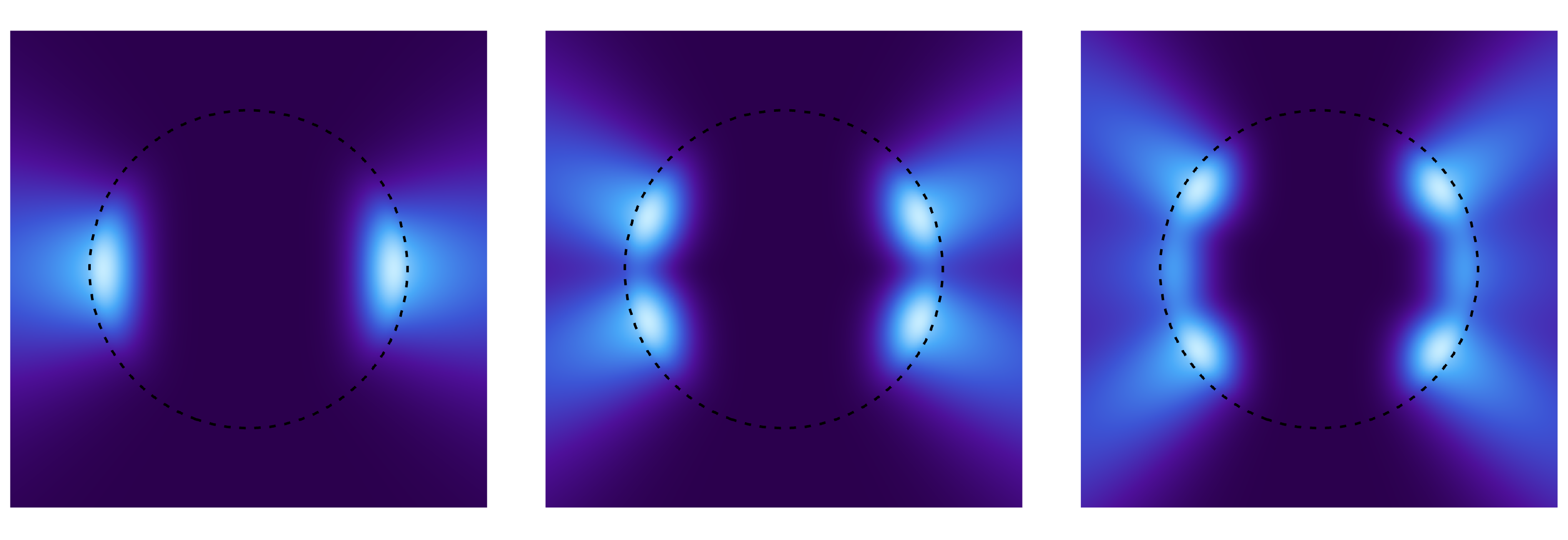}
    \caption{Side view of the electric energy densities for the $\TE$-modes characterized by, from left to right, $(l,m,n)=(9,9,1),(9,8,1),(9,7,1)$ (or equivalent, $(9,-9,1),(9,-8,1),(9,-7,1)$). Color scale as in Figure \ref{fig:plot3d}. For decreasing (increasing) $m$, the WGMs split into multiple so-called lobes up to a maximum at $m=0$.}
    \label{fig:fieldsm}
\end{figure}

Finally, let us discuss the modes associated to the resonances. The Debye potentials associated with the resonances $x_{l,n}^\sigma$ are given by
\begin{subequations}
    \begin{align}
        u_{1,l \, m}^\TE(x_{l \,n}^\text{TE} \, r/r_0) &= 
        a_{l\,m}^\text{E} \frac{j_l(n_1 \, x_{l \, n}^\text{TE} \, r/r_0)}{j_l(n_1 \, x_{l \, n}^\text{TE})} , \\[6pt]
        u_{2,l \, m}^\TE(x_{l\, n}^\text{TE} \,  r/r_0) &= a_{l\,m}^\text{E} \, \frac{h_l(n_2 \, x_{l \, n}^\text{TE}  \, r/r_0)}{ h_l(n_2 \, x_{l \, n}^\text{TE})},
    \end{align}
\end{subequations}
and
\begin{subequations}
    \begin{align}
        u_{1,l \, m}^\TM(x_{l \,n}^\text{TM} \, r/r_0) &= 
        a_{l\,m}^\text{M} \frac{j_l(n_1 \, x_{l \, n}^\text{TM} \, r/r_0)}{n_1 \, j_l(n_1 \, x_{l \, n}^\text{TM})} , \\[6pt]
        u_{2,l \, m}^\TM(x_{l\, n}^\text{TM} \,  r/r_0) &= a_{l\,m}^\text{M} \, \frac{h_l(n_2 \, x_{l \, n}^\text{TM}  \, r/r_0)}{n_2 \, h_l(n_2 \, x_{l \, n}^\text{TM})},
    \end{align}
\end{subequations}
and the associated electromagnetic modes can be obtained by inserting them into \eqref{eq:EBDebye}. The $\TE$-modes are visualized in Figures \ref{fig:plot3d}, \ref{fig:fieldsl} and \ref{fig:fieldsm}. These Debye potentials show the general complication of assuming outgoing waves only. If one considers the far field and again splitting $x_{l \, n}^\sigma = x_r + i \, x_i$ with $x_i<0$ results in
\begin{align}
    u_{2,l \,m}^\sigma(x_{l\, n}^\sigma \, r/r_0) &\propto h_l(n_2 \, x_{l \, n}^\sigma \, r/r_0) \notag \\[4pt]
    &\simeq \exp(-i n_2 \,  x_r \, r/r_0) \exp(-n_2 \, x_i \, r/r_0), && r\to\infty .
\end{align}
As the second factor tends to infinity, the Debye potentials as well as the associated electromagnetic modes are not normalizable in a standard sense and the coefficients $a_{l \, m}^\sigma$ stay undetermined. However, when deriving the formula for the resonances, there was no need for any normalization or any restriction on the coefficients. We will exploit this fact in further considerations.
\newpage

\section{Generalization of the Problem}
\label{sec:deformed}
In the previous section we introduced the general physical fields which solve Maxwell's equations for an arbitrarily shaped dielectric body within another dielectric medium. By imposing outgoing waves only, we found the resonances of a perfectly spherical body. This work is dedicated to apply this approach to dielectric bodies the form of which deviate from a perfect spherical one. \par
In order to find the resonances of such non-spherical bodies, we analyze the geometry of this generalized problem in Section~\ref{sec:geometry}. With that, we can discuss in Section \ref{sec:bcpt} the perturbative approach we choose to find the resonances of our problem: The \emph{\emph{B}oundary \emph{C}ondition \emph{P}erturbation \emph{T}heory} (BCPT). To be able to employ the BCPT later on, we rewrite the boundary condition in Section \ref{sec:reformulatebc}. As the perfectly spherical body serves as unperturbed problem, it will be the starting point for our perturbative treatment and thus we solve this problem again in Section \ref{sec:unperturbed} using the newly derived boundary conditions.
\subsection{The Geometry}
\label{sec:geometry}
As the main aim of this section is to derive suitable boundary conditions, let us recall the appropriate boundary conditions \eqref{eq:generalBoundaryCondition} for this problem
\begin{align*}
    \bn \times \left. \left( \bE_2 - \bE_1 \right) \right|_{\partial A} = 0, \und \bn \times \left. \left( \bB_2 - \bB_1 \right) \right|_{\partial A} = 0.
\end{align*}
From the boundary conditions it is apparent that they depend on the geometry of the dielectric body $A$, which is encoded in the form of the dielectric interface $\partial A$ and its normal $\bn$. \par
\begin{figure}[tb]
    \centering
    \includegraphics[width=0.6\textwidth]{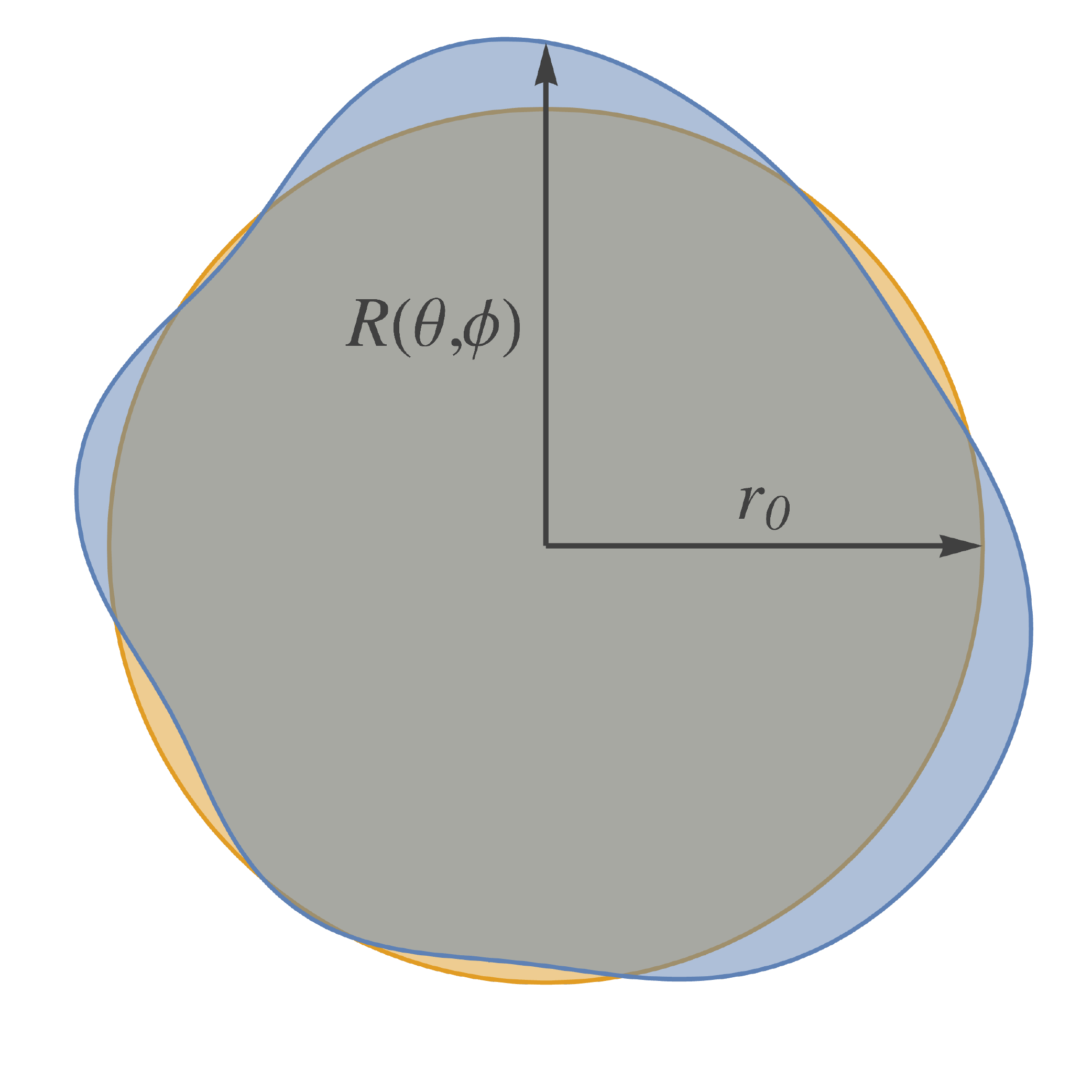}
    \caption{2D representation of an arbitrary deformation. The orange disc depicts a sphere with radius $r_0$. The blue region depicts a deformed sphere characterized by the surface profile function $R(\theta,\phi)$.}
    \label{fig:scheme2d}
\end{figure}

To be able to do calculations, we have to choose a suitable parametrization for the dielectric interface $\partial A$. One possibility is to introduce the \emph{surface profile function}
\begin{align}
    \label{eq:Rh}
    R(\theta,\phi) = r_0 \big( 1 + g(\theta,\phi) \big),
\end{align}
where $r_0$ is a constant and the \emph{deformation function} $g$ is a smooth, single-valued function of $\theta$ and $\phi$ defined on the unit sphere $S$. Figure \ref{fig:scheme2d} illustrates the surface profile function. The form of this parametrization seems suitable for a perturbative approach, because one can interpret $r_0$ as the radius of a sphere and $g$ creates a supposedly small deviation from the perfect sphere. Using this parametrization it makes sense to denote the body $A$ as a \emph{deformed sphere}. \par
Before we continue, let us state that the form of \eqref{eq:Rh} is not the most general \hyphenation{pa-ra-me-tr-i-za-tion} parametrization, as we assume a smooth, single-valued $g$. However, for the physical applications we have in mind, both requirements are met if one sets the center of the deformed sphere to the origin. In Section \ref{sec:applicability} we reason that if we drop one or both of these requirements, the perturbation theory we employ will most likely fail. 

\par
In view of a perturbation theory, it is useful to introduce a supposedly small expansion parameter. In this setup, we define the \emph{deformation strength} $\epsilon$ as
\begin{align}
    \epsilon \equiv \max\{ \abs{g(\theta,\phi)} \}_S,
\end{align}
and if we require $g$ to be small, this translates to
\begin{align}
    \label{eq:epsilonsmall}
    \epsilon \ll 1 .
\end{align}
Using the deformation strength, \eqref{eq:Rh} can be rewritten as
\begin{align}
    \label{eq:Rf}
    R(\theta,\phi) = r_0 \big( 1 + \epsilon f(\theta,\phi) \big),
\end{align}
where $f(\theta,\phi) \equiv g(\theta,\phi)/ \max\{ g(\theta,\phi)\}_S \leq 1$. We will show in Section \ref{sec:applicability} that \eqref{eq:epsilonsmall} is not sufficient to apply the perturbation theory. \par
Using the surface profile function, we can rewrite the boundary conditions as
\begin{align}
    \label{eq:BoundaryConditionsParametrized}
    \bn \times \left. \left( \bE_2 - \bE_1 \right) \right|_{r = R(\theta,\phi)} = 0, \und \bn \times \left. \left( \bB_2 - \bB_1 \right) \right|_{r = R(\theta,\phi)} = 0.
\end{align}
To progress further, we want to determine the normal $\bn$ of the deformed sphere $A$. Let us write down the interface implicitly as
\begin{align}
    F(r,\theta,\phi) = 0 ,
\end{align}
where
\begin{align}
    F(r,\theta,\phi) &= r - R(\theta,\phi) \notag \\
    &= (r - r_0) - \epsilon \, r_0 \, f(\theta,\phi) \label{eq:FExplicit} .
\end{align}
Here we again see that for $\epsilon = 0$, we have the implicit definition of a sphere with radius~$r_0$. By acting with the nabla operator \eqref{eq:nablaSpherical} on \eqref{eq:FExplicit} \cite{weisstein}, we find the normal 
\begin{align}
    \bn &= \bnabla F(r,\theta,\phi) \notag \\
    &\equiv \er - \bn_\parallel \label{eq:nExplicit} ,
\end{align}
where we introduced
\begin{align}
    \label{eq:nParallelExplicit}
    \bn_\parallel = \frac{\epsilon}{1 + \epsilon f(\theta,\phi)} \left( \etheta \frac{\partial f(\theta,\phi)}{\partial \theta} + \ephi \frac{1}{ \sin \theta} \frac{\partial f(\theta,\phi)}{\partial \phi} \right).
\end{align}
It needs to be understood that the subscript $\parallel$ denotes quantities that are parallel to the surface of \emph{undeformed} sphere and are therefore orthogonal to $\er$. Let us finally introduce an $\epsilon$-independent vector $\be_\parallel$, defined as
\begin{align}
    \label{eq:eParallel}
    \be_\parallel \equiv \etheta \frac{\partial f(\theta,\phi)}{\partial \theta} + \ephi \frac{1}{ \sin \theta} \frac{\partial f(\theta,\phi)}{\partial \phi} ,
\end{align}
so \eqref{eq:nParallelExplicit} can be rewritten as
\begin{align}
    \bn_\parallel = \frac{\epsilon}{1 + \epsilon f(\theta,\phi)} \, \be_\parallel \label{eq:nParalleleParallel} .
\end{align}
\subsection{Boundary Condition Perturbation Theory}
\label{sec:bcpt}
To motivate our further steps, let us discuss how we want to solve our problem. In the previous section we chose a parametrization for a deformed dielectric sphere in dependence of a supposedly small parameter $\epsilon$. \par
The method to solve our problem is BCPT, originally developed by Lord Rayleigh in his book \emph{The Theory of Sound}~\cite{rayleigh94}, where he investigated how the modes of a circular membrane, and therefore the pitch of the emitted sound, changes if one slightly deforms the membrane\footnote{When considering the deformed circular membrane, one has to solve Helmholtz's equation in polar coordinates for a fixed boundary in order to determine the (scalar) sound waves as well as the (real) eigenvalue of the system. In our case, we have to consider two three-dimensional vector fields, which we can reduce to the Debye potentials satisfying Helmholtz's equation. However, the three-dimensional character as well as the boundary condition introduces lots of complications. For simpler boundary conditions, cf. \cite{panda14}.}.\par
The idea behind BCPT is pretty clear. By expanding the \emph{boundary condition} around the supposedly small parameter $\epsilon$, one gets a chain of equations for all orders in $\epsilon$. Here, the zeroth order equation corresponds to the boundary condition of the unperturbed problem, which solution is analytically known. By successively solving the chain of equations, one finds a perturbative solution of the problem. \par
Let us motivate BCPT by applying it to the current form of our boundary conditions \eqref{eq:BoundaryConditionsParametrized}. By expanding the normal $\bn$ from \eqref{eq:nExplicit} as
\begin{align*}
    \bn = \er - \epsilon \, \be_\parallel + \order{\epsilon^2},
\end{align*}
and using
\begin{align*}
    g(R(\theta,\phi)) = g(r_0,\theta,\phi) + \epsilon \, r_0 \, f(\theta,\phi) \, g'(r_0,\theta,\phi) + \order{\epsilon^2},
\end{align*}
for the field evaluated at the boundary, we can rewrite the boundary condition for the electric field as
\begin{align*}
    0 &= \bn \times \left. \left( \bE_2 - \bE_1 \right) \right|_{r = R(\theta,\phi)}\\
    &= \er \times \left. \left( \bE_2 - \bE_1 \right) \right|_{r = r_0} \\
    &\hphantom{=} - \epsilon \left[ \be_\parallel \times \left. \left( \bE_2 - \bE_1 \right) \right|_{r = r_0} - r_0 \, f(\theta,\phi) \, \left. \left(\er \times \left( \bE_2 - \bE_1 \right)\right)' \right|_{r = r_0} \right] + \order{\epsilon^2}.
\end{align*}
Here we immediately see that the zeroth order boundary condition corresponds, as expected, to the boundary condition of the undeformed sphere discussed in Section \ref{sec:perfectsolution}. Now one might be able to use the general zeroth-order result and insert it in the first-order equation,
\begin{align}
    \label{eq:firstorder}
    0 = \be_\parallel \times \left. \left( \bE_2 - \bE_1 \right) \right|_{r = r_0} - r_0 \, f(\theta,\phi) \, \left. \left(\er \times \left( \bE_2 - \bE_1 \right)\right)' \right|_{r = r_0},
\end{align}
to refine the result. It turns out that solving the problem in this form, if it is even possible, is inconvenient, and we need to adapt the boundary condition to fit our needs in Section~\ref{sec:reformulatebc}.
\subsubsection{Applicability of BCPT}
\label{sec:applicability}
However, this outline reveals a problem. Recalling the definition of $\be_\parallel$,
\begin{align*}
    \be_\parallel = \etheta \frac{\partial f(\theta,\phi)}{\partial \theta} + \ephi \frac{1}{ \sin \theta} \frac{\partial f(\theta,\phi)}{\partial \phi},
\end{align*}
from \eqref{eq:eParallel}, we see that the derivatives of $f$ with respect to $\theta$ and $\phi$ might be huge, a situation that is denoted as a \emph{strongly winding boundary condition} \cite{badel19}. This has the effect that the first term in \eqref{eq:firstorder} might be of the same order of magnitude as the zeroth order, and the perturbative approach breaks down. \par
This rises the question, when BCPT is applicable and when it is not. Let us derive a criterion, for which the perturbative expansion is valid. We define
\begin{align}
    \beta &= \max_S \, \|\be_\parallel\|\\
    &= \max_S \left\{ \sqrt{\left(\frac{\partial f(\theta,\phi)}{\partial \theta}\right)^2 + \frac{1}{ \sin^2 \theta} \left(\frac{\partial f(\theta,\phi)}{\partial \phi} \right)^2} \right\} \notag,
\end{align}
where $\|.\|$ denotes the norm of a vector and $\max_S$ the maximum over the sphere $S$. This parameter encodes both troublesome derivatives. As we want the first term in \eqref{eq:firstorder} to be of order unity, we have to require
\begin{align}
    \label{eq:betaCriterion}
    \beta \sim 1
\end{align}
in order for the BCPT to hold. We can also give this criterion a geometrical meaning by considering the angle $\gamma$ between the normal vector $\er$ of the undeformed sphere and the normal vector $\bn$ of the deformed sphere. We compute
\begin{align*}
    \cos \gamma(\theta,\phi) &= \er \cdot \frac{\bn}{\| {\bn} \|} = \frac{1}{\sqrt{1 + \|\bn_\parallel\|^2}} \\
    &= \left\{ 1 + \frac{\epsilon^2 \, \|\be_\parallel\|^2}{[1+ \epsilon \, f(\theta,\phi)]^2} \right\}^{-\frac{1}{2}} \\
    &\simeq 1 - \frac{\epsilon^2  \, \|\be_\parallel\|^2}{2} \\
    &\geq 1 - \frac{(\epsilon \, \beta)^2}{2},
\end{align*}
where we used properties of the normal $\bn$ from Section \ref{sec:geometry} and neglected terms of higher orders in $\epsilon$ in the third line. Therefore, $\epsilon \, \beta$ encodes the lower bound of $\cos \gamma$.  By imposing \eqref{eq:betaCriterion} we finally get
\begin{align}
    \cos \gamma(\theta,\phi) \simeq 1,
\end{align}
or in other words, $\bn$ is almost parallel to $\er$. Therefore, our criterion $\beta \sim 1$ is equivalent to the requirement of \emph{local paraxiality}.\par
This also brings us back to the choice of our parametrization of the boundary of Section~\ref{sec:geometry}. There we choose the deformation function $f$ to be smooth and single-valued due to our physical intuition of the problem. Now we can further reason, why we chose these requirements. If we do not require smoothness of $f$, $\beta$ is ill-defined. Likewise to this, a multi-valued $f$ will violate the local paraxiality. This does not exclude the possibility to get the BCPT working when dropping one or both of the requirements, but strongly suggests that this needs further considerations. \par 
For the remainder of this work, we will assume that \eqref{eq:betaCriterion} holds and thus consider~$\epsilon$ as bona fide expansion parameter.
\subsection{Reformulation of the Boundary Conditions}
\label{sec:reformulatebc}
We now want to rewrite the boundary conditions in a way which will be suitable to employ the BCPT. We will do this in two steps. The first one is to remove a redundancy in the boundary conditions. The second step is to use the explicit form of the electromagnetic fields from Section \ref{sec:phfields} to express the boundary condition in terms of the field coefficients.
\subsubsection{Removing Redundancy}
\label{sec:removeRedundancy}
For the perfect spherical boundary condition in Section \ref{sec:perfectsolution}, we expressed the electric and magnetic field in terms of the vector spherical harmonics and also had $\bn = \er$. This enabled us to use
\begin{align*}
    \er \times \bY{l}{m} = 0 \komma  \er \times \bPsi{l}{m} = \bPhi{l}{m}, \und \er \times \bPhi{l}{m} = - \bPsi{l}{m},
\end{align*}
from \eqref{eq:ercrossvsh} to simplify the boundary condition. This had the effect that the $Y$-components of the fields do not contribute to the actual boundary condition. This gives us the hint that there might be some redundancy in the boundary condition. \par
Going to the perturbed problem, the normal $\bn$ in the boundary condition of the deformed sphere additionally contains $\etheta$ and $\ephi$. Thus, one also has to determine the cross products of those basis vectors with the vector spherical harmonics, which cannot be expressed as simple as in \eqref{eq:ercrossvsh}. \par
To resolve both of these points, one can write the electromagnetic fields in spherical coordinates \eqref{eq:scexpansion}. Let us choose the electric field as example, i.e.
\begin{align}
    \bE_\alpha = E_{\alpha \, r} \, \er + E_{\alpha \, \theta} \, \etheta + E_{\alpha \, \phi} \, \ephi,
\end{align}
and also rewrite the normal vector \eqref{eq:nExplicit} as
\begin{align}
    \bn = \er - n_\theta \, \etheta - n_\phi \, \ephi.
\end{align}
Using this, one can immediately rewrite the boundary condition \eqref{eq:BoundaryConditionsParametrized} for the three components as
\begin{subequations}
    \label{eq:BoundarySpericalCoordinates}
    \begin{align}
        n_\phi E_{1 \, \theta} - n_\theta E_{1 \, \phi} &= n_\phi E_{2 \, \theta} - n_\theta E_{2 \, \phi}, \label{eq:BoundarySpericalCoordinatesr} \\
        n_\phi E_{1 \, r} + E_{1 \, \phi} &= n_\phi E_{2 \, r} + E_{2 \, \phi} , \label{eq:BoundarySpericalCoordinatestheta} \\
        n_\theta E_{1 \, r} + E_{1 \, \theta} &= n_\theta E_{2 \, r} + E_{2 \, \theta} , \label{eq:BoundarySpericalCoordinatesphi}
    \end{align}
\end{subequations}
where we used \eqref{eq:sccrossproducts}, and all coefficients $E_{\alpha \, i}$ need to be evaluated at $(R(\theta,\phi),\theta,\phi)$. Now it is not hard to show that only two of these three equations are independent. For example
\begin{align}
    \eqref{eq:BoundarySpericalCoordinatesr} =  n_\phi \,  \eqref{eq:BoundarySpericalCoordinatesphi} -n_\theta \, \eqref{eq:BoundarySpericalCoordinatestheta}.
\end{align}
Let us choose \eqref{eq:BoundarySpericalCoordinatestheta} and \eqref{eq:BoundarySpericalCoordinatesphi} as independent equations. Following \cite{erma69}, we multiply \eqref{eq:BoundarySpericalCoordinatestheta} by $\ephi$, \eqref{eq:BoundarySpericalCoordinatesphi} by $\etheta$ and summing these two equations, we obtain a single vector equation,
\begin{align}
    \big( \bE_{1 \, \parallel} - \bE_{2 \, \parallel} \big) + \big( E_{1 \,r} - E_{2 \,r} \big) \bn_\parallel  &= 0,
\end{align}
which needs to be evaluated at the boundary. Here, $\bn_\parallel$ corresponds to the definition in \eqref{eq:nParallelExplicit}, and we introduced
\begin{align}
    \label{eq:bEParallel}
    \bE_{\alpha \, \parallel} = E_{\alpha \, \theta} \, \etheta + E_{\alpha \, \phi} \, \ephi,
\end{align}
similar to $\bn_\parallel$, parallel to the surface of the undeformed sphere. With these calculations we on the one hand removed redundancy from the boundary conditions\footnote{So this was no special feature of the boundary conditions of the undeformed sphere. In general, the electromagnetic boundary conditions are redundant \cite{jackson99}.}, on the other hand, avoided evaluating the cross products. \par
Now we want to build the bridge back to the multipole expansion. As $\bY{l}{m} \propto \er$ and $\bPsi{l}{m}$ as well as $\bPhi{l}{m}$ are orthogonal to it \eqref{eq:VSHscalarProducts}, $\bY{l}{m}$ encodes the radial part of the field $E_{\alpha \, r}$ and $\bPsi{l}{m}$ together with $\bPhi{l}{m}$ encode the parallel part $\bE_{\alpha \, \parallel}$. \par
As these considerations similarly hold for the magnetic field, we summarize both boundary conditions for later reference as
\begin{subequations}
    \label{eq:BoundaryConditionParallel}
    \begin{align}
        \big( \bE_{1 \, \parallel} - \bE_{2 \, \parallel} \big) + \big( E_{1 \,r} - E_{2 \,r} \big) \bn_\parallel  &= 0, \label{eq:BoundaryConditionParallelE} \\
        \big( \bB_{1 \, \parallel} - \bB_{2 \, \parallel} \big) + \big( B_{1 \,r} - B_{2 \,r} \big) \bn_\parallel &= 0. \label{eq:BoundaryConditionParallelB}
    \end{align}    
\end{subequations}
\subsubsection{Derivation of a Matrix Equation}
\label{sec:bcMatrix}
Now that we removed redundancy from the boundary condition and found \eqref{eq:BoundaryConditionParallel}, we can insert the general form of the electromagnetic fields into it and therefore find a boundary condition connecting the field coefficients. \par
Let us recall the multipole expansions of the fields inside (\ref{eq:Einside},\ref{eq:cBinside})
\begin{align}
    \bE_1 &= \sum_{l=1}^\infty \sum_{m=-l}^l \left\{ a_{l\,m}^\text{E} \, A_{1 \, l}^\Phi(r) \, \bPhi{l}{m} + \frac{a_{l \, m}^\text{M}}{n_1} \left[ A_{1 \, l}^Y(r) \, \bY{l}{m} + A_{1 \, l}^\Psi(r) \, \bPsi{l}{m} \right] \right\} \label{eq:E1A(r)}, \\[6pt]
    c_1 \,\bB_1 &= \sum_{l=1}^\infty \sum_{m=-l}^l \left\{ \frac{a_{l \, m}^\text{M}}{n_1} \, A_{1 \, l}^\Phi(r) \bPhi{l}{m} - a_{l \, m}^\text{E} \left[ A_{1 \, l}^Y(r) \, \bY{l}{m} + A_{1 \, l}^\Psi(r) \, \bPsi{l}{m} \right] \right\},
\end{align}
and outside (\ref{eq:Eoutside},\ref{eq:cBoutside})
\begin{align}
    \bE_2 &= \sum_{l=1}^\infty \sum_{m=-l}^l \left\{ b_{l\,m}^\text{E} \, A_{2 \, l}^\Phi(r) \, \bPhi{l}{m} + \frac{b_{l \, m}^\text{M}}{n_2} \left[ A_{2 \, l}^Y(r) \, \bY{l}{m} + A_{2 \, l}^\Psi(r) \, \bPsi{l}{m} \right] \right\}, \\[6pt]
    c_2 \,\bB_2 &= \sum_{l=1}^\infty \sum_{m=-l}^l \left\{ \frac{b_{l \, m}^\text{M}}{n_2} \, A_{2 \, l}^\Phi(r) \bPhi{l}{m} - b_{l \, m}^\text{E} \left[ A_{2 \, l}^Y(r) \, \bY{l}{m} + A_{2 \, l}^\Psi(r) \, \bPsi{l}{m} \right] \right\}, \label{eq:cB2A(r)}
\end{align}
of the dielectric sphere. Here we introduced the radial functions $A_{\alpha \, l}^X(r) \equiv A_{\alpha \, l}^X(k_\alpha \,r)$, with $X = Y, \Psi$ and $\Phi$, defined as
\begin{align}
    \label{eq:Ar}
    \begin{split}
        \begin{aligned}
            A_{1 \, l}^Y(r) &= -i \, l(l+1) \frac{j_l(k_1r)}{(k_1r)j_l(k_1r_0)}, & A_{2 \, l}^Y(r) &= -i \, l(l+1) \frac{h_l(k_2r)}{(k_2r)h_l(k_2r_0)}, \\[4pt]
            A_{1 \, l}^\Psi(r) &= -i \, \frac{[(k_1r)j_l(k_1r)]'}{(k_1r)j_l(k_1r_0)}, & A_{2 \, l}^\Psi(r) &= -i \, \frac{[(k_2r)h_l(k_2r)]'}{(k_2r)h_l(k_2r_0)}, \\[4pt]
            A_{1 \, l}^\Phi(r) &= \frac{j_l(k_1r)}{j_l(k_1r_0)}, & A_{2 \, l}^\Phi(r) &= \frac{h_l(k_2r)}{h_l(k_2r_0)}.
        \end{aligned}
    \end{split}
\end{align}
Recalling the boundary conditions \eqref{eq:BoundaryConditionParallel} from Section \ref{sec:removeRedundancy}, we need to know the parallel part of the fields $\bX_{\alpha \, \parallel}$ and the radial part $X_{\alpha \, r}$. Following the argument of the previous section, we exemplary find the electric field inside the deformed sphere
\begin{align*}
    \bE_{1 \, \parallel} &= \sum_{l=1}^\infty \sum_{m=-l}^l \left\{ a_{l \, m}^\text{E} A_{1 \, l}^\Phi(r) \bPhi{l}{m} + \frac{a_{l \, m}^\text{M}}{n_1} A_{1 \, l}^\Psi(r) \bPsi{l}{m} \right\}, \\[6pt]
    E_{1 \,r} &= \sum_{l=1}^\infty \sum_{m=-l}^l \left\{ \frac{a_{l \, m}^\text{M}}{n_1} A_{1 \, l}^Y(r) \sY{l}{m} \right\}.
\end{align*}
Hence, collecting all terms, the boundary condition for the electric field \eqref{eq:BoundaryConditionParallelE} reads
\begin{align}
    \label{eq:EboundaryIntermediate1}
    0 = \sum_{l \, m} \bigg\{ &\bigg[a_{l \, m}^\text{E} \left(A_{1 \, l}^\Phi \bPhi{l}{m} \right) - b_{l \, m}^\text{E} \left( A_{2 \, l}^\Phi \bPhi{l}{m} \right) \bigg] \notag \\[4pt]
    &+ \left[ \frac{a_{l \, m}^\text{M}}{n_1} \left(A_{1 \, l}^\Psi \bPsi{l}{m} + A_{1 \, l}^Y \, \sY{l}{m} \, \bn_\parallel \right) - \frac{b_{l \, m}^\text{M}}{n_2} \left( A_{2 \, l}^\Psi \bPsi{l}{m} + A_{2 \, l}^Y \, \sY{l}{m} \, \bn_\parallel \right) \right]  \bigg\} ,
\end{align}
which needs to be evaluated at $r=R(\theta,\phi)$, and here and hereafter,
\begin{align}
    \label{eq:sumlmShort}
    \sum_{l \, m} && \text{stands for} && \sum_{l=1}^\infty \sum_{m=-l}^l.
\end{align}
Likewise, the boundary condition for the magnetic field \eqref{eq:BoundaryConditionParallelB} reads
\begin{align}
    \label{eq:BboundaryIntermediate1}
    0 \! = \! \sum_{l \, m} \bigg\{ &\left[ a_{l \, m}^\text{M} \left(A_{1 \, l}^\Phi \bPhi{l}{m} \right) - b_{l \, m}^\text{M} \left( A_{2 \, l}^\Phi \bPhi{l}{m} \right) \right] \notag \\[2pt]
    &- \left[ a_{l \, m}^\text{E} \, n_1 \left(A_{1 \, l}^\Psi \bPsi{l}{m} + A_{1 \, l}^Y \, \sY{l}{m} \, \bn_\parallel \right) - b_{l \, m}^\text{E} \, n_2 \left( A_{2 \, l}^\Psi \bPsi{l}{m} + A_{2 \, l}^Y \, \sY{l}{m} \, \bn_\parallel \right)\right]  \bigg\}.
\end{align}
Let us define vector quantities, suggested by the round brackets of both previous boundary conditions, as
\begin{subequations}
    \label{eq:bAbB}
    \begin{align}
        \bA_{\alpha , l \, m}(\theta,\phi) &= A_{\alpha \, l}^\Phi(\theta,\phi) \bPhi{l}{m}(\theta,\phi) \label{eq:bA}, \\[4pt]
        \bB_{\alpha , l \, m}(\theta,\phi) &= A_{\alpha \, l}^\Psi(\theta,\phi) \bPsi{l}{m}(\theta,\phi) + A_{\alpha \, l}^Y(\theta,\phi) \sY{l}{m}(\theta,\phi) \bn_\parallel (\theta,\phi), \label{eq:bB}
    \end{align}
\end{subequations}
where $A_{\alpha \, l}^X(\theta,\phi) = A_{\alpha \, l}^X(R(\theta,\phi))$. Again we employ a multipole expansion to move the entire angular dependence to the vector spherical harmonics at the cost of an infinite sum. As we have no radial component in this equation anymore, the vector spherical harmonic $\bY{l}{m}$ does not occur and we find
\begin{subequations}
    \label{eq:bAbBVSH}
    \begin{align}
        \bA_{\alpha , l \, m}(\theta,\phi) &= \sum_{l' \, m'} \left\{ [A_\alpha^\Psi]_{l \, m}^{l' \, m'} \bPsi{l'}{m'} + [A_\alpha^\Phi]_{l \, m}^{l' \, m'} \bPhi{l'}{m'} \right\}, \\[4pt]
        \bB_{\alpha , l \, m}(\theta,\phi) &= \sum_{l' \, m'} \left\{ [B_\alpha^\Psi]_{l \, m}^{l' \, m'} \bPsi{l'}{m'} + [B_\alpha^\Phi]_{l \, m}^{l' \, m'} \bPhi{l'}{m'} \right\},
    \end{align}
\end{subequations}
where the coefficients can be determined using \eqref{eq:VSHexpansionCoefficients} via
\begin{subequations}
    \label{eq:ABmatrices}
    \begin{align}
        [A_\alpha^V]_{l \, m}^{l' \, m'} &= \frac{1}{l' (l' + 1)} \intdOmega \mathbf{V}_{l' \, m'}^* \cdot \bA_{\alpha , l \, m} \label{eq:Amatrices}, \\[4pt]
        [B_\alpha^V]_{l \, m}^{l' \, m'} &= \frac{1}{l' (l' + 1)} \intdOmega \mathbf{V}_{l' \, m'}^* \cdot \bB_{\alpha , l \, m} \label{eq:Bmatrices},
    \end{align}
\end{subequations}
with $V$ being either $\Psi$ or $\Phi$. Again, as $\bPsi{0}{0}$ and $\bPhi{0}{0}$ vanish, the $l'=0$ term does not contribute to \eqref{eq:bAbBVSH} and thus does not need to be determined by \eqref{eq:ABmatrices}. Substituting \eqref{eq:bAbBVSH} into the boundary conditions \eqref{eq:EboundaryIntermediate1} and \eqref{eq:BboundaryIntermediate1}, we find
\begin{multline}
    \label{eq:EboundaryIntermediate2}
        0 = \sum_{l' \, m'} \bigg( \bPsi{l'}{m'} \sum_{l \, m} \bigg\{ a_{l \, m}^\text{E} [A_1^\Psi]_{l \, m}^{l' \, m'}  - b_{l \, m}^\text{E} [A_2^\Psi]_{l \, m}^{l' \, m'}  + a_{l \, m}^\text{M} \frac{[B_1^\Psi]_{l \, m}^{l' \, m'}}{n_1}  - b_{l \, m}^\text{M} \frac{[B_2^\Psi]_{l \, m}^{l' \, m'}}{n_2}\bigg\} \\[6pt]
        +\bPhi{l'}{m'} \sum_{l \, m} \bigg\{ a_{l \, m}^\text{E} [A_1^\Phi]_{l \, m}^{l' \, m'}  - b_{l \, m}^\text{E} [A_2^\Phi]_{l \, m}^{l' \, m'}  + a_{l \, m}^\text{M} \frac{[B_1^\Phi]_{l \, m}^{l' \, m'}}{n_1}  - b_{l \, m}^\text{M} \frac{[B_2^\Phi]_{l \, m}^{l' \, m'}}{n_2} \bigg\} \bigg),
\end{multline}
and
\begin{multline}
    \label{eq:BboundaryIntermediate2}
    0 =  \sum_{l' \, m'}  \bigg( \bPsi{l'}{m'} \sum_{l \, m}  \bigg\{ a_{l \, m}^\text{M} [A_1^\Psi]_{l \, m}^{l' \, m'} - b_{l \, m}^\text{M} [A_2^\Psi]_{l \, m}^{l' \, m'} - a_{l \, m}^\text{E} n_1 [B_1^\Psi]_{l \, m}^{l' \, m'} + b_{l \, m}^\text{E} n_2 [B_2^\Psi]_{l \, m}^{l' \, m'}\bigg\} \\[6pt]
    +\bPhi{l'}{m'} \sum_{l \, m}  \bigg\{ a_{l \, m}^\text{M} [A_1^\Phi]_{l \, m}^{l' \, m'}  - b_{l \, m}^\text{M} [A_2^\Phi]_{l \, m}^{l' \, m'}  - a_{l \, m}^\text{E} n_1 [B_1^\Phi]_{l \, m}^{l' \, m'}  + b_{l \, m}^\text{E} n_2 [B_2^\Phi]_{l \, m}^{l' \, m'} \bigg\} \bigg).
\end{multline}
From the orthogonality of the vector spherical harmonics \eqref{eq:VSHorthogonality} it follows that we need to satisfy for each $l'$ and $m'$ the following four homogeneous equations:
\begin{subequations}
    \begin{align}
        0 &= \sum_{l \, m} \bigg\{ a_{l \, m}^\text{E} [A_1^\Phi]_{l \, m}^{l' \, m'} - b_{l \, m}^\text{E} [A_2^\Phi]_{l \, m}^{l' \, m'} + a_{l \, m}^\text{M} \frac{[B_1^\Phi]_{l \, m}^{l' \, m'}}{n_1} - b_{l \, m}^\text{M} \frac{[B_2^\Phi]_{l \, m}^{l' \, m'}}{n_2} \bigg\}, \\[4pt]
        0 &= \sum_{l \, m} \bigg\{ - a_{l \, m}^\text{E} n_1 [B_1^\Psi]_{l \, m}^{l' \, m'} + b_{l \, m}^\text{E} n_2 [B_2^\Psi]_{l \, m}^{l' \, m'} + a_{l \, m}^\text{M} [A_1^\Psi]_{l \, m}^{l' \, m'} - b_{l \, m}^\text{M} [A_2^\Psi]_{l \, m}^{l' \, m'} \bigg\}, \\[4pt]
        0 &= \sum_{l \, m} \bigg\{ a_{l \, m}^\text{E} [A_1^\Psi]_{l \, m}^{l' \, m'} - b_{l \, m}^\text{E} [A_2^\Psi]_{l \, m}^{l' \, m'} + a_{l \, m}^\text{M} \frac{[B_1^\Psi]_{l \, m}^{l' \, m'}}{n_1} - b_{l \, m}^\text{M} \frac{[B_2^\Psi]_{l \, m}^{l' \, m'}}{n_2}\bigg\}, \\[4pt]
        0 &= \sum_{l \, m} \bigg\{ - a_{l \, m}^\text{E} n_1 [B_1^\Phi]_{l \, m}^{l' \, m'} + b_{l \, m}^\text{E} n_2 [B_2^\Phi]_{l \, m}^{l' \, m'} + a_{l \, m}^\text{M} [A_1^\Phi]_{l \, m}^{l' \, m'} - b_{l \, m}^\text{M} [A_2^\Phi]_{l \, m}^{l' \, m'} \bigg\}.
    \end{align}
\end{subequations}
We can now rewrite these equations in the suggestive matrix form
\begin{align}
    \label{eq:bcMatrix}
    \sum_{l \, m} \bM_{l \, m}^{l' \, m'} \cdot \bpsi_{l \, m} = 0,
\end{align}
where we introduced the \emph{perturbation matrix}
\begin{align}
\label{eq:bMdef}
    \bM_{l \, m}^{l' \, m'} =
    \begin{bmatrix}
        [A_1^\Phi]_{l \, m}^{l' \, m'} &  -[A_2^\Phi]_{l \, m}^{l' \, m'} & [B_1^\Phi]_{l \, m}^{l' \, m'}/n_1 & -[B_2^\Phi]_{l \, m}^{l' \, m'}/n_2 \\[6pt]
        -n_1 [B_1^\Psi]_{l \, m}^{l' \, m'} & n_2 [B_2^\Psi]_{l \, m}^{l' \, m'} & [A_1^\Psi]_{l \, m}^{l' \, m'} & -[A_2^\Psi]_{l \, m}^{l' \, m'} \\[6pt]
        [A_1^\Psi]_{l \, m}^{l' \, m'} &  -[A_2^\Psi]_{l \, m}^{l' \, m'} & [B_1^\Psi]_{l \, m}^{l' \, m'}/n_1 & -[B_2^\Psi]_{l \, m}^{l' \, m'}/n_2 \\[6pt]
        -n_1 [B_1^\Phi]_{l \, m}^{l' \, m'} & n_2 [B_2^\Phi]_{l \, m}^{l' \, m'} & [A_1^\Phi]_{l \, m}^{l' \, m'} & -[A_2^\Phi]_{l \, m}^{l' \, m'}
    \end{bmatrix},
\end{align}
as well as the (field-) \emph{coefficients vector}
\begin{align}
\label{eq:bpsidef}    
    \bpsi_{l \, m} =
    \begin{bmatrix}
        a_{l \, m}^\text{E} \\[6pt]
        b_{l \, m}^\text{E} \\[6pt]
        a_{l \, m}^\text{M} \\[6pt]
        b_{l \, m}^\text{M}
    \end{bmatrix} .
\end{align}
We want to emphasize that the matrix formulation of the boundary condition in \eqref{eq:bcMatrix} is \emph{exact}. Admittedly, it contains infinite sums and, as we have to consider all $l'$ and $m'$, infinitely many equations. However, this boundary condition is favorable compared to previous formulations, as it does not include any radial or angular dependence and is a single, non-redundant equation. Additionally, it is a set of linear equations and we have a wealth of mathematical tools to solve problems of this kind.
\subsection{The Unperturbed Problem}
\label{sec:unperturbed}
To get further convinced of the usefulness of our new boundary condition, let us consider the unperturbed problem. As the spherical boundary is described by $R(\theta,\phi) = r_0$ and therefore $\bn_\parallel = 0$, we can substitute this into \eqref{eq:bAbB} to find
\begin{align}
    \bA_{\alpha , l \, m}^u(\theta,\phi) = A_{\alpha \, l}^\Phi(r_0) \bPhi{l}{m}(\theta,\phi), \und
    \bB_{\alpha , l \, m}^u(\theta,\phi) = A_{\alpha \, l}^\Psi(r_0) \bPsi{l}{m}(\theta,\phi),
\end{align}
where we added the superscript $u$ to distinguish quantities of the unperturbed problem from the general ones. As the entire angular dependence is encoded in the vector spherical harmonics, we can employ their orthogonality \eqref{eq:VSHorthogonality} to determine the perturbation matrix elements using \eqref{eq:ABmatrices}. A straight-forward calculation gives
\begin{subequations}
    \label{eq:ABmatricesUnperturbed}
    \begin{align}
        [A_\alpha^{\Phi \, u}]_{l \, m}^{l' \, m'} &= \delta_{l \, l'} \delta_{m \, m'} A_{\alpha \, l}^\Phi(r_0), \\[4pt]
        [A_\alpha^{\Psi \, u}]_{l \, m}^{l' \, m'} &= 0, \\[4pt]
        [B_\alpha^{\Phi \, u}]_{l \, m}^{l' \, m'} &= 0, \\[4pt]
        [B_\alpha^{\Psi \, u}]_{l \, m}^{l' \, m'} &= \delta_{l \, l'} \delta_{m \, m'} A_{\alpha \, l}^\Psi(r_0).
    \end{align}
\end{subequations}
By going back to the definition of the radial functions \eqref{eq:Ar} we find $A_{\alpha \, l}^\Phi(r_0) = 1$ and recalling that the radial functions actually depend on $k_\alpha \, r$, we again introduce the dimensionless wave number $x^u = k_0 \, r_0$. For later convenience, let us define $R_{\alpha \, l}^\Psi(x^u) \equiv A_{\alpha \, l}^\Psi(r_0)$ to rewrite the matrix equation \eqref{eq:bcMatrix} in the case of an undeformed sphere as
\begin{align}
    \label{eq:MpsiunperturbedExplicit}
    \begin{bmatrix}
        1 &  -1 & 0 & 0 \\[6pt]
        -n_1 R_{1 \, l}^\Psi(x^u) & n_2 R_{2 \, l}^\Psi(x^u) & 0 & 0 \\[6pt]
        0 & 0 & R_{1 \, l}^\Psi(x^u)/n_1 & -R_{2 \, l}^\Psi(x^u)/n_2 \\[6pt]
        0 & 0 & 1 & -1
    \end{bmatrix}
    \cdot
    \begin{bmatrix}
        a_{l \, m}^{\text{E} \, u} \\[6pt]
        b_{l \, m}^{\text{E} \, u} \\[6pt]
        a_{l \, m}^{\text{M} \, u} \\[6pt]
        b_{l \, m}^{\text{M} \, u}
    \end{bmatrix}
    = 0 ,
\end{align}
where we performed the infinite sum over $l$ and $m$ using the Kronecker deltas in \eqref{eq:ABmatricesUnperturbed} and renamed the indices $l' \rightarrow l$ and $m' \rightarrow m$. We again write it compactly as
\begin{align}
    \label{eq:Mpsiunperturbed}
    \bM_l(x^u) \cdot \bpsi_{l \, m}^u = 0.
\end{align}
This linear equation allows two types of solutions, the trivial one, characterized by a non-vanishing determinant of the system, and the non-trivial solutions, characterized by a vanishing determinant. To calculate the determinant, we notice that the perturbation matrix is block-diagonal, i.e.,
\begin{align}
    \label{eq:bMDirectSum}
    \bM_l(x^u) = \bM_l^\text{E}(x^u) \oplus \bM_l^\text{M}(x^u).
\end{align}
With this property, it is clear that the boundary condition does not mix $a_{l \, m}^{\text{E} \, u}$ and $b_{l \, m}^{\text{E} \, u}$ with $a_{l \, m}^{\text{M} \, u}$ and $b_{l \, m}^{\text{M} \, u}$, i.e., the boundary condition does not mix $\text{TE}$- and $\text{TM}$- modes. This property also allows the factorization of the determinant via
\begin{align}
    \label{eq:detMdetMEdetMM}
    \det \bM_l(x^u) = \det \bM_l^\text{E}(x^u) \, \det \bM_l^\text{M}(x^u).
\end{align}
To calculate the determinant of both $2 \times 2$ matrices, we recall the definitions of $R_{\alpha \, l}^\Psi(x^u)$ and the equations \eqref{eq:fTEfTM} to find
\begin{align}
    \det \bM_l^\text{E}(x^u) = i \, f_l^\text{TE}(x^u)/x^u, \und \det \bM_l^\text{M}(x^u) = -i \, x^u \, f_l^\text{TM}(x^u).
\end{align}
Hence the determinant of the joint system reads
\begin{align}
    \det \bM_l(x^u) = f_l^\text{TE}(x^u) \, f_l^\text{TM}(x^u).
\end{align}
This sets us in the same situation as in Section \ref{sec:perfectsolution}: On the one hand, we get the trivial solution of \eqref{eq:Mpsiunperturbed} if $x^u$ is no eigenvalue of the perfectly spherical body. On the other hand, the non-trivial solutions are the $\text{TE}$- or $\text{TM}$-resonances characterized by $x^u = x_{l \, n}^\sigma$. Thus, we find full agreement of both approaches.
\newpage

\section[Resonances of a Deformed Dielectric Sphere: TE-Case]{Resonances of a Deformed Dielectric Sphere: \\ TE-Case}
\label{sec:solutionTE}
Equipped with the boundary condition encoded in the matrix equation \eqref{eq:bcMatrix}, we can conveniently employ BCPT and determine the resonances of slightly deformed dielectric spheres. Since the treatment of $\TE$-resonances allow simplifications compared to the $\TM$-resonances, we determine the corrections of $\TE$-resonances in this section, and the correction of the $\TM$-resonances in Section \ref{sec:solutionTM}. \par
To determine the corrections to the $\TE$-resonances, we first introduce a quantum-like notation and do some initial discussions in Section \ref{sec:quantum}. With this, we are able to solve the first-order equations conveniently in Section \ref{sec:first}. As the result of the first-order equations allow two distinct higher-order approaches, we study them separately in in Sections \ref{sec:secondNonDegenerate} and \ref{sec:secondDegenerate} and find the resonances of the perturbed system up to and including second-order terms.
\subsection{Introduction of a Quantum-like Notation}
\label{sec:quantum}
To solve the full boundary condition \eqref{eq:bcMatrix} in an efficient and clear manner, it is convenient to adopt a quantum-like notation to represent the coefficients and the variables of this equation. The use of such a notation is possible because one can always associate a linear operator to a matrix and vice versa. However, we should always keep in mind that we are dealing with a \emph{purely} classical physics problem, even though we are using quantum-like notation.
\subsubsection{Construction of the Hilbert Space}
\label{sec:hilbert}
To begin with, let us introduce the states $\ket{l \, m}$ with $l = 0,1, \ldots, \infty$ and $m = -l, {-l+1},$ $\ldots, l$. By hypothesis, they are orthonormal
\begin{align}
    \label{eq:lmOrthogonality}
    \brak{l' \, m'}{l \, m} = \delta_{l' \, l} \delta_{m' \, m},
\end{align}
and form an orthonormal basis of the infinite-dimensional Hilbert space that we denote as $\mathcal{H}_\infty$. The resolution of identity for this space reads
\begin{align}
     \hId_\infty = \sum_{l=0}^\infty \sum_{m=-l}^l \proj{l \, m}{l \,m}, 
\end{align}
and expresses its completeness. Here and hereafter, the subscript $\infty$ denotes operators in $\mathcal{H}_\infty$ and the caret symbol will mark operators in any infinite-dimensional Hilbert space. \par
Next we define the four vectors $\ket{i}$ with $i=1,2,3,4$. We also impose that they are orthonormal,
\begin{align}
    \label{eq:idef}
    \brak{i'}{i} = \delta_{i' \, i},
\end{align}
and span a four-dimensional Hilbert space denoted as $\mathcal{H}_4$. The completeness of this basis is again granted by the resolution of identity
\begin{align}
    I_4 = \sum_{i=1}^4 \proj{i}{i},
\end{align}
where the identity operator $I_4$ can be represented by the $4 \times 4$ identity matrix. \par
Finally we define the tensor product Hilbert space
\begin{align}
    \mathcal{H} = \mathcal{H}_\infty \otimes \mathcal{H}_4,
\end{align}
which is by definition spanned by the vectors
\begin{align}
    \label{eq:ketlmi}
    \ket{l \, m \, i} = \ket{l \,m } \otimes \ket{i}.
\end{align}
The completeness relation for $\mathcal{H}$ then reads:
\begin{align}
    \label{eq:Id}
    \hat{\mathcal{I}} = \hId_\infty \otimes I_4 = \sum_{l=0}^\infty \sum_{m=-l}^l \sum_{i=1}^4 \proj{l \, m \, i}{l \, m \, i},
\end{align}
where here and hereafter operators in $\mathcal{H}$ will be denoted by calligraphic letters and the caret again indicates the infiniteness of the associated Hilbert space. From now on we will write the triple sums in a more compact way as in \eqref{eq:sumlmShort}, so
\begin{align}
    \sum_{l \, m \, i} && \text{stands for} && \sum_{l=0}^\infty \sum_{m=-l}^l \sum_{i=1}^4.
\end{align}
\subsubsection{Derivation of an Operator Equation}
\label{sec:bcOperator}
Equipped with this paraphernalia, we can rewrite our full boundary condition \eqref{eq:bcMatrix} in terms of states and operators in $\mathcal{H}$. \par 
To do so, we first introduce the vector state $\ket{\psi}$ to be represented by the coefficients vector $\bpsi_{l \, m}$ defined in \eqref{eq:bpsidef}, i.e.,
\begin{align}
    \ket{\psi} \doteq \bpsi_{l \, m},
\end{align}
where we use the symbol $\doteq$ to denote a vector or matrix representation. As usual, we can go from the abstract state $\ket{\psi}$ to the vector representation by multiplying a basis state $\bra{l \, m \, i}$ from the left, and we define
\begin{align}
    \label{eq:psilmidef}
    \psi_{l \, m\, i} = \brak{l \, m \, i}{\psi}.
\end{align}
Here, $i$ denotes the component of the vector $\bpsi_{l \, m}$, as
\begin{align}
    \label{eq:psiexplicit}
    \bpsi_{l \, m} = 
    \begin{bmatrix}
        \psi_{l \, m\, 1} \\[4pt]
        \psi_{l \, m\, 2} \\[4pt]
        \psi_{l \, m\, 3} \\[4pt]
        \psi_{l \, m\, 4}
    \end{bmatrix}
    =
    \begin{bmatrix}
        a_{l \, m}^\text{E} \\[4pt]
        b_{l \, m}^\text{E} \\[4pt]
        a_{l \, m}^\text{M} \\[4pt]
        b_{l \, m}^\text{M}
    \end{bmatrix}.
\end{align}
\par
Secondly, we want to define the \emph{perturbation operator} $\hmM$ so that by acting with it on $\ket{\psi}$ we reproduce \eqref{eq:bcMatrix}. Therefore, we define the matrix elements of the perturbation operator in terms of the matrix elements of the perturbation matrix as
\begin{align}
    \label{eq:Mdef}
    \bra{l' \, m' \, i'} \hmM \ket{l \, m \, i} = [\bM_{l \, m}^{l' \, m'}]_{i'\, i},
\end{align}
where, in comparison with \eqref{eq:bMdef}, we have for example
\begin{align*}
    [\bM_{l \, m}^{l' \, m'}]_{1 1} = [A_1^\Phi]_{l \, m}^{l' \, m'},
    \und
    [\bM_{l\, m}^{l' \, m'}]_{3 4} = -[B_2^\Psi]_{l \, m}^{l' \, m'}/n_2. 
\end{align*}
Using this notation, we can equivalently rewrite \eqref{eq:bcMatrix} as
\begin{align}
    \label{eq:MpsiFull}
    \hmM \ket{\psi} = 0.
\end{align}
We can show this by inserting an identity \eqref{eq:Id} left and right of $\hmM$ and find
\begin{align*}
    \hmM \ket{\psi} &= \sum_{l' \, m' \, i'} \sum_{l \, m \, i} \proj{l' \, m' \, i'}{l' \, m' \, i'} \hmM \proj{l \, m \, i}{l \, m \,i} \psi \rangle \\[6pt]
    &= \sum_{l' \, m' \, i'} 
    \underbrace{
    \left(\sum_{l \, m \, i} [\bM_{l \, m}^{l' \, m'}]_{i' \, i} \ \psi_{l \, m \, i} \right)
    }_{\text{$=0$ from \eqref{eq:bcMatrix}}}
    \ket{l' \, m' \, i'} \\
    &= 0 .
\end{align*}
\subsubsection{Application of BCPT}
\label{sec:applicationBCPT}
Now we want to apply BCPT using this quantum-like notation. In Section \ref{sec:bcpt} we discussed that the idea is to expand the boundary condition, and therefore all quantities in \eqref{eq:MpsiFull}, in powers of a supposedly small parameter $\epsilon$. To this end, we start by doing some purely \emph{formal} manipulations, without worrying about the convergence of the power series. \par
Let us start by expanding the state $\ket{\psi} \equiv \ket{\psi(\epsilon)}$ as
\begin{align}
    \label{eq:psiExpansion}
    \ket{\psi(\epsilon)} = \ket{\psi^\z} + \epsilon \, \ket{\psi^\one} + \epsilon^2 \, \ket{\psi^\two} + \dots \, .
\end{align}
Similarly we want to expand the perturbation operator $\hmM \equiv \hmM(x,\epsilon)$, where the eigenvalue $x$ encodes the optical resonance of the perturbed system. In the light of customary perturbation theories we first expand $x$ in powers of $\epsilon$ as
\begin{align}
    \label{eq:xExpansion}
    x(\epsilon) = x^\z + \epsilon \, x^\one + \epsilon^2 \, x^\two + \dots \,,
\end{align}
and introduce the expansion of the perturbation operator
\begin{align}
    \label{eq:MExpansion}
    \hmM (x(\epsilon),\epsilon) &= \hmM^\z + \epsilon \, \hmM^\one + \epsilon^2 \, \hmM^\two + \dots \,.
\end{align}
Since $x(0) = x^\z$ from \eqref{eq:xExpansion}, we notice that the zeroth-order perturbation operator can be rewritten as
\begin{align}
    \label{eq:M0}
    \hmM^\z = \hmM(x^{(0)},0) .
\end{align}
\par
Now that we expanded the state $\ket{\psi}$ in \eqref{eq:psiExpansion} as well as the perturbation operator $\hmM$ in \eqref{eq:MExpansion}, we can insert them into the boundary condition \eqref{eq:MpsiFull}. We find
\begin{align}
    0 &= \hphantom{\big(} \hmM(x(\epsilon),\epsilon) \, \ket{\psi(\epsilon)} \notag \\[6pt]
    &= \left(\hmM^\z + \epsilon \, \hmM^\one + \epsilon^2 \, \hmM^\two + \dots \right) \left( \ket{\psi^\z} + \epsilon \, \ket{\psi^\one} + \epsilon^2 \, \ket{\psi^\two} + \dots \right) \notag \\[6pt]
    &= \hphantom{\big(}
    \begin{aligned}[t]
        \hmM^\z \ket{\psi^\z} &+ \epsilon\hphantom{^2} \left( \hmM^\z \ket{\psi^\one} + \hmM^\one \ket{\psi^\z} \right) \notag \\[4pt]
        &+ \epsilon^2 \left( \hmM^\z \ket{\psi^\two} + \hmM^\one \ket{\psi^\one} + \hmM^\two \ket{\psi^\z} \right) + \dots \,.
    \end{aligned}
\end{align}
We can solve this equation perturbatively by requiring that each term in this equation needs to vanish separately. Thus we find the chain of equations
\begin{subequations}
    \begin{align}
        &\epsilon^0: & \hmM^\z \ket{\psi^\z} & = 0 \label{eq:Mpsi0}, \\[4pt]
        &\epsilon^1: & \hmM^\z \ket{\psi^\one} + \hmM^\one \ket{\psi^\z} &= 0 \label{eq:Mpsi1}, \\[4pt]
        &\epsilon^2: & \hmM^\z \ket{\psi^\two} + \hmM^\one \ket{\psi^\one} + \hmM^\two \ket{\psi^\z} &= 0, \label{eq:Mpsi2} \\
        &\vdots & &\ \, \vdots \notag
    \end{align}
\end{subequations}
which we need to solve successively to determine the resonance $x$ as well as the field coefficients encoded by $\ket{\psi}$ up to the desired order. \par
This sets in a similar situation as in quantum mechanics, where the analogous perturbative approach is called \emph{Rayleigh-Schrödinger perturbation theory}, also known as \emph{time-independent perturbation theory} or \emph{stationary state perturbation theory} (cf., e.g., \cite{sakurai14}). This is due to the fact that the underlying Hilbert space structure is compatible. However, besides this mathematical similarity, the underlying physics encoded in the operators and states is completely different.
\subsubsection{Properties of the Perturbation Operator}
\label{sec:properties}
Let us remark up to now this is only a formal expansion and we did not use any properties of the perturbation operator. However, as we defined it in terms of the matrix elements of the perturbation matrix in \eqref{eq:Mdef}, and the matrix elements are given by \eqref{eq:bMdef}, one can explicitly calculate them. As these computations are rather lengthy and not very insightful, they are presented with full details in Appendix \ref{app:mat}. To progress, we want to summarize the most important results of these computations. The first noticeable one is that we actually find an expansion as depicted in \eqref{eq:MExpansion}. The second result is that we can rewrite the zeroth-, first- and second-order perturbation operators as
\begin{subequations}
    \label{eq:MDecomposition}
    \begin{align}
        \hmM^\z &= \hmD^\z \label{eq:M0Decomposition}, \\[4pt]
        \hmM^{(\nu)} &= \hmV^{(\nu)} + x^{(\nu)} \hmD^{(\nu)}, \label{eq:MnuDecomposition}
    \end{align}
\end{subequations}
where $\nu=1,2$ denotes first- and second-order operators. These newly introduced operators have further properties. The first one addresses the operators $\hmD^{(n)}$, where $n=0,1,2$ denotes all orders considered in this work. They are diagonal with respect to the basis vectors $\ket{l \,m}$, that is,
\begin{align}
    \label{eq:Dndef}
    \hmD^{(n)} \ket{l \, m \, i} = \ket{l \, m} \otimes M_l^{(n)} \ket{i},
\end{align}
and the operator $M_l^{(n)}$ can be represented by a $4 \times 4$ matrix. By multiplying $\hmD^{(n)}$ with $\bra{l' \, m' \, i'}$ from the left, one finds its matrix elements
\begin{align}
    \label{eq:Dnmat}
    \bra{l' \, m' \, i'} \hmD^{(n)} \ket{l \, m \, i}
    &= \delta_{l \, l'} \delta_{m \, m'} \bra{i'} M_l^{(n)} \ket{i},
\end{align}
where especially $\bra{i'} M_l^\z \ket{i}$ is represented by $\bM_l(x^\z)$ defined in \eqref{eq:Mpsiunperturbed}. \par
The second property of the operators $\hmV^{(\nu)}$ and $\hmD^{(\nu)}$ is that they are independent of $x^{(\nu)}$, which we write as
\begin{align}
    \label{eq:VDindependent}
    \frac{\mathrm{d}\hmV^{(\nu)}}{\mathrm{d}x^{(\nu)}} = 0,
    \und
    \frac{\mathrm{d}\hmD^{(\nu)}}{\mathrm{d}x^{(\nu)}} = 0 .
\end{align}
Simply speaking, this property states that the entire $x^{(\nu)}$-dependence of $\hmM^{(\nu)}$ in \eqref{eq:MnuDecomposition} is explicitly written down.
\subsection{Zeroth-Order Perturbation Theory}
\label{sec:zeroth}
Let us now use our quantum-like notation by solving the zeroth order equation \eqref{eq:Mpsi0}. We can rewrite this equation using \eqref{eq:M0Decomposition} as
\begin{align}
    \label{eq:D0psi0}
    \hmD^\z \ket{\psi^\z} = 0.
\end{align}
By inserting two resolutions of identity \eqref{eq:Id} left and right of $\hmD^\z$ we find
\begin{align}
    0 &= \sum_{l' \, m' \, i'} \sum_{l \,m \, i} \ket{l'\, m' \, i'} \mean{l' \, m' \, i'}{\hmD^\z }{l \, m \, i} \, \brak{l \, m \, i}{\psi^\z} \notag \\
    &= \sum_{l' \, m' \, i'} \sum_{l \,m \, i} \ket{l'\, m' \, i'} \, \delta_{l \, l'} \delta_{m \, m'} \, \bra{i'} M_l^{(0)} \, \ket{i} \, \psi_{l \, m\, i}^\z \notag \\
    &= \sum_{l' \, m' \, i'} \left[ \sum_{i} \bra{i'} M_l^\z \ket{i} \, \psi_{l \, m\, i}^\z \right] \ket{l' \, m' \, i'} \label{eq:100},
\end{align}
where we used \eqref{eq:Dndef}, defined $\psi_{l \, m\, i}^\z = \brak{l \, m \, i}{\psi^\z}$ in the second line and carried out the summation over $l$ and $m$ in the last one. Since the $\ket{l \, m \, i}$ form a complete basis of $\mathcal{H}$, all the coefficients in the square bracket need to be identically zero. That is
\begin{align}
    \sum_{i = 1}^4 \bra{i'} M_l^\z \ket{i} \, \psi_{l \, m\, i}^\z = 0.
\end{align}
Of cause, this is the same result that we would had obtained if we just multiplied \eqref{eq:D0psi0} from the left by $\bra{l' \, m' \, i'}$ and we will use this property heavily later on. As previously stated, $M_l^\z$ in basis $\ket{i}$ has the matrix representation $\bM_l(x^\z)$ defined by \eqref{eq:Mpsiunperturbed} and $\psi_{l \, m\, i}^\z$ is represented by the vector $\bpsi_{l \, m}^\z$. Hence the previous equation is equivalent to
\begin{align}
    \label{eq:M0psi0}
    \bM_l(x^\z) \cdot \bpsi_{l \, m}^\z = 0.
\end{align}
But this sets us exactly in the same situation as for the unperturbed problem, discussed in Section \ref{sec:unperturbed}, if we require $x^\z \equiv x^u$, i.e., the zeroth-order eigenvalue has to be a valid eigenvalue of the unperturbed problem. This confirms our physical expectation that the zeroth-order perturbation theory reproduces the unperturbed problem. Thus, we can now interpret $x^\z$ as an \emph{unperturbed eigenvalue} and the quantities $x^\one$ and $x^\two$ as its \emph{first- and second-order corrections} respectively. \par
To continue, let us write \eqref{eq:M0psi0} in its full form as
\begin{align}
    \label{eq:M0psi0lmat}
    \begin{bmatrix}
        1 &  -1 & 0 & 0 \\[6pt]
        -n_1 R_{1 \, l}^\Psi(x^\z) & n_2 R_{2 \, l}^\Psi(x^\z) & 0 & 0 \\[6pt]
        0 & 0 & R_{1 \, l}^\Psi (x^\z) /n_1 & -R_{2 \, l}^\Psi (x^\z) /n_2 \\[6pt]
        0 & 0 & 1 & -1
    \end{bmatrix}
    \cdot
    \begin{bmatrix}
        a_{l \, m}^{\text{E} \, \z} \\[6pt]
        b_{l \, m}^{\text{E} \, \z} \\[6pt]
        a_{l \, m}^{\text{M} \, \z} \\[6pt]
        b_{l \, m}^{\text{M} \, \z}
    \end{bmatrix}
    = 0 .
\end{align}
We recall from Section \ref{sec:perfectsolution} that the solutions of this equation are characterized by the resonances $x_{l \, n}^\sigma$, where $\sigma=\TE,\TM$, $l=1,2,\dots$ and $n=1,2,\dots$\,. Let us now choose the specific $\TE$-mode $x^\z = x_{\lz \, n_0}^\TE$. \par
From the aforementioned section we know that \eqref{eq:M0psi0lmat} is trivially solvable if $l \neq l_0$. Together with the fact that the resonances are non-degenerate with respect to $l$, we find $\bpsi_{l \, m}^\z = 0$ for $l \neq \lz$. Thus, the last step is to solve \eqref{eq:M0psi0lmat} for $l = \lz$. To do so, we first find that the $\text{TE}$-modes are characterized by
\begin{align*}
    f_\lz^\TE(x^\z)= -n_1 R_{1 \, \lz}^\Psi(x^\z) + n_2 R_{2 \, \lz}^\Psi(x^\z) = 0,
\end{align*}
where we recalled \eqref{eq:fTEfTMvanish} and the definition of $R_{\alpha \, \lz}^\Psi(x^\z)$ from Section \ref{sec:unperturbed}. Now it is useful to introduce
\begin{align}
    z = n_1 R_{1 \, \lz}^\Psi(x^\z) = n_2 R_{2 \, \lz}^\Psi(x^\z),
\end{align}
to rewrite \eqref{eq:M0psi0lmat} for $l=\lz$ compactly as
\begin{align}
    \label{eq:M0zpsi0mat}
    \begin{bmatrix}
        1 &  -1 & 0 & 0 \\[10pt]
        -z & z & 0 & 0 \\[10pt]
        0 & 0 & z/n_1^2 & -z/n_2^2 \\[10pt]
        0 & 0 & 1 & -1
    \end{bmatrix}
    \cdot
    \begin{bmatrix}
        a_{\lz \, m}^{\text{E} \, \z} \\[10pt]
        b_{\lz \, m}^{\text{E} \, \z} \\[10pt]
        a_{\lz \, m}^{\text{M} \, \z} \\[10pt]
        b_{\lz \, m}^{\text{M} \, \z}
    \end{bmatrix}
    = 0 .
\end{align}
From the first line we find that $a_{\lz \, m}^{\text{E} \, \z} = b_{\lz \, m}^{\text{E} \, \z}$ and then the second line gives a trivial identity. The remaining two lines give $a_{\lz \, m}^{\text{M} \, \z} = 0 = b_{\lz \, m}^{\text{M} \, \z}$, as the determinant of the magnetic block is non-zero due to the fact that the resonances are non-degenerate with respect to the polarization $\sigma$ as discussed in Section \ref{sec:perfectsolution}. \par
Now that we fully solved the zeroth order, let us summarize the results. By choosing an unperturbed eigenvalue $x^\z = x_{\lz \, n_0}^\text{TE}$, we found the field coefficients
\begin{align}
    \label{eq:psi0}
    \bpsi_{\lz \, m}^\z = a_{\lz \, m}^{\text{E} \, \z}
    \begin{bmatrix}
    1 \\ 1 \\ 0 \\ 0
    \end{bmatrix},
    \und
    \bpsi_{l \, m}^\z =
    \begin{bmatrix}
    0 \\ 0 \\ 0 \\ 0
    \end{bmatrix},
    \qquad l \neq l_0 .
\end{align}
Up to now, the $2\,\lz+1$ coefficients $a_{\lz \, m}^{\text{E} \, \z}$ are completely arbitrary real or complex numbers.
\subsubsection{Change of Basis}
The four-dimensional basis $\ket{i}$ defined by \eqref{eq:idef}, although natural, is not the most suitable basis for further considerations. In the previous paragraphs we noticed that we have a vanishing determinant in the electric block which results in the trivial identity in the second line of \eqref{eq:M0zpsi0mat}. To exploit this property we diagonalize the electric block. Let us consider the operator $M_\lz^\text{E}$ in $\mathcal{H}_2$, which encodes the electric block. In basis $\ket{i}$, $i=1,2$, it is given by
\begin{align}
    \label{eq:MEi}
    \mean{i'}{M_\lz^\text{E}}{i} \doteq \bM_\lz^\text{E}(x^\z) = 
    \begin{bmatrix}
        1 &  -1 \\[4pt]
        -z & z \\
    \end{bmatrix},
\end{align}
where $\bM_\lz^\text{E}$ was defined in \eqref{eq:bMDirectSum}. First of all we notice that this matrix, and therefore also the operators $M_\lz^\text{E}$ and $M_\lz^\z$, is non-Hermitian. To handle this in a convenient way, we employ bi-orthogonal states, which is a method heavily used in quantum mechanics when considering open systems, characterized by non-Hermitian Hamiltonians \cite{sternheim72,rotter09,brody13}. \par
The idea of this method is to introduce two sets of eigenvectors of $M_\lz^\text{E}$, the \emph{right}-eigenvectors $\ket{\alpha_i}$ and the \emph{left}-eigenvectors $\bra{\ta_i}$, so that
\begin{subequations}
    \begin{align}
        M_\lz^\text{E} \, \ket{\alpha_i} &= \lambda_i \, \ket{\alpha_i},\\
        \bra{\ta_i} \, M_\lz^\text{E} &= \lambda_i \, \bra{\ta_i}.
    \end{align}
\end{subequations}
In the case of non-Hermitian operators, one has in general $\ket{\alpha_i}^\dagger = \bra{\alpha_i} \neq \bra{\ta_i}$. It is not hard to show that these eigenvalue equations are satisfied by
\begin{subequations}
    \label{eq:alphadef}
    \begin{align}
        \ket{\alpha_1} &\doteq
        \begin{bmatrix}
            1 \\ 1
        \end{bmatrix},
        & \bra{\ta_1} &\doteq 
        \frac{1}{1+z}
        \begin{bmatrix}
        z & 1
        \end{bmatrix}, \label{eq:alpha1def} \\[6pt]
        \ket{\alpha_2} &\doteq \frac{1}{1+z} 
        \begin{bmatrix}
            1 \\ -z
        \end{bmatrix},
        & \bra{\ta_2} &\doteq
        \begin{bmatrix}
        1 & -1
        \end{bmatrix},
    \end{align}
\end{subequations}
with $\lambda_1 = 0$ and $\lambda_2 = z+1$. By writing these vectors we have chosen the standard normalization for bi-orthogonal vectors, that is
\begin{align}
    \brak{\ta_{i'}}{\alpha_i} = \delta_{i' \, i}, && i,i' = 1,2,
\end{align}
and by inspection we also find
\begin{align}
    \sum_{i=1}^2 \proj{\alpha_i}{\ta_i} = I_2,
\end{align}
where the identity operator $I_2$ is represented by the $2 \times 2$ identity matrix. Thus we found that our bi-orthogonal states form a complete basis of $\mathcal{H}_2$. \par
We now want to promote this new basis to a complete basis of $\mathcal{H}_4$. Therefore we simply extend the vector representation of our state by padding it with two zeros, e.g.,
\begin{align}
    \bra{\ta_2} &\doteq
    \begin{bmatrix}
    1 & -1 & 0 & 0
    \end{bmatrix}.
\end{align}
They get completed by defining\footnote{An alternative possibility would be to fully diagonalize $M_\lz^\z$ defined by \eqref{eq:M0zpsi0mat}, but this is not necessary and not practically convenient.}
\begin{align}
    \ket{\alpha_i} = \ket{i}, \und  \bra{\ta_i} = \bra{i}, && i=3,4.
\end{align}
It is an elementary algebra exercise to verify that these basis vectors also satisfy the standard normalization condition for bi-orthogonal vectors,
\begin{align}
    \brak{\ta_{i'}}{\alpha_i} = \delta_{i' i}, && i,i'=1,2,3,4,
\end{align}
and form a complete basis of $\mathcal{H}_4$, i.e.,
\begin{align}
    \sum_{i=1}^4 \proj{\alpha_i}{\ta_i} = I_4.
\end{align}
By construction, we diagonalized the electric block of \eqref{eq:M0zpsi0mat} and kept the magnetic block, so in the new basis we have
\begin{align}
    \label{eq:alphaM0alpha}
    \bra{\ta_{i'}} M_\lz^\z \ket{\alpha_i} \doteq
    \begin{bmatrix}
      0 & 0 & 0 & 0 \\
      0 & 1+z & 0 & 0 \\
      0 & 0 & z/n_1^2 & -z/n_2^2 \\
      0 & 0 & 1 & -1
    \end{bmatrix},
\end{align}
where the determinant of the lower $3 \times 3$ matrix is non-vanishing. \par
With our new basis we can also rewrite our zeroth-order result. To do so, we rewrite our zeroth-order state as
\begin{align}
    \label{eq:psi0intermediate}
    \ket{\psi^\z} &= \sum_{l \, m \, i} \ket{l \, m\, i} \brak{l \, m\, i}{\psi^\z} \\
    &= \sum_{m = -\lz}^\lz \sum_{i=1}^2  a_{\lz \, m}^{\text{E} \, \z} \, \ket{\lz \, m \, i} \notag \\
    &= \sum_{m = -\lz}^\lz  a_{\lz \, m}^{\text{E} \, \z} \, \ket{\lz \, m} \otimes \sum_{i=1}^2 \ket{i},
\end{align}
where we used \eqref{eq:psi0} in the second line and \eqref{eq:ketlmi} in the last one. However, we notice that
\begin{align}
    \sum_{i=1}^2 \ket{i} = \ket{\alpha_1},
\end{align}
and we can conveniently write \eqref{eq:psi0intermediate} as
\begin{align}
    \label{eq:psi0final}
    \ket{\psi^\z} = \sum_{m = -\lz}^\lz a_{\lz \, m}^{\text{E} \, \z} \, \ket{\lz \, m \, \alpha_1},
\end{align}
where we introduced 
\begin{align}
    \label{eq:lmalphadef}
    \ket{\lz \, m \, \alpha_i} = \ket{\lz \, m} \otimes \ket{\alpha_i}.
\end{align}
It needs to be understood that the states $\ket{\lz \, m \, \alpha_1}$ are the fundamental solutions of our unperturbed problem, that is,
\begin{align}
    \label{eq:D0l0malpha1}
    \hmD^\z \, \ket{\lz \, m \, \alpha_1} = 0,
\end{align}
and likewise, the bi-orthogonal conjugate expression $\bra{\lz \, m \, \ta_1}$ satisfies the conjugate zeroth-order equation
\begin{align}
    \label{eq:l0mta1D0}
    \bra{\lz \, m \, \ta_1} \, \hmD^\z = 0.
\end{align}
\subsubsection{Some Preparatory Remarks}
\label{sec:degeneracy}
In the previous section we found that all $\ket{\lz \, m \, \alpha_1}$ are solutions of the zeroth-order equation \eqref{eq:D0l0malpha1}, and the zeroth-order result is an undetermined linear combination of those states. The reason for this is that the unperturbed resonance $x_{\lz \, n_0}^\text{TE}$ is $(2\,\lz+1)$-fold \emph{degenerate} with respect to $m$. We need to take this fact into consideration when solving the higher-order equations. \par
To treat the degeneracy\footnote{There is no non-degenerate case as $\lz > 0$.} in a convenient and clear manner, let us consider the states
\begin{align}
    \ket{\lz \, m} = \ket{\lz , -\lz}, \ket{\lz , -\lz +1}, \dots, \ket{\lz , \lz},
\end{align}
in $\mathcal{H}_\infty$ associated to $x_{\lz \, n_0}^\text{TE}$.
These states span a $2\,\lz+1$-dimensional subspace of $\mathcal{H}_\infty$ which we denote as \emph{degenerate subspace $\mathcal{H}_\lz$}. With this, we can split $\ket{\psi(\epsilon)}$ into a part contained in $\mathcal{H}_\lz \otimes \mathcal{H}_4$ and a part contained in its complementary subspace $\mathcal{H} \setminus \mathcal{H}_\lz \otimes \mathcal{H}_4$ via
\begin{align}
    \label{eq:psiDecomposition1}
    \ket{\psi(\epsilon)} &= \underbrace{
    \sum_{m = -\lz}^{\lz} \sum_{i=1}^4  \brak{\lz \, m \, i}{\psi(\epsilon)} \, \ket{\lz \, m \, i}
    }_{\text{in $\mathcal{H}_\lz \otimes \mathcal{H}_4$} }
    + \underbrace{
    {\sum_{l \, m \, i}}' \brak{l \, m \, i}{\psi(\epsilon)} \, \ket{l \, m \, i}
    }_{\text{in $\mathcal{H} \setminus \mathcal{H}_\lz \otimes \mathcal{H}_4$ }},
\end{align}
where here and hereafter
\begin{align}
    {\sum_{l \, m \, i}}' && \text{stands for} && \sum_{\substack{
        l = 0 \\[2pt]
        l \neq \lz
    }}^\infty \sum_{m=-l}^l \sum_{i=1}^4.
\end{align}
Mathematically speaking, in \eqref{eq:psiDecomposition1} we split $\ket{\psi(\epsilon)}$ into a projection on the degenerate subspace and a projection on the complementary subspace. \par
From our zeroth-order considerations we found it convenient to use the basis $\ket{\lz \, m \, \alpha_i}$ in $\mathcal{H}_\lz \otimes \mathcal{H}_4$, so we rewrite \eqref{eq:psiDecomposition1} as
\begin{align}
    \label{eq:psiDecomposition2}
    \ket{\psi(\epsilon)} = \sum_{m = -\lz}^{\lz} \sum_{i=1}^4  \brak{\lz \, m \, \ta_i}{\psi(\epsilon)} \, \ket{\lz \, m \, \alpha_i} +
    {\sum_{l \, m \, i}}' \brak{l \, m \, i}{\psi(\epsilon)} \, \ket{l \, m \, i}.
\end{align}
To get a series expansion of $\ket{\psi(\epsilon)}$ as in \eqref{eq:psiExpansion}, we introduce a shorthand notation for the scalar products and expand them. In the degenerate subspace, we have
\begin{align}
    \psi_{m \, \alpha_i}(\epsilon) &\equiv \brak{\lz \, m \, \ta_i}{\psi(\epsilon)} \notag \\[6pt]
    &= \psi_{m \, \alpha_i}^\z + \epsilon \, \psi_{m \, \alpha_i}^\one + \epsilon^2 \, \psi_{m \, \alpha_i}^\two + \dots \,,
\end{align}
and in the non-degenerate subspace we have
\begin{align}
    \psi_{l \, m \, i}(\epsilon) &= \brak{l \, m \, i}{\psi(\epsilon)} \notag \\[6pt]
    &= \psi_{l \, m \, i}^\z + \epsilon \, \psi_{l \, m \, i}^\one + \epsilon^2 \, \psi_{l \, m \, i}^\two + \dots \,,
\end{align}
with $l \neq \lz$. With this definitions, we immediately find the zeroth-order results
\begin{align}
    \psi_{m \, \alpha_i}^\z = \delta_{i \, 1} \, a_{\lz \, m}^{\text{E} \, \z},
    \und
    \psi_{l \, m \, i}^\z = 0, \qquad l \neq \lz.
\end{align}
In order to solve the higher-order equations, we collect our previous results and find
\begin{align}
    \label{eq:psinDecomposition}
    \ket{\psi^{(n)}} = \sum_{m = -\lz}^{\lz} \sum_{i=1}^4 \psi_{m \, \alpha_i}^{(n)} \, \ket{\lz \, m \, \alpha_i}
    + {\sum_{l \, m \, i}}' \psi_{l \, m \, i}^{(n)} \, \ket{l \, m \, i} .
\end{align}
With this, we are fully equipped to solve the first-order equation.
\subsection{First-Order Perturbation Theory}
\label{sec:first}
Let us start by recalling the equation under consideration \eqref{eq:Mpsi1}. Together with \eqref{eq:MDecomposition}, it reads
\begin{align}
    \label{eq:first}
    \hmD^\z \, \ket{\psi^\one} + \left[ \hmV^\one + x^\one \hmD^\one \right]\ket{\psi^\z} = 0.
\end{align}
In order to solve this equation, we project onto the degenerate and non-degenerate subspace as discussed in the previous section.
\subsubsection{First-Order Eigenvalue Corrections}
\label{sec:firstordereigenvalue}
Let us start by considering this problem in the degenerate subspace. Therefore we multiply \eqref{eq:first} from the left with $\bra{\lz \, m' \, \ta_{i'}}$ to find
\begin{align}
    \label{eq:1stDegSub1}
    \bra{\lz \, m' \, \ta_{i'}} \hmD^\z \, \ket{\psi^\one} + \bra{\lz \, m' \, \ta_{i'}} \left[ \hmV^\one + x^\one \hmD^\one \right]\ket{\psi^\z} = 0,
\end{align}
which needs to hold for all $m'$ and $i'$. Recalling the zeroth-order result \eqref{eq:D0l0malpha1}, we find that the first term vanishes for $i' = 1$. Let us consider this specific case first. Inserting our zeroth-order state $\ket{\psi^\z}$ from \eqref{eq:psi0final} yields
\begin{align}
    \label{eq:eigenIntermediate}
    \sum_{m = -\lz}^\lz \left\{ \bra{\lz \, m' \, \ta_1} \hmV^\one \ket{\lz \, m \, \alpha_1} + x^\one \bra{\lz \, m' \, \ta_1}\hmD^\one \ket{\lz \, m \, \alpha_1} \right\} a_{\lz \, m}^{\text{E} \, \z} = 0 .
\end{align}
Now we can use \eqref{eq:Dnmat} to exploit the diagonal form of $\hmD^\one$ and do some straight forward manipulations to get\footnote{The division by $x^\z$ is always possible as $x^\z \neq 0$.}
\begin{align}
    \label{eq:eigenIntermedite2}
    \sum_{m = -\lz}^\lz \frac{1}{x^\z} \frac{\bra{\lz \, m' \, \ta_1} \hmV^\one \ket{\lz \, m \, \alpha_1}}{\bra{\ta_1}M_\lz^\one \ket{\alpha_1}} a_{\lz \, m}^{\text{E} \, \z}  = - \frac{x^\one}{x^\z} a_{\lz \, m'}^{\text{E} \, \z}.
\end{align}
This is nothing but an eigenvalue equation, which we rewrite as
\begin{align}
    \label{eq:eigen1}
    \bV \bolda = \Delta^\one \bolda.
\end{align}
Here we introduced the $(2 \, \lz + 1) \times (2 \, \lz + 1)$ matrix $\bV$ with elements
\begin{align}
    \label{eq:Vmatdef}
    \bV_{m' \, m} = \frac{1}{x^\z} \frac{\bra{\lz \, m' \, \ta_1} \hmV^\one \ket{\lz \, m \, \alpha_1}}{\bra{\ta_1}M_\lz^\one \ket{\alpha_1}} ,
\end{align}
and going through the explicit calculations in Section \ref{sec:TEfirstExplicit}, it turns out that $\bV$ is a Hermitian matrix. Furthermore, we defined the $(2 \, \lz + 1)$-component vector $\bolda$ with components
\begin{align}
    \label{eq:badef}
    a_m \equiv a_{\lz \, m}^{\text{E} \, \z},
\end{align}
and the, due to the Hermicity of $\bV$, real eigenvalues $\Delta^\one$ as
\begin{align}
    \label{eq:Delta1def}
    \Delta^\one = - \frac{x^\one}{x^\z} .
\end{align}
In this formulation it is clear that we find $2 \, \lz +1$ eigenpairs $ (\Delta^\one,\bolda)$
solving \eqref{eq:eigen1}, and we label them with $\mu$ as
\begin{align}
    (\Delta^\one(\mu),\bolda(\mu)), && \mu = -\lz,-\lz+1,\dots,\lz.
\end{align}
Thus we finally write \eqref{eq:eigen1} as
\begin{align}
    \label{eq:eigen2}
    \bV \bolda(\mu) = \Delta^\one(\mu) \bolda(\mu).
\end{align}
With this, we found the first-order eigenvalue corrections. \par
To proceed, we notice that the deformation is encoded in $\bV$ and that we can consider two cases separately. The first one is the non-degenerate case, where $\Delta^\one(\mu) \neq \Delta^\one(\mu')$ for all $\mu \neq \mu'$, i.e., all first-order eigenvalue corrections $x_\mu^\one$ are different. The second case is the degenerate case, where we have $\Delta^\one(\mu) = \Delta^\one(\mu')$ for some $\mu \neq \mu'$. Until Section \ref{sec:secondDegenerate}, let us assume that \emph{the perturbation fully removed the degeneracy}, i.e., we consider the non-degenerate case.
\subsubsection{Normalization and another Basis Change}
\label{sec:normAndBasisChange}
The next step is to discuss the eigenvectors $\bolda(\mu)$ of the Hermitian matrix $\bV$. First we recall that $a_m(\mu) \equiv a_{\lz \, m}^{\text{E} \, \z}(\mu)$ from \eqref{eq:badef}, which means that the first-order eigenvalue equation \eqref{eq:eigen2} determines the zeroth-order coefficients $a_{\lz \, m}^{\text{E} \, \z}$. \par
Furthermore we did not rely on any normalization of the electromagnetic fields, and thus on any constraint on the $a_{\lz \, m}^{\text{E} \, \z}$, to determine the unperturbed eigenvalues as well as the corresponding first-order eigenvalue corrections. Due to the Hermicity of $\bV$, the $\bolda(\mu)$ are orthogonal and it will be advantageous to choose them orthonormal as
\begin{align}
    \label{eq:baOrthogonality}
    \big(\bolda(\mu') , \bolda(\mu) \big) = \sum_{m = -\lz}^\lz a_m^*(\mu') \, a_m(\mu) = \delta_{\mu' \mu},
\end{align}
and we will see soon, why this helps us to keep the calculations close to the quantum-mechanical Rayleigh-Schr\"odinger perturbation theory. \par
To proceed, we notice that the $\bolda(\mu)$ dictate a new basis of $\mathcal{H}_\lz \otimes \mathcal{H}_4$. Let us consider the  $2 \, \lz + 1$ states  
\begin{align}
    \ket{\varphi_\mu} = \sum_{m = -\lz}^\lz a_m(\mu) \, \ket{\lz \, m},
\end{align}
in the degenerate subspace $\mathcal{H}_\lz$. Using the orthogonality of the states $\ket{\lz \, m}$ from \eqref{eq:lmOrthogonality} as well as the orthogonality of the $\bolda(\mu)$ from \eqref{eq:baOrthogonality}, it immediately follows that these states are orthonormal
\begin{align}
    \label{eq:varphiOrthogonality}
    \brak{\varphi_{\mu'}}{\varphi_\mu} = \delta_{\mu'\mu},
\end{align}
and due to the completeness of the $\ket{\lz \, m}$ in $\mathcal{H}_\lz$, we also find that these states are complete
\begin{align}
    I_\lz = \sum_{\mu = -\lz}^\lz \proj{\varphi_\mu}{\varphi_\mu},
\end{align}
where $I_\lz$ denotes the identity operator in $\mathcal{H}_\lz$. Thus, the states $\ket{\varphi_\mu}$ form a complete and orthonormal basis of $\mathcal{H}_\lz$. \par
As always, we can use this basis to build a basis for $\mathcal{H}_\lz \otimes \mathcal{H}_4$ via
\begin{align}
    \ket{\varphi_\mu \, \alpha_i} &= \ket{\varphi_\mu} \otimes \ket{\alpha_i} \\
    &= \sum_{m = -\lz}^\lz a_m(\mu) \, \ket{\lz \, m \, \alpha_i},
\end{align}
and we also define its conjugate $\bra{\varphi_\mu \, \ta_i} = \bra{\varphi_\mu} \otimes \bra{\ta_i}$. With these definitions it is easy to show that the $\ket{\varphi_\mu \, \alpha_i}$ form a complete and orthonormal basis of $\mathcal{H}_\lz \otimes \mathcal{H}_4$. \par
This new basis is natural in the sense that it has useful properties to conveniently treat the problem. First of all, the states $\ket{\varphi_\mu \, \alpha_1}$ are solutions of the zeroth-order equation \eqref{eq:D0psi0}, i.e., 
\begin{align}
    \label{eq:D0varphimu}
    \hmD^\z \, \ket{\varphi_\mu \, \alpha_1} = 0,
    \und
    \bra{\varphi_\mu \, \ta_1} \, \hmD^\z = 0 .
\end{align}
This just clearly expresses the fact that we do not know \emph{ab initio} which is the correct linear combination of $\ket{\lz \, m \, \alpha_1}$ to consider to find the desired eigenvalue. After we solved the first-order equation however, we know that
\begin{equation}
    \lim_{\epsilon \to 0} \ket{\psi(\epsilon)} = \ket{\varphi_\mu \, \alpha_1},
\end{equation}
depending on which eigenvalue correction $x^\one \equiv x_\mu^\one$ we consider\footnote{We could also introduce an index $\mu$ to $\ket{\psi(\epsilon)}$ and $\ket{\psi^\z}$ to label this this dependence, however, we choose not to do so in order to keep a more compact notation.}. \par
Furthermore, the $\ket{\varphi_\mu \, \alpha_1}$ are not only the correct zeroth-order states to begin with, but they also diagonalize $\hmV$ via
\begin{align}
    \label{eq:VDiagonalize1}
    \bra{\tilde \varphi_{\mu'} \, \ta_1} \, \hmV^\one \, \ket{\varphi_\mu \, \alpha_1} = -\delta_{\mu' \mu} \, x_\mu^\one \, \mean{\ta_1}{M_\lz^\one}{\alpha_1}.
\end{align}
Now let us consider a specific eigenvalue correction $x_{\mu_0}^\one$. Thus, the corresponding zeroth-order state is
\begin{align}
    \label{eq:psi0varphimu0}
    \ket{\psi^\z} = \ket{\varphi_{\mu_0} \, \alpha_1},
\end{align}
and its bi-orthogonal conjugate zeroth-order state is
\begin{align}
    \label{eq:psi0TEStar}
    \bra{\tilde \psi^\z} &=  \sum_{m=-\lz}^\lz a_m^*(\mu_0) \, \bra{\lz \, m \, \ta_1} \notag \\[6pt]
    &= \bra{\varphi_{\mu_0} \, \ta_1}.
\end{align}
This unperturbed state is normalized according to
\begin{align}
    1 &= \brak{\tilde \psi^\z}{\psi^\z} \label{eq:zerothNormalization} \\[4pt]
    &= \sum_{m = -\lz}^\lz \left|a_{\lz \, m}^{\text{E} \, \z}\right|^2.
\end{align}
This is the result of our choice, that the $\bolda(\mu)$ are orthonormal, but has no straightforward physical meaning. This is different from quantum mechanical perturbation theories, where the wavefunction needs to be properly normalized to have its well-known probabilistic interpretation. \par
Before we continue, let us use our new basis for $\mathcal{H}_\lz \otimes \mathcal{H}_4$ by rewriting \eqref{eq:psinDecomposition} as
\begin{align}
    \label{eq:psinDecompositionvarphi}
    \ket{\psi^{(n)}} 
    = \sum_{\mu = -\lz}^{\lz} \sum_{i=1}^4  \brak{\varphi_\mu \, \ta_i}{\psi^{(n)}} \, \ket{\varphi_\mu \, \alpha_i}
    + {\sum_{l \, m \, i}}' \brak{l \, m \, i}{\psi^{(n)}} \, \ket{l \, m \, i},
\end{align}
and we likewise introduce the expansion coefficients
\begin{align}
    \label{eq:varphinCoefficients}
    \varphi_{\mu \, \alpha_i}^{(n)} = \brak{\varphi_\mu \, \ta_i}{\psi^{(n)}}.
\end{align}
Using this change of basis, the zeroth-order result reads
\begin{align}
    \varphi_{\mu \, \alpha_i}^\z = \delta_{\mu \, \mu_0} \delta_{i \, 1},
    \und
    \psi_{l \, m \, i}^\z = 0, \qquad l \neq \lz.
\end{align}
\subsubsection{First-Order State Corrections}
In order to solve the second-order equation in the next section, we also need to determine the first-order state corrections encoded in $\ket{\psi^\one}$. Therefore we need to fully solve the first-order equation \eqref{eq:first}. \par
Let us start to solve the first-order equation in the degenerate subspace first. Therefore we multiply it with $\bra{\varphi_{\mu'} \, \ta_{i'}}$ from the left and recall that we consider the eigenvalue correction $x_{\mu_0}^\one$ setting the zeroth-order state as in \eqref{eq:psi0varphimu0}. We find
\begin{align}
    \label{eq:1stOrderDegVarphi}
    \bra{\varphi_{\mu'} \, \ta_{i'}} \hmD^\z \, \ket{\psi^\one} + \bra{\varphi_{\mu'} \, \ta_{i'}} \hmV^\one \ket{\varphi_{\mu_0} \, \alpha_1} + x_{\mu_0}^\one \bra{\varphi_{\mu'} \, \ta_{i'}}\hmD^\one \ket{\varphi_{\mu_0} \, \alpha_1} = 0 .
\end{align}
\par Before we can solve this equation, we need to notice that the matrix elements of $\hmD^{(n)}$ in the new basis are given by
\begin{align}
    \bra{\varphi_{\mu'} \, \ta_{i'}} \hmD^{(n)} \, \ket{\varphi_\mu \, \alpha_i} = \delta_{\mu'\mu} \, \mean{\ta_{i'}}{M_\lz^{(n)}}{\alpha_i},
\end{align}
which directly follow from the diagonality of $\hmD^{(n)}$ in \eqref{eq:Dndef} as well as the orthonormality of the $\ket{\varphi_\mu}$ in \eqref{eq:varphiOrthogonality}. With this, we can now determine all terms in \eqref{eq:1stOrderDegVarphi} separately. \par
Starting with the first term using \eqref{eq:psinDecompositionvarphi}, we find
\begin{align}
    \bra{\varphi_{\mu'} \, \ta_{i'}} \hmD^\z \, \ket{\psi^\one}
    &= \sum_{\mu = -\lz}^\lz \sum_{i=1}^4 \bra{\varphi_{\mu'} \, \ta_{i'}} \hmD^\z \, \ket{\varphi_\mu \, \alpha_i} \, \varphi_{\mu \, \alpha_i}^\one \notag \\
    &= \sum_{i=2}^4  \mean{\ta_{i'}}{M_\lz^\z}{\alpha_i} \, \varphi_{\mu' \, \alpha_i}^\one,
\end{align}
where in the second line, we dropped the $i=1$ term due to \eqref{eq:alphaM0alpha}. The second term of \eqref{eq:1stOrderDegVarphi} equals
\begin{align}
    \begin{split}
    \bra{\varphi_{\mu'} \, \ta_{i'}} \hmV^\one \ket{\varphi_{\mu_0} \, \alpha_1} &= -\delta_{i'1} \, \delta_{\mu' \mu_0} \, x_{\mu_0}^\one \, \mean{\ta_1}{M_\lz^\one}{\alpha_1} \\[6pt]
    &\hphantom{=}+ \left[1 - \delta_{i'1} \right] \bra{\varphi_{\mu'} \, \ta_{i'}} \hmV^\one \ket{\varphi_{\mu_0} \, \alpha_1},
    \end{split}
\end{align}
where we simplified the $i' = 1$ term using \eqref{eq:VDiagonalize1} and get no simplifications for $i' \neq 1$. The third and last term is proportional to
\begin{align}
    \bra{\varphi_{\mu'} \, \ta_{i'}}\hmD^\one \ket{\varphi_{\mu_0} \, \alpha_1} = \delta_{\mu' \mu_0} \mean{\ta_{i'}}{M_\lz^\one}{\alpha_1} .
\end{align}
\par Now that we determined all three terms separately, we can collect them and find after some elementary manipulations
\begin{align*}
    \sum_{i=2}^4  \mean{\ta_{i'}}{M_\lz^\z}{\alpha_i} \, \varphi_{\mu' \, \alpha_i}^\one
    = [\delta_{i' \, 1} - 1] \left[ \bra{\varphi_{\mu'} \, \alpha_{i'}} \hmV^\one \ket{\varphi_{\mu_0} \, \alpha_1} + \delta_{\mu' \mu_0} \, x_{\mu_0}^\one \, \mean{\ta_{i'}}{M_\lz^\one}{\alpha_1}\right].
\end{align*}
First of all we notice that we get a trivial identity for the $i'=1$ case using \eqref{eq:alphaM0alpha} for the left hand side. This just confirms that we are doing things consistently, as we already used this case to construct the basis $\ket{\varphi_{\mu} \, \alpha_i}$. Additionally we notice that the coefficients $\varphi_{\mu' \, \alpha_1}^\one$ are left undetermined in the first-order perturbation theory. \par
To find the remaining coefficients in the degenerate subspace, let us write down the case $i' \neq 1$ separately as
\begin{align}
    \label{eq:varphi1nondeg}
    \sum_{i=2}^4  \mean{\ta_{i'}}{M_\lz^\z}{\alpha_i} \, \varphi_{\mu' \, \alpha_i}^\one
    = -\left[ \bra{\varphi_{\mu'} \, \alpha_{i'}} \hmV^\one \ket{\varphi_{\mu_0} \, \alpha_1} + \delta_{\mu' \mu_0} \, x_{\mu_0}^\one \, \mean{\ta_{i'}}{M_\lz^\one}{\alpha_1}\right].
\end{align}
By noticing that $\mean{\ta_{i'}}{M_\lz^\z}{\alpha_i}$ can be represented by the invertible $3 \times 3$ matrix depicted in \eqref{eq:alphaM0alpha}, we can solve for the $\varphi_{\mu \, \alpha_i}^\one$ with $i \neq 1$ by inversion. \par
In contrast to these calculations, the determination of the coefficients in the non-degenerate subspace seems trivial. First of all, we project \eqref{eq:first} on the non-degenerate subspace and find for all $l' \neq \lz$, $m'$ and $i'$
\begin{align}
    \bra{l' \, m' \, i'} \hmD^\z \, \ket{\psi^\one} + \bra{l' \, m' \, i'} \hmV^\one \, \ket{\varphi_{\mu_0} \, \alpha_1} + x^\one \bra{l' \, m' \, i'} \hmD^\one \, \ket{\varphi_{\mu_0} \, \alpha_1} = 0.
\end{align}
Again we consider all terms separately. Using \eqref{eq:psinDecompositionvarphi}, the first term is given by
\begin{align}
    \bra{l' \, m' \, i'} \hmD^\z \, \ket{\psi^\one} = \sum_{i=1}^4 \mean{i'}{M_{l'}^\z}{i} \psi_{l' \, m' \, i}^\one.
\end{align}
For the second term, no simplification can be made and the last term vanishes due to the diagonality of $\hmD^\one$. Therefore we find
\begin{align}
    \sum_{i=1}^4 \mean{i'}{M_{l'}^\z}{i} \psi_{l' \, m' \, i}^\one = - \bra{l' \, m' \, i'} \hmV^\one \, \ket{\varphi_{\mu_0} \, \alpha_1}.
\end{align}
Similar to the previous case, we can represent $\mean{i'}{M_{l'}^\z}{i}$ as invertible, this time $4 \times 4$, matrix $\bM_{l'}(x^\z)$ defined in \eqref{eq:M0psi0} and again solve for the $\psi_{l \, m \, i}^\one$ by inversion.

\subsection{Second-Order Perturbation Theory}
\label{sec:secondNonDegenerate}
Now we want to determine the second-order correction to the eigenvalue $x_{\mu_0}^\two$, still only for $\TE$-resonances under the assumption that the deformation removed the degeneracy at first order discussed in Section \ref{sec:firstordereigenvalue}. \par
First of all, let us bring the second-order equation \eqref{eq:Mpsi2} using \eqref{eq:MDecomposition} into our familiar form
\begin{align}
    \label{eq:second}
    \hmD^\z \ket{\psi^\two} +  \left[ \hmV^\one  + x_{\mu_0}^\one \hmD^\one  \right] \ket{\psi^\one} + \left[ \hmV^\two  + x_{\mu_0}^\two \,\hmD^\two \right] \ket{\psi^\z } = 0.
\end{align}
\subsubsection{Second-Order Eigenvalue Corrections}
Likewise to our first-order calculations, we multiply \eqref{eq:second} with $\bra{\varphi_{\mu'} \, \ta_1}$ from the left and find
\begin{multline}
    \label{eq:secondDegSub}
    \bra{\varphi_{\mu'} \, \ta_1}\hmD^\z \ket{\psi^\two} +  \bra{\varphi_{\mu'} \, \ta_1} \left[ \hmV^\one + x_{\mu_0}^\one \hmD^\one  \right] \ket{\psi^\one} \\[6pt]
    + \bra{\varphi_{\mu'} \, \ta_1} \left[ \hmV^\two  + x_{\mu_0}^\two \,\hmD^\two \right] \ket{\varphi_{\mu_0} \, \alpha_1} = 0.
\end{multline}
Again we factor out this equation and determine all five contributions separately. \par
To determine the first term, we notice that the $\bra{\varphi_{\mu'} \, \ta_1}$ are solutions of the conjugate zeroth-order equation depicted in \eqref{eq:D0varphimu} and thus, the first term vanishes. For the second term, we simply decompose the first-order state correction as in \eqref{eq:psinDecompositionvarphi} to write this term as
\begin{align}
    \mean{\varphi_{\mu'} \, \ta_1}{\hmV^\one}{\psi^\one}
    &= \sum_{\mu = -\lz}^{\lz} \sum_{i=1}^4 \mean{\varphi_{\mu'}\, \ta_1}{\hmV^\one}{\varphi_\mu \, \alpha_i} \, \varphi_{\mu \, \alpha_i}^\one
    + {\sum_{l \, m \, i}}' \mean{\varphi_{\mu'}\, \ta_1}{\hmV^\one}{l \, m \, i} \, \psi_{l \, m \, i}^\one \notag \\
    &= -x_{\mu'}^\one \, \mean{\ta_1}{M_\lz^\one}{\alpha_1} \, \varphi_{\mu' \, \alpha_1}^\one + \sum_{\mu = -\lz}^{\lz} \sum_{i=2}^4 \mean{\varphi_{\mu'}\, \ta_1}{\hmV^\one}{\varphi_\mu \, \alpha_i} \, \varphi_{\mu \, \alpha_i}^\one \notag \\
    &\hphantom{=} + {\sum_{l \, m \, i}}' \mean{\varphi_{\mu'}\, \ta_1}{\hmV^\one}{l \, m \, i} \, \psi_{l \, m \, i}^\one ,
\end{align}
where we separated the sum into its $i=1$ and $i\neq 1$ contribution in the second line and used \eqref{eq:VDiagonalize1}. The third term without its prefactor $x_{\mu_0}^\one$ is given by
\begin{align}
    \mean{\varphi_{\mu'}\, \ta_1}{\hmD^\one}{\psi^\one}
    &= \sum_{\mu = -\lz}^\lz \sum_{i=1}^4 \delta_{\mu'\mu} \mean{\ta_1}{M_\lz^\one}{\alpha_i} \, \varphi_{\mu \, \alpha_i}^\one \\
    &= \mean{\ta_1}{M_\lz^\one}{\alpha_1} \, \varphi_{\mu' \, \alpha_1}^\one 
    + \sum_{\mu = -\lz}^\lz \sum_{i=2}^4 \delta_{\mu'\mu} \, \mean{\ta_1}{M_\lz^\one}{\alpha_i} \, \varphi_{\mu \, \alpha_i}^\one,
\end{align}
where we again split the sum into two contributions and did not carry out the summation over $\mu$ for later convenience. The fourth term allows no simplification and the fifth and final one is proportional to
\begin{align}
    \mean{\varphi_{\mu'}\, \ta_1}{\hmD^\two}{\varphi_{\mu_0} \, \alpha_1} = \delta_{\mu'\mu_0} \mean{\ta_1}{M_\lz^\two}{\alpha_1}.
\end{align}
Collecting all pieces together and rearranging the terms we find
\begin{multline}
\label{eq:200}
    0 = \left[x_{\mu_0}^\one - x_{\mu'}^\one \right] \, \mean{\ta_1}{M_\lz^\one}{\alpha_1} \, \varphi_{\mu' \, \alpha_1}^\one
    + {\sum_{l \, m \, i}}' \mean{\varphi_{\mu'}\, \ta_1}{\hmV^\one}{l \, m \, i} \, \psi_{l \, m \, i}^\one \\
    + \sum_{\mu = -\lz}^{\lz} \sum_{i=2}^4 \left[\mean{\varphi_{\mu'} \, \ta_1}{\hmV^\one}{\varphi_\mu \, \alpha_i} + \delta_{\mu'\mu} x_{\mu_0}^\one \mean{\ta_1}{M_\lz^\one}{\alpha_i}\right] \, \varphi_{\mu \, \alpha_i}^\one \\
    + \left[ \mean{\varphi_{\mu'} \, \ta_1}{\hmV^\two}{\varphi_{\mu_0} \, \alpha_1}
    + x_{\mu_0}^\two \delta_{\mu'\mu_0} \mean{\ta_1}{M_\lz^\two}{\alpha_1} \right] .
\end{multline}
By choosing $\mu' = \mu_0$, we can solve this equation the second-order eigenvalue correction via
\begin{multline}
\label{eq:x2nondeg}
    x_{\mu_0}^\two = - \frac{1}{\mean{\ta_1}{M_\lz^\two}{\alpha_1}} \bigg\{
    \sum_{\mu = -\lz}^{\lz} \sum_{i=2}^4 \left[\mean{\varphi_{\mu_0} \, \ta_1}{\hmV^\one}{\varphi_\mu \, \alpha_i} + \delta_{\mu_0\mu} x_{\mu_0}^\one \mean{\ta_1}{M_\lz^\one}{\alpha_i}\right] \, \varphi_{\mu \, \alpha_i}^\one \\
    + {\sum_{l \, m \, i}}' \mean{\varphi_{\mu_0}\, \ta_1}{\hmV^\one}{l \, m \, i} \, \psi_{l \, m \, i}^\one
    + \mean{\varphi_{\mu_0} \, \ta_1}{\hmV^\two}{\varphi_{\mu_0} \, \alpha_1} \bigg\}.
\end{multline}
As all quantities on the right hand side of this equation are either known first-order coefficients or known matrix elements, this equation actually gives us the second-order eigenvalue correction. \par
Finally, let us introduce likewise to our first-order considerations
\begin{align}
    \Delta^\two(\mu) = -\frac{x_\mu^\two}{x^\z},
\end{align}
in order to rewrite the eigenvalue of the perturbed system \eqref{eq:xExpansion} up to second order as
\begin{align}
    \label{eq:xExpansionNonDeg}
    x(\mu) = x_{\lz \, n_0}^\TE \left(1 - \epsilon \, \Delta^\one(\mu) - \epsilon^2 \, \Delta^\two(\mu) + \dots \right),
\end{align}
where both $\Delta^\one$ and $\Delta^\two$ have an implicit dependence on the unperturbed resonance. This concludes our search for the perturbed $\TE$-eigenvalues $x$ if the perturbation removes the degeneracy at first order.
\subsubsection{Further Remarks}
\label{sec:further1}
At this point it should be clear how to determine higher-order corrections. However, there is one detail left to discuss. \par
In order to determine the third-order eigenvalue correction, and similarly for higher orders, one actually has to determine the up to now unknown coefficients $\varphi_{\mu' \, \alpha_1}^\one$. Therefore we recall that we still have to solve \eqref{eq:200} for the case $\mu' \neq \mu_0$. In the case that the deformation fully removes the perturbation, the prefactor $[x_{\mu_0}^\one - x_{\mu'}^\one]$ does not vanish. Therefore we can solve this equation via
\begin{multline}
    \varphi_{\mu' \, \alpha_1}^\one = -\frac{1}{\left[x_{\mu_0}^\one - x_{\mu'}^\one \right] \, \mean{\ta_1}{M_\lz^\one}{\alpha_1}} \\[6pt]
     \times \bigg\{
    \sum_{\mu = -\lz}^{\lz} \sum_{i=2}^4 \bigg[\mean{\varphi_{\mu'} \, \ta_1}{\hmV^\one}{\varphi_\mu \, \alpha_i}+ \delta_{\mu'\mu} x_{\mu_0}^\one \mean{\ta_1}{M_\lz^\one}{\alpha_i} \bigg] \, \varphi_{\mu \, \alpha_i}^\one \\[2pt]
    + {\sum_{l \, m \, i}}' \mean{\varphi_{\mu'}\, \ta_1}{\hmV^\one}{l \, m \, i} \, \psi_{l \, m \, i}^\one
    + \mean{\varphi_{\mu'} \, \ta_1}{\hmV^\two}{\varphi_{\mu_0} \, \alpha_1} \bigg\},
\end{multline}
where again all quantities on the right-hand side are already known. However, there is still the undetermined first-order quantity $\varphi_{\mu_0 \, \alpha_1}^\one$. To find it, we impose the normalization of the perturbed state as
\begin{align}
    1 &= \brak{\tilde \psi(\epsilon)}{\psi(\epsilon)} \\
    &= \brak{\tilde \psi^\z}{\psi^\z} + \epsilon \left| \brak{\tilde \psi^\z}{\psi^\one}\right|^2 + \dots \notag \,,
\end{align}
and together with our zeroth-order normalization $\brak{\tilde \psi^\z}{\psi^\z} = 1$ from \eqref{eq:zerothNormalization}, we therefore need
\begin{align}
    \label{eq:intermediatenorm}
    \brak{\tilde \psi^\z}{\psi^{(\nu)}} = 0, && \alpha = 1,2,\dots \,.
\end{align}
Now we can insert our zeroth-order state as well as our first-order state correction from \eqref{eq:psinDecompositionvarphi} to find
\begin{align}
    0 & = \brak{\tilde \psi^\z}{\psi^\one} \notag \\
    &= \sum_{\mu=-\lz}^\lz \sum_{i=1}^4 \brak{\varphi_{\mu_0} \, \ta_1}{\varphi_\mu \, \alpha_i} \, \varphi_{\mu \, \alpha_i}^\one \notag \\
    &= \varphi_{\mu_0 \, \alpha_1}^\one,
\end{align}
where we used the orthogonality of our states. At this point, we determined all first-order quantities. \par 
Let us remark that the normalization \eqref{eq:intermediatenorm} is in quantum mechanics known as the intermediate normalization and the wavefunction renormalization has to be applied in order to keep the probabilistic interpretation of the wavefunction. However, as discussed in Section \ref{sec:normAndBasisChange}, the overall normalization of the state $\ket{\psi(\epsilon)}$ is arbitrary and we can choose it for our convenience. \par
Now we want to discuss the first-order correction to $\ket{\psi^\z}$. First let us recall the rather complicated spectrum of the perturbation operator $\hmM$ in $\mathcal{H}_\lz \otimes \mathcal{H}_4$. By construction, this Hilbert space is spanned by the $(2 \, \lz + 1) \times 4$ vectors $\ket{\lz \, m \, \alpha_i}$. Let us write them down as
\begin{align*}
    \textcolor{blue}{\left\{ \ket{l_0 \,m \, \alpha_1} \right\} =
    \left\{
        \begin{array}{c}
            \ket{l_0, -l_0} \\
            \vdots \\
            \ket{l_0, l_0} \\
        \end{array}
    \right\} \otimes \ket{\alpha_1},} \quad
\left\{ \ket{l_0 \,m , \alpha_i} \right\} = 
    \left\{
        \begin{array}{c}
            \ket{l_0, -l_0} \\
            \vdots \\
            \ket{l_0, l_0} \\
        \end{array}
    \right\} \otimes \ket{\alpha_i}, \qquad i=2,3,4.
\end{align*}
We knew from our zeroth-order calculation that our zeroth-order state $\ket{\psi^\z}$ had to be a linear combination of the $2 \, \lz + 1$ basis states $\ket{\lz \, m \, \alpha_1}$ colored {\color{blue}blue} in the equation above. Apart from that, $\mathcal{H}_\lz \otimes \mathcal{H}_4$ also contains the $(2 \, \lz + 1) \times 3$ basis states $\ket{\lz \, m \, \alpha_i}$, $i=2,3,4$, colored black in the equation above. The fact that these states also lie in $\mathcal{H}_\lz \otimes \mathcal{H}_4$ is a choice, as we denote $\mathcal{H}_\lz$ as the degenerate subspace and not $\mathcal{H}_\lz \otimes \text{span}{\{\ket{\alpha_1}\}}$. \par
When we changed the basis, we were simply rearranging the vectors in $\mathcal{H}_\lz$ without affecting the space $\mathcal{H}_4$, so we have
\begin{align*}
\left\{ \ket{\varphi_\mu \, \alpha_1} \right\} =
    \left\{
        \begin{array}{c}
            \ket{\varphi_{-l_0}} \\
            \vdots \\
            \ket{\varphi_{\mu_0-1}} \\[4pt]
            \textcolor{blue}{ \ket{\varphi_{\mu_0}} }\\[4pt]
            \ket{\varphi_{\mu_0+1}} \\
            \vdots \\
            \ket{\varphi_{l_0}} \\
        \end{array}
    \right\} \otimes \ket{\alpha_1}, \quad
\left\{ \ket{\varphi_{\mu} \, \alpha_i} \right\} =
    \left\{
        \begin{array}{c}
            \ket{\varphi_{-l_0}} \\
            \vdots \\
            \ket{\varphi_{l_0}} \\
        \end{array}
    \right\} \otimes \ket{\alpha_i}, \qquad i=2,3,4,
\end{align*}
where we highlighted in {\color{blue}blue} the zeroth-order state $\ket{\varphi_{\mu_0} \, \alpha_1}$ of which we calculate the perturbative corrections. By going to the first order in perturbation theory, we found that our zeroth-order state $\ket{\varphi_{\mu_0} \, \alpha_1}$ gets a correction. Especially all states in black in the above equation get multiplied by their corresponding $\varphi_{\mu \, \alpha_i}^\one$.\par
Furthermore, $\ket{\varphi_{\mu_0} \, \alpha_1}$ gets corrections from the states $\ket{l \, m \, i}$, $l \neq \lz$, which are in the non-degenerate subspace. As their basis does not get changed, they simply get multiplied by $\psi_{l \, m \, i}^\one$. \par 
To conclude this discussion, we want to write the perturbed state $\ket{\psi(\epsilon)}$ up to first order in its full form as
\begin{multline}
    \ket{\psi(\epsilon)} = \ket{\varphi_{\mu_0} \, \alpha_1} + \epsilon \Bigg[\sum_{\substack{\mu = -\lz \\ \mu \neq \mu_0}}^\lz \varphi_{\mu \, \alpha_1}^\one \, \ket{\varphi_\mu \, \alpha_1} + \sum_{i=2}^4 \sum_{\mu = -\lz}^\lz \varphi_{\mu \, \alpha_i}^\one \, \ket{\varphi_\mu \, \alpha_i} \\
    + \sum_{\substack{l = 0 \\l \neq \lz}}^\infty \sum_{m=-l}^l \sum_{i=1}^4 \psi_{l \, m \,i}^\one \, \ket{l \, m \, i} \Bigg] + \order{\epsilon^2}.
\end{multline}
\subsection{Degenerate Second-Order Perturbation Theory}
\label{sec:secondDegenerate}
Now that we have computed the second-order eigenvalue correction under the assumption that all first-order eigenvalue corrections $x^\one$ are all different, we want to drop this assumption and generalize our approach to geometries which do not have this property. For example, an oblate spheroid has rotational symmetry and keeps a two-fold degeneracy at all orders \cite{lai90-2}.
\subsubsection{Modification of the First-Order Eigenvalue Equation}
To find this generalization, we first of all need to reconsider the first-order eigenvalue equation \eqref{eq:eigen2}. The standard technique to handle the degenerate eigenvalues is to go from a single label indexation $\mu$ to a two label indexation $(\mu , p)$. This gives us the flexibility to rewrite the first-order eigenvalue equation as
\begin{align}
    \label{eq:eigen3}
    \bV \, \bolda(\mu,p) = \Delta^\one(\mu) \, \bolda(\mu,p), && \mu = 1,\ldots,M , \ p = 1,\dots,P_\mu,
\end{align}
where $\mu$ labels the $M$ different first-order eigenvalues $\Delta^\one(\mu)$ so that $\Delta^\one(\mu) \neq \Delta^\one(\mu')$ for all $\mu \neq \mu'$, which is equivalent to $x_\mu^\one \neq x_{\mu'}^\one$ for all $\mu \neq \mu'$, and $p$ labels the corresponding eigenvectors $\bolda(\mu,p)$ with a $P_\mu$-fold multiplicity. These $P_\mu$ are potentially different for each $\mu$, but need to satisfy
\begin{align}
    \sum_{\mu=1}^M P_\mu = 2 \, \lz + 1.
\end{align}
In the present case, the perturbation does not fully remove the degeneracy, but reduces the $(2 \, \lz + 1)$-fold degeneracy of the zeroth-order eigenvalue $x^\z$ to a $P_\mu$-fold degeneracy at first order. After noticing this, the next steps are completely analogue to the ones in Section \ref{sec:firstordereigenvalue}. \par
First of all, we choose the eigenvectors orthonormal
\begin{align}
    \big(\bolda(\mu',p') , \bolda(\mu,p) \big) = \sum_{m=-\lz}^{\lz} a_m^*(\mu',p') \, a_m(\mu,p) = \delta_{\mu' \mu} \delta_{p' p},
\end{align}
and likewise, they dictate a new orthonormal and complete basis of $\mathcal{H}_\lz \otimes \mathcal{H}_4$ via
\begin{align}
    \label{eq:statevarphidegenerate}
    \ket{\varphi_{\mu \, p} \, \alpha_i} = \sum_{m=-\lz}^{\lz} a_m(\mu,p) \, \ket{\lz, m,\alpha_i},
\end{align}
similar to the $\ket{\varphi_\mu \, \alpha_i}$ previously. Again, for $i=1$ these states are solutions of the zeroth-order equation
\begin{align}
    \hmD^\z \, \ket{\varphi_{\mu \, p} \, \alpha_1} = 0, \und \bra{\varphi_{\mu \, p} \, \ta_1} \, \hmD^\z = 0,
\end{align}
as well as they diagonalize $\hmV^\one$ via
\begin{align}
    \mean{\varphi_{\mu' \, p'} \, \ta_1}{\hmV^\one}{\varphi_{\mu \, p} \, \alpha_1} = -\delta_{\mu'\mu} \, \delta_{p'p} \, x_{\mu}^\one \, \mean{\ta_1}{M_\lz^\one}{\alpha_1}.
\end{align}
Using this new basis, we can decompose our full state $\ket{\psi(\epsilon)}$ similar to \eqref{eq:psinDecompositionvarphi} into its part contained in $\mathcal{H}_\lz \otimes \mathcal{H}_4$ and its part contained in the complementary subspace via
\begin{align}
    \label{eq:psiDecompositionDeg}
    \ket{\psi(\epsilon)} 
    = \sum_{\mu = 1}^M \sum_{p = 1}^{P_\mu} \sum_{i=1}^4  \brak{\varphi_{\mu \, p} \, \ta_i}{\psi(\epsilon)} \, \ket{\varphi_{\mu \, p} \, \alpha_i}
    + {\sum_{l \, m \, i}}' \brak{l \, m \, i}{\psi(\epsilon)} \, \ket{l \, m \, i},
\end{align}
and we introduce the shorthand notations
\begin{align}
    \varphi_{\mu \, p \, \alpha_i}(\epsilon) = \brak{\varphi_{\mu \, p} \, \ta_i}{\psi(\epsilon)},
    \und
    \varphi_{\mu \, p \, \alpha_i}^{(n)} = \brak{\varphi_{\mu \, p} \, \ta_i}{\psi^{(n)}}.
\end{align}
Even though we found a basis to handle our problem more conveniently, this time our zeroth-order state $\ket{\psi^\z}$ is not yet completely determined. By fixing $\mu = \mu_0$ and checking the limit
\begin{align}
    \lim_{\epsilon \to 0} \ket{\psi(\epsilon)} = \ket{\psi^\z},
\end{align}
using \eqref{eq:psiDecompositionDeg} we find the zeroth order-state
\begin{align}
    \label{eq:zerothDegenerate}
    \ket{\psi^\z} = \sum_{p=1}^{P_{\mu_0}} \varphi_{\mu_0 \, p \, \alpha_1}^\z \, \ket{\varphi_{\mu_0 \,  p} \, \alpha_1},
\end{align}
where the $P_{\mu_0}$ coefficients $\varphi_{\mu_0 \, p \, \alpha_1}^\z$ are still undetermined. In this form it should be clear that the degenerate case we currently consider is a generalization of the non-degenerate case: If we pick an eigenvalue labeled by $\mu=\mu_0$ which is non-degenerate, i.e., $P_{\mu_0} = 1$, then the sum collapses and we are in the same situation as in the non-degenerate case.
\subsubsection{The First-Order State Corrections}
In order to determine the second-order eigenvalue corrections, we have to determine the first-order corrections to the perturbed state. The only difference to non-degenerate case is that we need to consider the zeroth-order state given by \eqref{eq:zerothDegenerate} as well as the new basis states in the degenerate subspace. \par
This time we start with the non-degenerate case. By multiplying our first-order equation \eqref{eq:first} with $\bra{l',m',i'}$, $l' \neq \lz$, we find
\begin{align}
    \label{eq:psidegeneratenondegenerate}
    \sum_{i=1}^4 \mean{i'}{M^\z_{l'}}{i} \psi_{l' m' i}^\one +  \sum_{p=1}^{P_{\mu_0}} \mean{l',m',i'}{\hat{\mathcal{V}}^\one}{\varphi_{\mu_0 p} \, \alpha_1} \varphi_{\mu_0 \, p \, \alpha_1}^\z  = 0.
\end{align}
We already understand that we can represent $\mean{i'}{M^\z_{l'}}{i}$ as an invertible $4 \times 4$ matrix. In anticipation of our second-order calculations, we explicitly solve this equation for the $\psi_{l' m' i'}^\one$. To do so, we multiply the previous equation by $\mean{i''}{(M^\z_{l'})^{-1}}{i'}$ and sum with respect to $i'$. We find
\begin{align}
    \psi_{l' m' i'}^\one &= \sum_{p=1}^{P_{\mu_0}} \left\{ - \sum_{i=1}^4 \mean{i'}{(M^\z_{l'})^{-1}}{i} \mean{l',m',i}{\hmV^\one}{\varphi_{\mu_0 p} \alpha_1} \right\} \varphi_{\mu_0 \, p \, \alpha_1}^\z,
\end{align}
where we relabeled $i'' \rightarrow i'$. The most important observation is that the $\psi_{l' m' i'}^\one$ are linear combinations of the, up to now, unknown zeroth-order coefficients $\varphi_{\mu_0 \, p \, \alpha_1}^\z$.\par
Secondly we want to solve the first-order equation in the degenerate subspace. Therefore we multiply \eqref{eq:first} from the left with $\bra{\varphi_{\mu' p'} \, \alpha_{i'}}$. Similar to our previous consideration, we find a trivial identity and 
\begin{multline}
    \label{eq:varphidegeneratedegenerate}
    \sum_{i=2}^4  \mean{\ta_{i'}}{M_\lz^\z}{\alpha_i} \, \varphi_{\mu' \, p' \, \alpha_i}^\one \\
    = -\sum_{p=1}^{P_{\mu_0}} \left\{
    \bra{\varphi_{\mu' \, p'} \, \ta_{i'}} \hmV^\one \ket{\varphi_{\mu_0 \, p} \, \alpha_1} + \delta_{\mu' \mu_0} \, \delta_{p' p} \, x_{\mu_0}^\one \, \mean{\ta_{i'}}{M_\lz^\one}{\alpha_1}
    \right\} \varphi_{\mu_0 \, p \, \alpha_1}^\z,
\end{multline}
where $i'\neq 1$, which is the analogue of \eqref{eq:varphi1nondeg}. Again, we can solve this equation by inversion for $\varphi_{\mu' \, p' \, \alpha_{i'}}^\one$, $i'\neq 1$, and find that they are a linear combination of the $\varphi_{\mu_0 \, p \, \alpha_1}^\z$.
\subsubsection{Second-Order Eigenvalue Corrections}
Finally we want to determine the second-order eigenvalue corrections $x_{\mu_0}^\two$. To find them, we multiply \eqref{eq:second} from the left with $\bra{\varphi_{\mu_0 \, p'} \, \ta_1}$ and find the analogue of \eqref{eq:200} with $\mu' = \mu_0$,
\begin{multline}
    \label{eq:secondOrderDeg}
    0 = \sum_{\mu=1}^M \sum_{p=1}^{P_\mu} \sum_{i=2}^4  \left[ \mean{\varphi_{\mu_0 \, p'} \, \ta_1}{\hat{\mathcal{V}}^\one}{\varphi_{\mu \, p} \, \alpha_i} + x_{\mu_0}^\one \, \delta_{p' p} \, \mean{ \ta_1}{M^\one_{\lz}}{\alpha_i} \right] \varphi_{\mu \, p \, \alpha_i}^\one \\
    +{\sum_{l \, m \, i}}' \mean{\varphi_{\mu_0 \, p'}\, \ta_1}{\hmV^\one}{l \, m \, i} \, \psi_{l \, m \, i}^\one + \sum_{p=1}^{P_{\mu_0}} \mean{\varphi_{\mu_0 \, p'} \, \ta_1}{\hmV^\two}{\varphi_{\mu_0 \, p} \, \alpha_1}  \varphi_{\mu_0 \, p \, \alpha_1}^\z \\[4pt]
    + x_{\mu_0}^\two \mean{\ta_1}{M_\lz^\two}{\alpha_1} \varphi_{\mu_0 \, p' \, \alpha_1}^\z.
\end{multline}
Let us recall from our first order considerations that $\varphi_{\mu \, p \, \alpha_i}^\one$ as well as $\psi_{l \, m \, i}^\one$ are just linear combinations of the unknown zeroth-order coefficients. Thus every single term in \eqref{eq:secondOrderDeg} depends on the unknown zeroth-order coefficients, and we can rewrite this equation as an eigenvalue equation
\begin{align}
    \label{eq:eigen2nd}
    \mathbf{W} \, \bvarphi = \Delta^\two \, \bvarphi,
\end{align}
where $\mathbf{W}$ is a $P_{\mu_0} \times P_{\mu_0}$-matrix, $\bvarphi$ a $P_{\mu_0}$-component vector and we similarly to $\Delta^\one$ define
\begin{align}
    \Delta^\two = -\frac{x_{\mu}^\two}{x^\z}.
\end{align}
Similar to our first-order considerations, we introduce the label $\nu$ to denote the $P_\mu$ eigenvalues and we write
\begin{align}
    \Delta^\two \equiv \Delta^\two(\mu,\nu).
\end{align}
With this, we can rewrite the eigenvalue of the perturbed system likewise to \eqref{eq:xExpansionNonDeg} as
\begin{align}
    \label{eq:xExpansionDeg}
    x(\mu,\nu) = x_{\lz \, n_0}^\TE \left( 1 - \epsilon \, \Delta^\one(\mu) - \epsilon^2 \, \Delta^\two(\mu, \nu) + \dots \right).
\end{align}
In this form it is clear that the zeroth-order eigenvalue is degenerate with respect to $\mu$ and $\nu$, the first-order correction removes the degeneracy with respect to $\mu$ and the second-order correction removes the degeneracy with respect to $\nu$. This concludes our search for perturbed eigenvalues up to and including second-order corrections.
\subsubsection{Further Remarks}
First of all we want to mention that the approach to find the second-order eigenvalue corrections is the same as in our first-order considerations. This is due to the fact that we found a new degenerate subspace, labeled by $\mu_0$, within the degenerate subspace labeled by $\lz$. Therefore we should not be surprised to find an eigenvalue equation as in \eqref{eq:eigen2nd}. \par
As in the first order, solving the eigenvalue equation could result in degenerate and non-degenerate eigenvalues, depending on the deformation encoded in $\mathbf{W}$, and the eigenvectors $\bvarphi(\nu)$ dictate for fixed $\lz$ and $\mu_0$ a new basis $\ket{\phi_{\nu}}$ or $\ket{\phi_{\nu \, q}}$, depending on whether or not the perturbation fully removes the degeneracy at second order. With this, one can similarly to our previous consideration determine the higher-order corrections to the eigenvalues and states. \par
To conclude this section, we want to discuss the spectrum of the perturbation operator in analogy to Section \ref{sec:further1}. For the degenerate case, we have
\begin{align*}
\left\{ \ket{\varphi_{\mu \, p} \, \alpha_1} \right\} =
    \left\{
        \begin{array}{c}
            \ket{\varphi_{1 , 1}} \\
            \vdots \\
            \ket{\varphi_{\mu_0-1 , P_{\mu_0-1}}}\\[8pt]
            \textcolor{blue}{\ket{\varphi_{\mu_0 , 1}}}\\
            \textcolor{blue}{\vdots} \\
            \textcolor{blue}{\ket{\varphi_{\mu_0 , P_{\mu_0}}}}\\[8pt]
            \ket{\varphi_{\mu_0+1 , 1}} \\
            \vdots \\
            \ket{\varphi_{M , P_M}} \\
        \end{array}
    \right\} \otimes \ket{\alpha_1}, \quad
\left\{ \ket{\varphi_{\mu} \, \alpha_i} \right\} =
    \left\{
        \begin{array}{c}
            \ket{\varphi_{1 , 1}} \\
            \vdots \\
            \ket{\varphi_{M , P_M}} \\
        \end{array}
    \right\} \otimes \ket{\alpha_i}, \qquad i=2,3,4.
\end{align*}
This time, the first-order perturbation theory does not result in a single state, but in a unknown linear combination of the basis states highlighted in {\color{blue}blue}. Again, we determined all factors multiplying the states marked in black in first order, however the quantities $ \varphi_{\mu_0 \, p \, \alpha_1}^\z$ are still undetermined.
\newpage

\section[Resonances of a Deformed Dielectric Sphere: TM-Case]{Resonances of a Deformed Dielectric Sphere: \\ TM-Case}
\label{sec:solutionTM}
After we determined the eigenvalue corrections for $\TE$-modes in Section \ref{sec:solutionTE}, we adapt our approach to determine the corrections for the $\TM$-eigenvalues. Therefore we recall our zeroth-order considerations for $\TE$-modes and adapt them for $\TM$-modes in Section \ref{sec:zerothTM}. After that, we determine the first- and second-order corrections of the $\TM$-eigenvalue in the most general case in Section \ref{sec:firstTM} and \ref{sec:secondTM}, respectively.
\subsection{Zeroth-Order Perturbation Theory}
\label{sec:zerothTM}
To set the stage for the $\TM$-eigenvalues, we recall the zeroth-order equation \eqref{eq:D0psi0}
\begin{align*}
    \hmD^\z \ket{\psi^\z} = 0,
\end{align*}
which can represented in the matrix equation \eqref{eq:M0psi0lmat}
\begin{align*}
    \begin{bmatrix}
        1 &  -1 & 0 & 0 \\[6pt]
        -n_1 R_{1 \, l}^\Psi(x^\z) & n_2 R_{2 \, l}^\Psi(x^\z) & 0 & 0 \\[6pt]
        0 & 0 & R_{1 \, l}^\Psi (x^\z) /n_1 & -R_{2 \, l}^\Psi (x^\z) /n_2 \\[6pt]
        0 & 0 & 1 & -1
    \end{bmatrix}
    \cdot
    \begin{bmatrix}
        a_{l \, m}^{\text{E} \, \z} \\[6pt]
        b_{l \, m}^{\text{E} \, \z} \\[6pt]
        a_{l \, m}^{\text{M} \, \z} \\[6pt]
        b_{l \, m}^{\text{M} \, \z}
    \end{bmatrix}
    = 0 .
\end{align*}
Let us choose the eigenvalue $x^\z \equiv x_{\lz \, n_0}^\TM$. As any $\TM$-eigenvalue satisfies 
\begin{align*}
    f_\lz^\TM(x^\z) = - R_{1\, \lz}(x^\z)/n_1 + R_{2\, \lz}(x^\z)/n_2 = 0,
\end{align*}
let us define
\begin{align}
    \label{eq:zTMdef}
    z = R_{1\, \lz}(x^\z)/n_1 = R_{2\, \lz}(x^\z)/n_2.
\end{align}
With this, we can rewrite the zeroth-order equation for the non-trivial case $l=\lz$ as
\begin{align}
    \label{eq:M0psi0l0TM}
    0 &= \bM_\lz(x^\z) \cdot \bpsi_{\lz \, m}^\z \\[6pt]
    &=
    \begin{bmatrix}
        1 &  -1 & 0 & 0 \\[10pt]
        -z n_1^2 & z n_2^2 & 0 & 0 \\[10pt]
        0 & 0 & z & -z \\[10pt]
        0 & 0 & 1 & -1
    \end{bmatrix}
    \cdot
    \begin{bmatrix}
        a_{\lz \, m}^{\text{E} \, \z} \\[10pt]
        b_{\lz \, m}^{\text{E} \, \z} \\[10pt]
        a_{\lz \, m}^{\text{M} \, \z} \\[10pt]
        b_{\lz \, m}^{\text{M} \, \z}
    \end{bmatrix}.
\end{align}
Analogous to the zeroth-order equation for $\TE$-modes, we notice that the magnetic block, i.e., the lower block-matrix, has a vanishing determinant. Furthermore, $\bM_\lz(x^\z)$ is again non-Hermitian. This means, we again have to employ a bi-orthogonal basis, which we denote as $\ket{\beta_i}$ and $\bra{\tb_i}$ for $\TM$-modes, to diagonalize the magnetic block. The vectors
\begin{align}
    \ket{\beta_3} \doteq \frac{1}{z-1} \begin{bmatrix} 0 \\ 0 \\ z \\ 1 \end{bmatrix},
    \und
    \ket{\beta_4} &\doteq \begin{bmatrix} 0 \\ 0 \\ 1 \\ 1 \end{bmatrix},
\end{align}
are the right-eigenvectors diagonalizing the magnetic block with eigenvalues $\lambda_3 = z-1$ and $\lambda_4 = 0$. The associated left-eigenvectors are given by
\begin{align}
    \bra{\tb_3} \doteq \begin{bmatrix} 0 & 0 & 1 & -1 \end{bmatrix},
    \und
    \bra{\tb_4} \doteq \frac{1}{z-1} \begin{bmatrix} 0 & 0 & -1 & z \end{bmatrix}.
\end{align}
In order to get a complete bi-orthogonal basis, we define
\begin{align}
    \ket{\beta_i} = \ket{i}, \und \bra{\tb_i} = \bra{i}, \qquad i=1,2,
\end{align}
and the completeness and orthogonality reads
\begin{align}
    \sum_{i=1}^4 \proj{\beta_i}{\tb_i} = I_4,
    \und
    \brak{\tb_{i'}}{\beta_i}=\delta_{i'i}, \qquad i,i' = 1,2,3,4.
\end{align}
In this basis, the operator $M_\lz^\z$ associated to $\bM_\lz(x^\z)$ from \eqref{eq:M0psi0l0TM} has the matrix-representation
\begin{align}
    \label{eq:betaM0beta}
    \mean{\tb_{i'}}{M_\lz^\z}{\beta_i} \doteq
    \begin{bmatrix}
        1       & -1        & 0     & 0 \\[4pt]
        -zn_1^2 & z n_2^2   & 0     & 0 \\[4pt]
        0       & 0         & z-1   & 0 \\[4pt]
        0       & 0         & 0     & 0
    \end{bmatrix},
\end{align}
and the upper $3 \times 3$ matrix has non-vanishing determinant. \par
With this we are completely in the same situation as for the $\TE$-modes. The fundamental solutions of the zeroth-order equation are given by
\begin{align}
    \label{eq:D0l0mb4}
    \hmD^\z \ket{\lz \, m \, \beta_4} = 0,
    \und
    \bra{\lz \, m \, \tb_4} \hmD^\z = 0.
\end{align}
Furthermore, the zeroth-order state is a linear combinations of those with the currently undetermined coefficients $a_{\lz \, m}^{\text{M} \, \z}$, i.e., it reads 
\begin{align}
    \label{eq:psi0TMDef}
    \ket{\psi^\z} = \sum_{m = -\lz}^\lz a_{\lz \, m}^{\text{M} \, \z} \, \ket{\lz \, m \, \beta_4}.
\end{align}
Finally, the bi-orthogonal conjugate expression reads
\begin{align}
    \label{eq:psi0TMStarDef}
    \bra{\tilde\psi^\z} = \sum_{m = -\lz}^\lz \tilde{a}_{\lz \, m}^{\text{M} \, \z} \, \bra{\lz \, m \, \tb_4}.
\end{align}
We will discuss the difference of this expression with the corresponding one for $\TE$-eigenvalues in the next section.

\subsection{First-Order Perturbation Theory}
\label{sec:firstTM}
To determine the first-order eigenvalue equation, we can start similarly to the $\TE$-case discussed in Section \ref{sec:first}. Therefore we recall the first-order equation \eqref{eq:Mpsi1} and use \eqref{eq:MDecomposition} to find again
\begin{align}
    \label{eq:firstTM}
    \hmD^\z \, \ket{\psi^\one} + \left[ \hmV^\one + x^\one \hmD^\one \right]\ket{\psi^\z} = 0.
\end{align}
As exercised multiple times before, it will be useful to separate the unknown first-order state correction $\ket{\psi^\one}$ into its parts contained in the degenerate and non-degenerate subspace as
\begin{align}
    \label{eq:psi1TM}
    \ket{\psi^\one} = \sum_{m = -\lz}^{\lz} \sum_{i=1}^4 \psi_{m \, \beta_i}^\one \, \ket{\lz \, m \, \beta_i}
    + {\sum_{l \, m \, i}}' \psi_{l \, m \, i}^\one \, \ket{l \, m \, i},
\end{align}
where
\begin{align}
    \psi_{m \, \beta_i}^\one = \brak{\lz \, m \, \tb_i}{\psi^\one}, 
    \und
    \psi_{l \, m \, i}^\one = \brak{l \, m \, i}{\psi^\one}, \qquad l\neq\lz.
\end{align}
\subsubsection{First-Order Eigenvalue Equation}
To find the first-order eigenvalue equation, we multiply \eqref{eq:firstTM} with $\bra{\lz \, m' \, \tb_4}$ from the left to find
\begin{align}
    \sum_{m=-\lz}^\lz \left\{ \mean{\lz \, m' \, \tb_4}{\hmV^\one}{\lz \, m \, \beta_4} + x^\one \mean{\lz \, m' \, \tb_4}{\hmD^\one}{\lz \, m \, \beta_4} \right\} a_{\lz \, m}^{\text{M} \, \z} = 0,
\end{align}
where we used \eqref{eq:D0l0mb4} to eliminate the first term in \eqref{eq:firstTM} and inserted the explicit form of the unperturbed state $\ket{\psi^\z}$ from \eqref{eq:psi0TMDef}. Using the diagonal form of $\hmD^\one$ from \eqref{eq:Dndef} and doing some straightforward manipulations we find
\begin{align}
    \sum_{m=-\lz}^\lz \frac{1}{x^\z} \frac{\mean{\lz \, m' \, \tb_4}{\hmV^\one}{\lz \, m \, \beta_4}}{\mean{\tb_4}{M_{\lz}^\one}{\beta_4}} \, a_{\lz \, m}^{\text{M} \, \z} = - \frac{x^\one}{x^\z} \, a_{\lz \, m'}^{\text{M} \, \z},
\end{align}
where we introduced the factor $1/x^\z$ to keep the formulation close to the one used in the $\TE$-case. Again, this is an eigenvalue equation, which we write in the suggestive matrix-vector form as
\begin{align}
    \bV \bolda = \Delta^\one \bolda.
\end{align}
In contrast to the $\TE$-case, $\bV$ is \emph{not} a Hermitian matrix or proportional to a Hermitian one. In order to diagonalize $\bV$, as well as its corresponding operator, we also need to find its left-eigenvectors. They naturally arise by considering the bi-orthogonal conjugate of \eqref{eq:firstTM},
\begin{align}
    \bra{\tilde \psi^\one} \, \hmD^\z + \bra{\tilde \psi^\z} \left[ \hmV^\one + x^\one \hmD^\one \right] = 0,
\end{align}
which can be treated as an independent equation. By multiplying this equation from the right with $\ket{\lz \, m' \, \beta_4}$ and recalling the bi-orthogonal conjugate zeroth-order state $\bra{\tilde \psi^\z}$ from \eqref{eq:psi0TMStarDef}, \eqref{eq:D0l0mb4} as well as \eqref{eq:Dndef}, we find
\begin{align}
    \sum_{m=-\lz}^\lz \tilde{a}_{\lz \, m}^{\text{M} \, \z} \frac{1}{x^\z} \frac{\mean{\lz \, m' \, \tb_4}{\hmV^\one}{\lz \, m \, \beta_4}}{\mean{\tb_4}{M_{\lz}^\one}{\beta_4}} = - \frac{x^\one}{x^\z} \, \tilde{a}_{\lz \, m'}^{\text{M} \, \z}.
\end{align}
This can again be written in the matrix-vector form
\begin{align}
    \boldta \, \bV = \Delta^\one \boldsymbol{\tilde{a}},
\end{align}
where the left-eigenvectors $\boldta$ have the components
\begin{align}
    \tilde a_m \equiv \tilde{a}_{\lz \, m}^{\text{M} \, \z}.
\end{align}
This bi-orthogonal treatment is the main difference to the $\TE$-modes, where the Hermicity of $\bV^\TE$ results in the extra properties
\begin{align}
    \boldta^\TE = \left[\bolda^\TE\right]^\dag,
    \und
    \tilde{a}_{\lz \, m}^{\text{E} \, \z} = \left[a_{\lz \, m}^{\text{E} \, \z}\right]^*,
\end{align}
where we introduced the superscript $\TE$ for distinction. These simplifications were the reason why we considered the $\TE$-modes first. \par
To proceed, we notice that depending on the deformation encoded in $\bV$, the most general form of the eigenvalue equation reads
\begin{align}
    \begin{aligned}
        \bV \bolda(\mu,p) &= \Delta^\one(\mu) \, \bolda(\mu,p), \\[4pt]
        \boldta(\mu,p) \, \bV &= \Delta^\one(\mu)\, \boldta(\mu,p),
    \end{aligned}
    &&
    \mu = 1, \dots, M, \ p = 1, \dots, P_\mu.
\end{align}
where $\mu$ denotes the $M$ pairwise different eigenvalues $\Delta^\one(\mu)$ and $p$ labels the associated $P_\mu$-fold degenerate right- and left-eigenvectors $\bolda$ and $\boldta$, respectively. Again, we choose these bi-orthogonal eigenvectors orthonormal as
\begin{align}
    \big(\boldta(\mu',p'), \bolda(\mu,p)\big) = \sum_{m=-\lz}^{\lz} \tilde a_m(\mu',p') \, a_m(\mu,p) = \delta_{\mu' \mu} \delta_{p' p}.
\end{align}
As in our previous considerations, these vectors dictate a basis via
\begin{align}
    \ket{\varphi_{\mu \, p} \, \beta_i} = \sum_{m=-\lz}^{\lz} a_m(\mu,p) \, \ket{\lz, m,\beta_i},
\end{align}
and the corresponding bi-orthogonal conjugate reads
\begin{align}
    \bra{\tilde \varphi_{\mu \, p} \, \tb_i} = \sum_{m=-\lz}^{\lz} \tilde a_m(\mu,p) \, \bra{\lz, m,\tb_i}.
\end{align}
It is an easy exercise to show that this basis is complete and orthogonal, i.e.,
\begin{align}
    \sum_{\mu=1}^M \sum_{p=1}^{P_\mu} \sum_{i=1}^4 \proj{\varphi_{\mu \, p} \, \beta_i}{\tilde \varphi_{\mu \, p} \, \tb_i} = \hat{\mathcal{I}},
    \und
    \brak{\tilde \varphi_{\mu' \, p'} \, \tb_{i'}}{\varphi_{\mu \, p} \, \beta_i} = \delta_{\mu'\mu} \delta_{p'p} \delta_{i'i}.
\end{align}
Furthermore, the basis states for $i=1$ are zeroth-order solutions
\begin{align}
    \hmD \, \ket{\varphi_{\mu \, p} \, \beta_4} = 0, &&
    \bra{\tilde \varphi_{\mu \, p} \, \tb_4} \, \hmD = 0,
\end{align}
as well as they diagonalize the operator $\hmV^\one$ via
\begin{align}
    \mean{\tilde \varphi_{\mu' \, p'} \, \tb_4}{\hmV^\one}{\varphi_{\mu \, p} \, \beta_4} &=
    -\delta_{\mu'\mu} \delta_{p'p} x_\mu^\one \mean{\tb_4}{M_\lz^\one}{\beta_4}.
\end{align}
As always, this basis can be used to rewrite the $n$th-order state correction as
\begin{align}
    \label{eq:psinTM}
    \ket{\psi^{(n)}} 
    = \sum_{\mu = 1}^M \sum_{p = 1}^{P_\mu} \sum_{i=1}^4 \varphi_{\mu \, p \, \beta_i}^{(n)} \, \ket{\varphi_{\mu \, p} \, \beta_i}
    + {\sum_{l \, m \, i}}' \psi_{l \, m \, i}^{(n)} \, \ket{l \, m \, i},
\end{align}
where we introduced the shorthand notation
\begin{align}
    \varphi_{\mu \, p \, \beta_i}^{(n)} = \brak{\tilde \varphi_{\mu \, p} \, \tb_i}{\psi^{(n)}}.
\end{align}
By selecting the eigenvalue correction $x_{\mu_0}^\one$, i.e., we fix $\mu=\mu_0$, we are considering the zeroth-order state
\begin{align}
    \ket{\psi^\z} = \sum_{p=1}^{P_{\mu_0}} \varphi_{\mu_0 \, p \, \beta_4}^\z \, \ket{\varphi_{\mu_0 \,  p} \, \beta_4}.
\end{align}
With this we are prepared to determine the first-order state corrections in order to proceed to the second order.
\subsubsection{First-Order State Corrections}
The first-order state corrections can be determined by solving \eqref{eq:firstTM} in the degenerate and the non-degenerate subspace separately.\par
Let us start with the degenerate subspace. To determine the $\varphi_{\mu' \, p' \, \beta_{i'}}^\one$ we multiply \eqref{eq:firstTM} from the left with $\bra{\tilde \varphi_{\mu' \, p'} \, \tb_{i'}}$ and find
\begin{multline}
    \sum_{i=1}^3  \mean{\tb_{i'}}{M_\lz^\z}{\beta_i} \, \varphi_{\mu' \, p' \, \beta_i}^\one \\
    = -\sum_{p=1}^{P_{\mu_0}} \left\{
    \bra{\tilde \varphi_{\mu' \, p'} \, \tb_{i'}} \hmV^\one \ket{\varphi_{\mu_0 \, p} \, \beta_4} + \delta_{\mu' \mu_0} \, \delta_{p' p} \, x_{\mu_0}^\one \, \mean{\tb_{i'}}{M_\lz^\one}{\beta_4}
    \right\} \varphi_{\mu_0 \, p \, \beta_4}^\z.
\end{multline}
As previously, this equation does not determine $\varphi_{\mu' \, p' \, \beta_4}^\one$ since the this equation gives a trivial identity for $i'=4$. For the remaining cases, we can interpret $\mean{\tb_{i'}}{M_\lz^\z}{\beta_i}$ as invertible $3 \times 3$-matrix, as depicted in \eqref{eq:betaM0beta}, and solve for the $\varphi_{\mu' \, p' \, \beta_{i'}}^\one$, where $i'\neq 4$, by inversion. \par
To determine the first-order coefficients in the non-degenerate subspace, we multiply \eqref{eq:firstTM} with $\bra{l',m',i'}$, where $l' \neq \lz$, from the left to find
\begin{align}
    \sum_{i=1}^4 \mean{i'}{M^\z_{l'}}{i} \, \psi_{l' m' i}^\one = -\sum_{p=1}^{P_{\mu_0}} \mean{l',m',i'}{\hat{\mathcal{V}}^\one}{\varphi_{\mu_0 p} \, \beta_4} \, \varphi_{\mu_0 \, p \, \beta_4}^\z.
\end{align}
Yet again, we can solve for the $\psi_{l'\, m'\, i}^\one$ by inversion.
\subsection{Second-Order Perturbation Theory}
\label{sec:secondTM}
With these preparations it is straightforward to determine the second-order eigenvalue corrections $x_{\mu_0}^\two$. Therefore we recall the second-order equation \eqref{eq:Mpsi2} and \eqref{eq:MDecomposition} to find the second-order equation
\begin{align}
    \label{eq:secondTM}
    \hmD^\z \ket{\psi^\two} +  \left[ \hmV^\one  + x_{\mu_0}^\one \hmD^\one  \right] \ket{\psi^\one} + \left[ \hmV^\two  + x_{\mu_0}^\two \,\hmD^\two \right] \ket{\psi^\z } = 0.
\end{align}
\subsubsection{Second-Order Eigenvalue Corrections}
By multiplying \eqref{eq:secondTM} from the left with $\bra{\tilde\varphi_{\mu_0 \, p'} \, \tb_4}$, we find
\begin{multline}
    \label{eq:secondOrderDegTM}
    0 = \sum_{\mu=1}^M \sum_{p=1}^{P_\mu} \sum_{i=1}^3  \left[ \mean{\tilde\varphi_{\mu_0 \, p'} \, \tb_4}{\hmV^\one}{\varphi_{\mu \, p} \, \beta_i} + x_{\mu_0}^\one \, \delta_{p' p} \, \mean{ \tb_1}{M^\one_{\lz}}{\beta_i} \right] \varphi_{\mu \, p \, \beta_i}^\one \\
    +{\sum_{l \, m \, i}}' \mean{\tilde \varphi_{\mu_0 \, p'}\, \tb_4}{\hmV^\one}{l \, m \, i} \, \psi_{l \, m \, i}^\one + \sum_{p=1}^{P_{\mu_0}} \mean{\tilde\varphi_{\mu_0 \, p'} \, \tb_4}{\hmV^\two}{\varphi_{\mu_0 \, p} \, \beta_4} \varphi_{\mu_0 \, p \, \beta_4}^\z \\[4pt]
    + x_{\mu_0}^\two \mean{\tb_4}{M_\lz^\two}{\beta_4} \varphi_{\mu_0 \, p' \, \beta_4}^\z.
\end{multline}
As the $\varphi_{\mu \, p \, \beta_i}^\one$ as well as the $\psi_{l \, m \, i}^\one$ are linear combinations of the undetermined zeroth-order coefficients $\varphi_{\mu_0 \, p \, \beta_4}^\z$, this equation can yet again be recast to an eigenvalue equation of the form
\begin{align}
    \label{eq:eigen2ndTM}
    \mathbf{W} \, \bvarphi = \Delta^\two \, \bvarphi,
\end{align}
where $\mathbf{W}$ is a $P_{\mu_0} \times P_{\mu_0}$-matrix, $\bvarphi$ a $P_{\mu_0}$-component vector and we similarly to $\Delta^\one$ define
\begin{align}
    \Delta^\two = -\frac{x_{\mu_0}^\two}{x^\z},
\end{align}
which encodes the second-order eigenvalue correction $x_{\mu_0}^\two$. As in our first-order considerations, we introduce $\nu$ to label the $P_\mu$ eigenvalues and write
\begin{align}
    \Delta^\two \equiv \Delta^\two(\mu,\nu).
\end{align}
As previously, depending on the deformation encoded in $\mathbf{W}$, the eigenvalue correction can again be degenerate, which has to be kept in mind if one is interested in higher order corrections. \par
To conclude this section, we want to write down the eigenvalue of the perturbed system $x(\epsilon)$ up to and including its second-order correction as
\begin{align}
    \label{eq:xExpansionDegTM}
    x(\mu,\nu) = x_{\lz \, n_0}^\TM \left( 1 - \epsilon \, \Delta^\one(\mu) - \epsilon^2 \, \Delta^\two(\mu, \nu) + \dots \right),
\end{align}
Thus we found the complete analogue of the expression for the $\TE$-modes \eqref{eq:xExpansionDeg}.
\subsubsection{Further Remarks}
To determine the eigenvalue corrections of the $\TM$-modes, we only had to do two major changes compared to the $\TE$-modes. The first one was to adapt the (partial) diagonalization of $M_\lz^\z$, where we introduced the bi-orthogonal basis $\{ \ket{\beta_i},\bra{\tb_i}\}$ in $\mathcal{H}_4$. We can express this change as set of substitutions
\begin{subequations}
    \label{eq:alphabetasubs}
    \begin{align}
        \ket{\alpha_1} \to \ket{\beta_4}, && \bra{\ta_1} \to \bra{\tb_4}, \\
        \ket{\alpha_2} \to \ket{\beta_3}, && \bra{\ta_2} \to \bra{\tb_3}, \\
        \ket{\alpha_3} \to \ket{\beta_1}, && \bra{\ta_3} \to \bra{\tb_1}, \\
        \ket{\alpha_4} \to \ket{\beta_2}, && \bra{\ta_4} \to \bra{\tb_2}.
    \end{align}
\end{subequations}
Additionally we need to substitute
\begin{align}
    \label{eq:aEaMsubs}
    a_{\lz \, m}^{\text{E} \, \z} \to a_{\lz \, m}^{\text{M} \, \z}.
\end{align}
\par
The second change was the bi-orthogonal treatment of the first-order eigenvalue equation. This however only had the effect, that we needed to consider the bi-orthogonal conjugate of the basis $\ket{\varphi_{\mu \, p}}$ in $\mathcal{H}_\lz$, which is $\bra{\tilde \varphi_{\mu \, p}}$, instead of $\bra{\varphi_{\mu \, p}}$ for $\TE$-modes. The $\TM$-approach is actually a more natural choice, as we always consider these states in the tensor product space $\mathcal{H}_\lz \otimes \mathcal{H}_4$, and we already need to treat the component $\mathcal{H}_4$ in a bi-orthogonal way. We again can express this change as substitutions via
\begin{align}
    \label{eq:varphisubs}
    \bra{\varphi_{\mu \, p}} \to \bra{\tilde \varphi_{\mu \, p}}.
\end{align}
Thus, by applying the substitutions \eqref{eq:alphabetasubs}, \eqref{eq:aEaMsubs} and \eqref{eq:varphisubs} to our calculations for the $\TE$-modes in Section \ref{sec:secondDegenerate}, we immediately find the results in this section. \par
As final remark, we want to point out that to determine higher order corrections, one does not only need to adapt the considerations in Section \ref{sec:further1} to a bi-orthogonal treatment, one also has to notice that $\mathbf{W}$ in \eqref{eq:eigen2ndTM} is again a non-Hermitian matrix, and itself needs another bi-orthogonal treatment. To be able to solve the bi-orthogonal conjugate second-order equation, one also has to fully solve the bi-orthogonal conjugate first-order equation. To determine the second-order eigenvalue corrections however, one does not need to carry out these calculations.

\newpage

\section{Explicit First-Order Eigenvalue Equations}
\label{sec:firstExplicit}
Now that we solved the problem of finding the eigenvalue corrections up to and including second-order corrections for $\TE$- as well as $\TM$-eigenvalues in our quantum-like notation, we want to transition back to a more explicit formulation.
\subsection{TE-Modes}
\label{sec:TEfirstExplicit}
The fastest way to determine the explicit eigenvalue equation for $\TE$-modes is to recall \eqref{eq:eigenIntermediate} and write it as
\begin{align}
    \label{eq:firstMopTE}
    \sum_{m = -\lz}^\lz \mean{\lz \, m' \, \ta_1}{\hmM^\one}{\lz \, m \, \alpha_1} \, a_{\lz \, m}^{\text{E} \, \z} = 0.
\end{align}
Thus we need to determine the matrix elements $\mean{\lz \, m' \, \ta_1}{\hmM^\one}{\lz \, m \, \alpha_1}$. Therefore we recall that the matrix elements of the perturbation operator are defined in terms of the perturbation matrix in \eqref{eq:Mdef}. Furthermore we also going to need the vector representations of $\ket{\alpha_1}$ and $\bra{\ta_1}$ from \eqref{eq:alpha1def}. With this we find
\begin{align}
    \label{eq:M1TESomething1}
    \mean{\lz \, m' \, \ta_1}{\hmM^\one}{\lz \, m \, \alpha_1}
    &= [\bM_{\lz \, m}^{\lz\, m' \, \one}]_{\ta_1 \, \alpha_1} \notag \\[6pt]
    &=\frac{1}{z+1} \begin{bmatrix} z & 1 \end{bmatrix}
    \begin{bmatrix}
        [A_1^{\Phi \, \one}]_{\lz \, m}^{\lz \, m'} &  -[A_2^{\Phi \, \one}]_{\lz \, m}^{\lz \, m'}\\[6pt]
        -n_1 [B_1^{\Psi \, \one}]_{\lz \, m}^{\lz \, m'} & n_2 [B_2^{\Psi \, \one}]_{\lz \, m}^{\lz \, m'}
    \end{bmatrix}
    \begin{bmatrix} 1 \\[6pt] 1 \end{bmatrix} \notag \\[6pt]
    &\equiv \frac{1}{z+1} \left\{ z \, [\Delta A^{\Phi \, \one}]_{\lz \, m}^{\lz \, m'} - [\Delta B^{\Psi \, \one}]_{\lz \, m}^{\lz \, m'}\right\},
\end{align}
where the superscript $\one$ denotes first-order quantities and we introduced
\begin{subequations}
    \begin{align}
    [\Delta A^{\Phi \, \one}]_{l \, m}^{l' \, m'} &= [A_1^{\Phi \, \one}]_{l \, m}^{l' \, m'} - [A_2^{\Phi \, \one}]_{l \, m}^{l' \, m'}, \\[4pt]
    [\Delta B^{\Psi \, \one}]_{l \, m}^{l' \, m'} &= n_1 [B_1^{\Psi \, \one}]_{l \, m}^{l' \, m'} - n_2 [B_2^{\Psi \, \one}]_{l \, m}^{l' \, m'}.
\end{align}
\end{subequations}
We want to emphasize, that the $\Delta$ in the previous equation is just a convenient notation and should not be confused with the Laplacian. In Appendix \ref{app:mat} we explicitly determined the $[A_\alpha^{\Phi \, (n)}]_{l \, m}^{l' \, m'}$ as well as $[B_\alpha^{\Psi \, (n)}]_{l \, m}^{l' \, m'}$ for $n=0,1,2$. Let us use the explicit form \eqref{eq:AMatrix1} to find
\begin{align}
    [\Delta A^{\Phi \, \one}]_{l \, m}^{l' \, m'} &= \left((n_1 x^\z)\frac{j_l'(n_1x^\z)}{j_l(n_1x^\z)} - (n_2 x^\z)\frac{h_l'(n_2x^\z)}{h_l(n_2x^\z)} \right) [\Phi f \Phi]_{l \,m}^{l'\,m'} \notag \\[6pt]
    &= f_l^\TE(x^\z) [\Phi f \Phi]_{l \,m}^{l'\,m'},
\end{align}
where we recalled the defining equations for $\TE$-modes from \eqref{eq:fTE} and introduced
\begin{align}
    \label{eq:PhifPhi}
    [\Phi f \Phi]_{l \,m}^{l'\,m'} = \frac{1}{l'(l'+1)} \intdOmega \bPhi{l'}{m'}^*(\theta,\phi) \cdot \bPhi{l}{m}(\theta,\phi) f(\theta,\phi)  .
\end{align}
Thus, in the case of $\TE$-modes, the quantity $[\Delta A^{\Phi \, \one}]_{\lz \, m}^{l' \, m'}$ vanishes. \par
Similarly, but exploiting more properties of Bessel's functions (cf. App. \ref{app:bessel}) in combination with the defining equations for $\TE$-modes, one finds
\begin{align}
    \label{eq:DeltaBPsi1}
    [\Delta B^{\Psi \, \one}]_{\lz \, m}^{l' \, m'} = i (n_1^2-n_2^2) \left( x^\z [\Psi f \Psi]_{\lz \,m}^{l'\,m'} + x^\one \delta_{l' \, \lz } \delta_{m' \, m}\right).
\end{align}
Collecting this previous results, we find
\begin{align}
    \mean{l' \, m' \, \ta_1}{\hmM^\one}{\lz \, m \, \alpha_1} &= -\frac{1}{z+1} i (n_1^2-n_2^2) \left( x^\z  [\Psi f \Psi]_{\lz \,m}^{l'\,m'} + x^\one \delta_{l' \, \lz } \delta_{m' \, m}\right) \notag \\[4pt]
    &\equiv \mean{l' \, m' \, \ta_1}{\hmV^\one}{\lz \, m \, \alpha_1} + x^\one \delta_{l' \, \lz } \delta_{m' \, m} \, \bra{\ta_1}M_\lz^\one \ket{\alpha_1},
\end{align}
where $[\Psi f \Psi]_{l \,m}^{l' \, m'}$ is defined analogous to \eqref{eq:PhifPhi}. Therefore, the matrix elements of $\bV^\TE$ defined in \eqref{eq:Vmatdef} are given by
\begin{align}
    \label{eq:VmatTE}
    \bV_{m'\, m}^\TE = \frac{1}{x^\z} \frac{\mean{\lz \, m' \, \ta_1}{\hmV^\one}{\lz \, m \, \alpha_1}}{\bra{\ta_1}M_\lz^\one \ket{\alpha_1}} = [\Psi f \Psi]_{\lz \,m}^{\lz\,m'},
\end{align}
and hence, the first-order eigenvalue equation \eqref{eq:eigen1} for $\TE$-modes reads
\begin{align}
    \label{eq:eigenTE1}
    \sum_{m=-\lz}^\lz [\Psi f \Psi]_{\lz \,m}^{\lz\,m'} \, a_{\lz \, m}^{\text{E} \, \z} = \Delta^\one \, a_{\lz \, m'}^{\text{E} \, \z}.
\end{align}
The form of this equation has some interesting implications. First of all, this equation does not include a term depending on $\be_\parallel$ defined in \eqref{eq:eParallel}, which includes the possibly troublesome derivatives of $f$ as discussed in Section \ref{sec:applicability}. This suggests that the criterion of local paraxiality might not be needed in order to apply BCPT in this case. Furthermore, we stated in Section \ref{sec:firstordereigenvalue} that $\bV^\TE$ is Hermitian, and therefore the eigenvalues $\Delta^\one$ are real and the eigenvectors can be chosen orthonormal. In this form, the Hermicity can be shown via
\begin{align}
    \overline{[\Psi f \Psi]_{\lz \,m'}^{\lz\,m}}
    &= \frac{1}{\lz(\lz+1)} \intdOmega \overline{\bPsi{\lz}{m}^*(\theta,\phi) \cdot \bPsi{\lz}{m'}(\theta,\phi) f(\theta,\phi)} \notag \\[4pt]
    &= \frac{1}{\lz(\lz+1)} \intdOmega \bPsi{\lz}{m}(\theta,\phi) \cdot \bPsi{\lz}{m'}^*(\theta,\phi) f(\theta,\phi) \notag \\[6pt]
    &= [\Psi f \Psi]_{\lz \,m}^{\lz\,m'},
\end{align}
where in the second line, we used that $f$ is a real function. This also justifies in retrospect, why we introduced the factor $1/x^\z$ in the definition of $\bV^\TE$.
\subsection{TM-Modes}
\label{sec:TMfirstExplicit}
Likewise to the $\TE$-modes, we can use \eqref{eq:firstMopTE} and do the substitutions \eqref{eq:alphabetasubs} and \eqref{eq:aEaMsubs} to immediately find the first-order eigenvalue equation
\begin{align}
    \label{eq:firstMopTM}
    \sum_{m = -\lz}^\lz \mean{\lz \, m' \, \tb_4}{\hmM^\one}{\lz \, m \, \beta_4} \, a_{\lz \, m}^{\text{M} \, \z} = 0.
\end{align}
Likewise to the previous case we can determine the matrix elements
\begin{align}
    \mean{\lz \, m' \, \tb_4}{\hmM^\one}{\lz \, m \, \beta_4}
    &\equiv \frac{1}{z-1} \left\{ - [\Box B^{\Psi \, \one}]_{\lz \, m}^{\lz \, m'} + z \, [\Delta A^{\Phi \, \one}]_{\lz \, m}^{\lz \, m'}\right\},
\end{align}
where in this case, $[\Delta A^{\Phi \, \one}]_{l \, m}^{l' \, m'} \propto f_l^\TE(x^\z) \neq 0$ as discussed in Section \ref{sec:perfectsolution}. Furthermore we introduced
\begin{align}
    [\Box B^{\Psi \, \one}]_{l \, m}^{l' \, m'} &= [B_1^{\Psi \, \one}]_{l \, m}^{l' \, m'}/n_1 - [B_2^{\Psi \, \one}]_{l \, m}^{l' \, m'}/n_2 \\[4pt]
    &\equiv \Box a_l^{\Psi \, \one} x^\one \delta_{l'\,l} \delta_{m'\,m}+ \Box c_l^{\Psi \, \one} [\Psi f \Psi]_{l \,m}^{l\,m'} +\Box b_l^{Y \, \one}[\Psi e_\parallel]_{l\,m}^{l'\,m'},
\end{align}
where again the $\Box$ is just a notation and should not be confused with the d'Alembertian, 
\begin{align}
    [\Psi e_\parallel]_{l\,m}^{l'\,m'} = \frac{1}{l'(l'+1)} \intdOmega \bPsi{l'}{m'}^*(\theta,\phi) \cdot \be_\parallel(\theta,\phi) \Y_{l\,m}(\theta,\phi),
\end{align}
and we defined
\begin{subequations}
    \begin{align}
        \Box a_l^{\Psi \, \one} &= a_{1 \, l}^{\Psi \, \one}/n_1 - a_{2 \, l}^{\Psi \, \one}/n_2, \\[4pt]
        \Box c_l^{\Psi \, \one} &= c_{1 \, l}^{\Psi \, \one}/n_1 - c_{2 \, l}^{\Psi \, \one}/n_2, \\[4pt]
        \Box b_l^{Y \, \one} &= b_{1 \, l}^{Y \, \one}/n_1 - b_{2 \, l}^{Y \, \one}/n_2.
    \end{align}
\end{subequations}
The coefficients $a_{\alpha \, l}^{\Psi \, \one}$, $c_{1 \, l}^{\Psi \, \one}$ as well as $b_{1 \, l}^{Y \, \one}$ are determined in \eqref{eq:R1coeff}. One finds
\begin{subequations}
    \begin{align}
        \label{eq:box1}
        \Box a_l^{\Psi \, \one} \!
        &= -i \left[\left( \frac{j_l''}{j_l}-\frac{h_l''}{h_l}\right) - \left( \left[\frac{j_l'}{j_l}\right]^2-\left[\frac{h_l'}{h_l}\right]^2\right) -\left( \frac{1}{(n_1x^\z)^2}-\frac{1}{(n_2x^\z)^2}\right) \right] \! , \\[6pt]
        \Box c_l^{\Psi \, \one} \!
        &= -i x^\z \! \!\left[ \left(\frac{j_l''}{j_l}-\frac{h_l''}{h_l}\right) \! + \! \left( \frac{j_l'}{n_1 x^\z j_l} \! - \! \frac{h_l'}{n_2 x^\z h_l}\right) \! - \! \left( \frac{1}{(n_1x^\z)^2}\!-\!\frac{1}{(n_2x^\z)^2}\right)\right] \! , \\[6pt]
        %
        %
        \Box b_l^{Y \, \one}
        &= -i x^\z \! \! \left[ l(l+1) \left( \frac{1}{(n_1x^\z)^2}-\frac{1}{(n_2x^\z)^2}\right) \right] \! .
    \end{align}
\end{subequations}
We want to mention that these terms can be further simplified using properties of Bessel's functions, however, for our considerations this form is sufficient. \par
With this, the matrix elements of $\bV^\TM$ defined in \eqref{eq:Vmatdef} are given by
\begin{align}
    \label{eq:VmatTM}
    \bV_{m'\, m}^\TM &= \frac{1}{x^\z}\frac{\mean{\lz \, m' \, \tb_4}{\hmV^\one}{\lz \, m \, \beta_4}}{\bra{\tb_4}M_\lz^\one \ket{\beta_4}} \notag \\[6pt]
    &= \frac{\Box c_\lz^{\Psi \, \one}}{x^\z \, \Box a_\lz^{\Psi \, \one}} [\Psi f \Psi]_{\lz \,m}^{\lz\,m'} + \frac{\Box b_\lz^{Y \, \one}}{x^\z \, \Box a_\lz^{\Psi \, \one}} [\Psi e_\parallel]_{\lz\,m}^{\lz\,m'} - \frac{z f_\lz^\TE(x^\z)}{x^\z \, \Box a_\lz^{\Psi \, \one}} [\Phi f \Phi]_{\lz \,m}^{\lz\,m'}.
\end{align}
In comparison with the $\TE$ result in \eqref{eq:VmatTE} we notice that we have three contributions for $\TM$-modes. Especially notable is the term containing $[\Psi e_\parallel]_{\lz\,m}^{\lz\,m'}$ encoding the possibly problematic derivatives of $f$. Furthermore we notice that $\bV^\TM$ is in general not Hermitian.
\newpage

\section{The Shrinking Sphere}
\label{sec:shrink}
In \cite{aiello19} an alternative approach to find the resonances of deformed spheres was discussed. One application proposed in this paper is the shrinking sphere, which is analytically solvable problem and helps to understand up to which order a given perturbative expansion is reasonable. In the following, we want to carry out the calculations to compare the predictions of our approach with the right result for this problem. \par
Thus we consider two spheres, $S_0$ with radius $r_0$ (the unperturbed sphere) and a smaller sphere $S_1$ with radius $r_1$ (the deformed body) with $r_1<r_0$ and define
\begin{align}
    \delta r = r_0 - r_1 > 0.
\end{align}
From Section \ref{sec:perfectsolution} we know that the resonances $k_i$, $i=1,2$, for both spheres $S_i$ are characterized by the eigenvalues $x_{l \, n}^\sigma$. As $k_i \equiv k_{l \, n}^\sigma(r_i) = x_{l \, n}^\sigma/r_i$, we can express the resonance of the smaller sphere $S_1$ in dependence of the larger one as
\begin{align}
    \label{eq:kshrinking}
    k_{l \, n}^\sigma(r_1)
    &= \frac{k_{l \, n}^\sigma(r_0)}{1-\delta r/r_0} \\[4pt]
    &= k_{l \, n}^\sigma(r_0) \left[ 1 + \frac{\delta r}{r_0} + \left( \frac{\delta r}{r_0} \right)^2 + \dots \right],
\end{align}
where we applied the geometric series. In order to compare this with our perturbative approach, we first notice that the surface profile function of the deformed body \eqref{eq:Rf} is given by $R(\theta,\phi) = r_1 = r_0 - \delta r$. With this we find our surface deformation strength $\epsilon = \delta r/ r_0$ the surface profile function $f(\theta,\phi) = -1$. Recalling \eqref{eq:xExpansionDeg} we find
\begin{align}
    x &= k_{l \, n}^\sigma(r_1) \, r_0 \\[4pt]
    &= x_{l \, n}^\sigma \left[ 1 - \frac{\delta r}{r_0} \Delta^\one - \left( \frac{\delta r}{r_0} \right)^2 \Delta^\two + \dots \right].
\end{align}
Thus, the perturbative approach matches the analytic expectation if $\Delta^\one = -1$ and $\Delta^\two = -1$.
\subsection{TE-Modes}
Let us start with the $\TE$-modes. First of all we need to recall \eqref{eq:eigenTE1}, and therefore we need to determine the matrix elements $[\Phi f \Phi]_{l\,m}^{l'\,m'}$. However, as $f(\theta,\phi) = -1$ we can use the orthogonality of the vector spherical harmonics \eqref{eq:VSHorthogonality} and find
\begin{align}
    \label{eq:shrink1}
    [\Phi f \Phi]_{l\,m}^{l'\,m'} = - \delta_{l'\, l} \delta_{m' \, m}.
\end{align}
Inserting this into \eqref{eq:eigenTE1} we immediately find $\Delta^\one = -1$ corresponding to our expectation. \par
To determine the second-order correction, we furthermore need to find the first-order state corrections. As $\Delta^\one$ is expectantly $(2 \, \lz +1)$-fold degenerate, that is, the perturbation does not remove the degeneracy, we need to apply the general approach discussed in Section~\ref{sec:secondDegenerate}. \par
First of all we need to adapt our states as in \eqref{eq:statevarphidegenerate}. However, as we only have one eigenvalue labeled by $\mu_0=1$, we drop this label and write
\begin{align}
    \label{eq:varphiShrinkingTE}
    \ket{\varphi_{p} \, \alpha_i} = \sum_{m=-\lz}^{\lz} a_m(p) \, \ket{\lz \,  m \, \alpha_i}, && p=-\lz,\dots,\lz,
\end{align}
where we also adapted the range of $p$ to get a formulation closer to the zeroth-order. Now we can determine the corrections in the degenerate subspace. To do so, we insert this expression into \eqref{eq:varphidegeneratedegenerate} and find
\begin{multline}
    \label{eq:somethingTEShrinking1}
    \sum_{i=2}^4  \mean{\ta_{i'}}{M_\lz^\z}{\alpha_i} \, \varphi_{p' \, \alpha_i}^\one \\
    = -\sum_{p=-\lz}^{\lz} \left\{ \sum_{m'=-\lz}^{\lz} a_{m'}^*(p') \sum_{m=-\lz}^{\lz} a_{m}(p)
    \mean{\lz \, m' \, \ta_{i'}}{\hmM^\one}{\lz \, m \, \alpha_1} \right\} \varphi_{p \, \alpha_1}^\z .
\end{multline}
Thus we need to determine the matrix elements $\mean{\lz \, m' \, \ta_{i'}}{\hmM^\one}{\lz \, m \, \alpha_1}$ for $i'=2,3,4$. Analogous to \eqref{eq:M1TESomething1} we find
\begin{align}
    \begin{bmatrix}
        \mean{\lz \, m' \, \ta_{2}}{\hmM^\one}{\lz \, m \, \alpha_1} \\[2pt]
        \mean{\lz \, m' \, \ta_{3}}{\hmM^\one}{\lz \, m \, \alpha_1} \\[2pt]
        \mean{\lz \, m' \, \ta_{4}}{\hmM^\one}{\lz \, m \, \alpha_1}
    \end{bmatrix}
    = \begin{bmatrix}
        - [\Delta B^{\Psi \, \one}]_{\lz \, m}^{\lz \, m'} \\[2pt]
        \hphantom{-{}}[\Delta A^{\Psi \, \one}]_{\lz \, m}^{\lz \, m'} \\[2pt]
        -[\Delta B^{\Phi \, \one}]_{\lz \, m}^{\lz \, m'}
    \end{bmatrix}.
\end{align}
In this case, we find that the first entry vanishes due to \eqref{eq:DeltaBPsi1}, and the explicit calculations for the other two entries show that these matrices also vanish. Inserting this into \eqref{eq:somethingTEShrinking1}, we find that the first-order coefficients $\varphi_{p' \, \alpha_{i'}}^\one$ for $i'=2,3,4$ vanish. \par
Next we want to determine the coefficients in the non-degenerate subspace. Similarly to our previous considerations, we need to recall \eqref{eq:psidegeneratenondegenerate} and therefore need to determine the matrix elements $\mean{l' \, m' \, i}{\hmM^\one}{\lz \, m \, \alpha_1}$ for $l' \neq \lz$. Due to the form of the deformation function $f$ one finds
\begin{align}
    \mean{l' \, m' \, i}{\hmM^\one}{\lz \, m \, \alpha_1} \propto \delta_{l' \, \lz} = 0,
\end{align}
and thus, also the remaining first-order coefficients $\psi_{l' \, m' \, i'}^\one$ vanish for $l' \neq \lz$. \par
As we now determined all needed first-order quantities we are able to consider the second-order eigenvalue equation \eqref{eq:eigen2nd}. As the first-order corrections of the states vanish, the eigenvalue equation reads
\begin{align}
    \label{eq:EigenvalueSomething100}
    \sum_{p=-\lz}^\lz \mean{\varphi_{\mu_0 \, p'} \, \ta_1}{\hmV^\two}{\varphi_{p} \, \alpha_1}  \varphi_{p \, \alpha_1}^\z = - x^\two \mean{\ta_1}{M_\lz^\two}{\alpha_1} \varphi_{p' \, \alpha_1}^\z.
\end{align}
However, as we are only interested in the eigenvalues, it is irrelevant in which basis of $\mathcal{H}_\lz$ we consider the problem. Thus we can choose for simplicity $a_m(p) = \delta_{m \, p}$ and together with \eqref{eq:varphiShrinkingTE} we simplify the eigenvalue equation to
\begin{align}
    \sum_{p=-\lz}^\lz \mean{\lz \, p' \, \ta_1}{\hmV^\two}{\lz \, p \, \alpha_1}  \varphi_{p \, \alpha_1}^\z = - x^\two \mean{\ta_1}{M_\lz^\two}{\alpha_1} \varphi_{p' \, \alpha_1}^\z.
\end{align}
Now we see that this equation has the same functional form as the first-order equation \eqref{eq:eigenIntermedite2}, except that we have to substitute the first-order operators by the second-order ones. Similar to the previous calculations, one can determine the second-order matrix elements and solve this eigenvalue equation. Numerical calculations show that also in second-order perturbation theory we get the expected result $\Delta^\two = -1$.
\subsection{TM-Modes}
To determine the first-order corrections for $\TM$-modes, we need, additionally to \eqref{eq:shrink1}, the matrix elements
\begin{align}
    \label{eq:shrink2}
    [\Psi f \Psi]_{l \,m}^{l'\,m'} = - \delta_{l'\, l} \delta_{m' \, m}, \und [\Psi \mathbf{e}_\parallel]_{l\,m}^{l'\,m'} = 0.
\end{align}
Inserting them into \eqref{eq:VmatTM} yields
\begin{align}
    \bV_{m'\, m}^\TM
    &= -\left[\frac{\Box c_\lz^{\Psi \, \one}}{x^\z \, \Box a_\lz^{\Psi \, \one}} - \frac{z f_\lz^\TE(x^\z)}{x^\z \, \Box a_\lz^{\Psi \, \one}}\right]\delta_{m'm},
\end{align}
and the corresponding eigenvalue equation \eqref{eq:eigen1} reads
\begin{align}
    \label{eq:eigenTM1Shrinking}
    - \sum_{m=-\lz}^\lz \left[\frac{\Box c_\lz^{\Psi \, \one}}{x^\z \, \Box a_\lz^{\Psi \, \one}} - \frac{z f_\lz^\TE(x^\z)}{x^\z \,\Box a_\lz^{\Psi \, \one}}\right]\delta_{m'm} \, a_{\lz \, m}^{\text{E} \, \z} = \Delta^\one \, a_{\lz \, m'}^{\text{E} \, \z}.
\end{align}
Hence our expectation is satisfied if the square bracket equals one. To show this, let us determine $z f_\lz^\TE(x^\z)$. Therefore we recall the definition of $z$ from \eqref{eq:zTMdef} and rewrite it as
\begin{subequations}
    \begin{align}
        z &= -ix^\z \left[ \frac{1}{(n_1x^\z)^2} + \frac{j_l'}{(n_1x^\z)j_l} \right]\\
        &= -ix^\z \left[ \frac{1}{(n_2x^\z)^2} + \frac{h_l'}{(n_1x^\z)h_l} \right].
    \end{align}
\end{subequations}
Using the definition of $f_l^\TE$ from \eqref{eq:fTE} we find
\begin{align}
    z f_\lz^\TE(x^\z)
    &= z \left[ (n_1x^\z)\frac{j_l'}{j_l} \right] - z\left[ (n_2x^\z)\frac{h_l'}{h_l}\right] \notag \\[6pt]
    &= -ix^\z \left[ \frac{1}{(n_1x^\z)^2} + \frac{j_l'}{(n_1x^\z)j_l} \right]\left[ (n_1x^\z)\frac{j_l'}{j_l} \right] \notag \\[4pt]
    &\hphantom{=} +ix^\z \left[ \frac{1}{(n_2x^\z)^2} + \frac{h_l'}{(n_2x^\z)h_l} \right]\left[ (n_2x^\z)\frac{h_l'}{h_l} \right] \notag \\[6pt]
    &= -ix^\z \left[ \left(\frac{j_l'}{(n_1x^\z)j_l}-\frac{h_l'}{(n_2x^\z)h_l} \right) + \left( \left[\frac{j_l'}{j_l}\right]^2 - \left[\frac{h_l'}{h_l}\right]^2 \right)\right].
\end{align}
Recalling \eqref{eq:box1} we then find
\begin{align}
    \Box c_\lz^{\Psi \, \one} - z f_\lz^\TE(x^\z) = x^\z \, \Box a_\lz^{\Psi \, \one}.
\end{align}
Thus, the factor in the square bracket in \eqref{eq:eigenTM1Shrinking} equals one and we finally find $\Delta^\one = -1$ also for $\TM$-modes. \par
To spare the reader from uninteresting details, we report that likewise to the $\TE$-modes, the second-order eigenvalue equation reduces for $\TM$-modes to
\begin{align}
    \sum_{p=-\lz}^\lz \mean{\varphi_{\mu_0 \, p'} \, \tb_4}{\hmV^\two}{\varphi_{p} \, \beta_4}  \varphi_{p \, \beta_4}^\z = - \frac{x^\two}{x^\z} \mean{\tb_4}{M_\lz^\two}{\beta_4} \varphi_{p' \, \beta_4}^\z,
\end{align}
and again, numerical calculations suggest the expected result $\Delta^\two=-1$.
\newpage

\section{Summary}
\label{sec:summary}
In this work we applied boundary condition perturbation theory to determine the \mbox{optical} resonances of slightly deformed dielectric spheres up to and including second-order corrections. \par
We began by considering the \emph{unperturbed} problem, i.e., we determined the optical resonances of a perfectly spherical body. As effectively open system, the resulting resonances $x^\z$ are complex numbers, where the real part relates to the frequency and the imaginary part to the linewidth of each resonance. These unperturbed resonances can be labeled as $x_{l \, n}^\sigma$, where $\sigma$ labels the two polarizations of light, namely the transverse electric (TE) and the transverse magnetic (TM) polarization, $l$ labels the angular momentum of the confined light and $n$ enumerates the infinitely many resonances associated to $\sigma$ and $l$. Furthermore we visualized and discussed the associated electromagnetic fields, the so-called Whispering Gallery Modes. \par
Having understood the unperturbed problem, we needed to generalize the procedure to determine the resonances for a more general class of bodies. We choose the \hyphenation{pa-ram-e-triza-tion} parametrization
\begin{align*}
    R(\theta,\phi) = r_0 \big( 1 + \epsilon f(\theta,\phi) \big),
\end{align*}
which can be interpreted as a sphere with radius $r_0$ modulated by a deformation function $\epsilon f(\theta,\phi)$, thus we denoted such bodies \emph{deformed} spheres. Using this parametrization, we derived a matrix equation encoding the boundary conditions, which allowed for a perturbative solution later on. \par
On the way to derive the matrix equation we found applicability criteria for the BCPT. These criteria are geometrically intuitive: The first is that the deformation strength $\epsilon$ in the previous equation needs to be small. The second criterion is the \emph{local paraxiality}, which states, that the angle $\gamma$ between the normal unit vector $\er$ of the undeformed sphere and the normal unit vector $\bf{\hat n}$ of the deformed sphere needs to be small, so
\begin{align*}
    \epsilon \ll 1, \und \cos\gamma = \er \cdot \hat{\bn} \simeq 1.
\end{align*}
\par
To treat the perturbative expansion in an efficient and clear manner we introduced a quantum-like notation. This allowed us to establish a expansion of the optical resonances
\begin{align*}
    x(\mu,\nu) = x_{l \, n}^\sigma + \epsilon\, x^\one(\mu) + \epsilon^2 \, x^\two(\mu,\nu) + \order{\epsilon^3},
\end{align*}
similar to quantum mechanical degenerate Rayleigh-Schrödinger perturbation theory. In the previous equation, $\mu$ and $\nu$ label the different perturbed eigenvalues, and $x^\one$ and $x^\two$ are the respective first- and second-order corrections to the unperturbed eigenvalue. Completely analogous to the quantum case, the eigenvalue corrections $x^\one$ and $x^\two$ are determined by intrinsic finite-dimensional eigenvalue equations. The approach we developed works analogous for $\TE$- and $\TM$-modes, but we found some simplifications for $\TE$-modes which we discussed. \par
Finally we applied our approach to determine the resonances of an analytically solvable problem, the shrinking sphere, and found full agreement up to and including second-order corrections, indicating the correctness of our approach. \par
We remark that, at the best of our knowledge, the second-order perturbative solutions for both $\TE$- and $\TM$-polarizations of the electromagnetic fields presented in this thesis, were never derived, in a correct form, before. Thus, the results presented are perfectly original and solve a long-standing open problem. \par
\newpage

\appendix
\section{Spherical Coordinates}
\label{app:sc}
In this work we use the physicists standard notation for spherical coordinates (cf., e.g., \cite{zangwill12}, Sec. 1.2.3), where the spherical coordinates $(r,\theta,\phi)$ in terms of the Cartesian ones $(x,y,z)$ are given by
\begin{align}
    r = \sqrt{x^2+y^2+z^2} \komma 
    \cos \theta = z/r \komma
    \tan \phi = x/y.
\end{align}
Here, the radius $r \in [0,\infty)$ is the distance from the origin, $\theta \in [0,\pi]$ is the polar angle and $\phi \in [0,2\pi)$ the azimuthal angle. Conversely one has
\begin{align}
        x = r \sin \theta \cos \phi \komma
        y = r \sin \theta \sin \phi \komma
        z = r \cos \theta .
\end{align}
In spherical coordinates one can express a generic vector field as
\begin{align}
    \label{eq:scexpansion}
    \bF = F_r \, \er + F_\theta \, \etheta + F_\phi \, \ephi ,
\end{align}
where the basis vectors are
\begin{align}
    \er = \begin{pmatrix} \sin \theta \cos \phi\\ \sin \theta \sin \phi \\ \cos \theta \end{pmatrix}
    \komma
    \etheta = \begin{pmatrix}\cos \theta \cos \phi \\ \cos \theta \sin \phi \\ -\sin \theta \end{pmatrix},
    \und
    \ephi = \begin{pmatrix} -\sin \phi \\ \cos \phi \\ 0 \end{pmatrix} .
\end{align}
They form an orthonormal trihedron, i.e., they are orthonormal
\begin{align}
    \label{eq:scscalarproducts}
    \hat{\mathbf{e}}_i \cdot \hat{\mathbf{e}}_j = \delta_{i \, j}, && i,j = r,\theta,\phi,
\end{align}
and additionally satisfy
\begin{align}
    \label{eq:sccrossproducts}
    \er \times \etheta = \ephi
    \komma
    \etheta \times \ephi = \er,
    \und
    \ephi \times \er = \etheta .
\end{align}
In this curvilinear coordinate system, the nabla operator is given by
\begin{align}
    \label{eq:nablaSpherical}
    \bnabla = \er \frac{\partial}{\partial r} + \frac{\etheta}{r} \frac{\partial}{\partial \theta} + \frac{\ephi}{r \sin \theta} \frac{\partial}{\partial \phi},
\end{align}
and the Laplacian $\Laplace = \bnabla \cdot \bnabla$ by
\begin{align}
    \Laplace &= \frac{1}{r^2} \frac{\partial}{\partial r} \left(r^2 \frac{\partial}{\partial r}\right)
    + \frac{1}{r^2 \sin\theta} \frac{\partial}{\partial \theta} \left(\sin\theta \frac{\partial}{\partial \theta} \right)
    + \frac{1}{r^2 \sin^2\theta} \frac{\partial^2}{\partial r^2} \label{eq:laplacian} \\[4pt]
    &= \frac{1}{r^2} \frac{\partial}{\partial r} \left(r^2 \frac{\partial}{\partial r}\right)
    - \frac{\Lsquared}{r^2} \label{eq:laplacianL},
\end{align}
with the squared angular momentum operator $\Lsquared = \bL \cdot \bL$ and $\bL = -i \, \br \times \bnabla$ as in \eqref{eq:bL}.
\section{Scalar Spherical Harmonics}
The scalar spherical harmonics $\sY{l}{m}(\theta,\phi)$ (cf., e.g., \cite{zangwill12}, Sec. 7.7) arise naturally by solving the angular part of Lapace's equation, so that they satisfy
\begin{align}
    \label{eq:LsquaredY}
    \Lsquared \sY{l}{m}(\theta,\phi) = l(l+1) \sY{l}{m}(\theta,\phi),
\end{align}
where $l = 0,1,2,\dots$ and $m=-l,-l+1,\dots,l$. The spherical harmonics have lots of useful properties. For example, they behave under complex conjugation as
\begin{align}
    \label{eq:sshComplexConjugation}
    \sY{l}{m}^*(\theta,\phi) &= (-1)^m \Y_{l,-m}(\theta,\phi).
\end{align}
More important, they are orthonormal
\begin{align}
    \intdOmega \sY{l'}{m'}^*(\theta,\phi) \sY{l}{m}(\theta,\phi)
    &= \int_0^{2\pi} \! \mathrm{d}\phi \int_0^\pi \! \mathrm{d}\theta \sin \theta \sY{l'}{m'}^*(\theta,\phi) \sY{l}{m}(\theta,\phi) \notag \\[4pt]
    &= \Kdelta{l'}{l} \Kdelta{m'}{m},
\end{align}
and complete
\begin{align}
    \sum_{l=0}^\infty \sum_{m=-l}^l \sY{l}{m}^*(\theta,\phi) \sY{l}{m}(\theta',\phi') =
    \delta(\theta - \theta') \, \delta (\phi - \phi'),
\end{align}
meaning that they form a basis of the Hilbert space of square-integrable functions on a sphere. This means, one can express every function $f$ in spherical coordinates as
\begin{align}
    \label{eq:LaplaceSeries}
    f(r, \theta, \phi) = \sum_{l=0}^\infty \sum_{m=-l}^l f_{l \, m}(r) \sY{l}{m}(\theta, \phi) ,
\end{align}
where the coefficients $f_{l \, m}(r)$ are given by
\begin{align}
    f_{l \, m}(r) = \intdOmega \sY{l}{m}^*(\theta, \phi) f(r, \theta, \phi).
\end{align}
This is called \emph{Laplace series} or more general denoted as a \emph{multipole expansion}. Simply speaking, this series allows us to decompose every scalar function $f$ into its radial and angular component at the cost of an infinite sum.
\section{Vector Spherical Harmonics}
\label{app:vsh}
The concept of a multipole expansion can be generalized to three-dimensional vector fields. We found the description in \cite{barrera85} most convenient for our purposes. In this paper, \emph{Barrera et al.} introduce three vector quantities
\begin{subequations}
    \label{eq:VSH}
    \begin{align}
        \bY{l}{m} &= \er \sY{l}{m} , \\[4pt]
        \bPsi{l}{m} &= r \, \bnabla \sY{l}{m} , \\[4pt]
        \bPhi{l}{m} &= \er \times \bPsi{l}{m} = \br \times \bnabla \sY{l}{m} ,
    \end{align}
\end{subequations}
which are called \emph{vector spherical harmonics} and are defined on the sphere. They inhere many properties of the scalar spherical harmonics. For example, their behavior under complex conjugation is completely analogous to \eqref{eq:sshComplexConjugation}, i.e.
\begin{align}
    \bY{l}{m}^* = (-1)^m \, \bY{l,-m}{\!} \komma \bPsi{l}{m}^* = (-1)^m \, \bPsi{l,-m}{\!}, \und \bPhi{l}{m}^* = (-1)^m \, \bPhi{l,-m}{\!}.
\end{align}
Especially important are the properties as Hilbert space functions. They can be shown to be orthogonal, where
\begin{align}
    \label{eq:VSHorthogonality}
    \begin{split}
        \intdOmega \bY{l'}{m'}^* \cdot \bY{l}{m} &= \delta_{l' \,l} \, \delta_{m' \,m} ,\\[4pt]
        \intdOmega \bPsi{l'}{m'}^* \cdot \bPsi{l}{m} &= l(l+1) \delta_{l' \, l} \, \delta_{m' \,m}, \\[4pt]
        \intdOmega \bPhi{l'}{m'}^* \cdot \bPhi{l}{m}  &= l(l+1) \delta_{l' \, l} \, \delta_{m' \, m},
    \end{split}
\end{align}
and all integrals over different vector spherical harmonics vanish. Furthermore, it can be proven that they form a complete basis \cite{kristensson16}. This allows us to expand any three-dimensional vector field $\bF$ as
\begin{align}
    \label{eq:VSHexpansion}
    \bF(r,\theta,\phi) = \sum_{l=0}^\infty \sum_{m=-l}^l\left\{ F_{l \, m}^Y(r) \bY{l}{m} + F_{l \, m}^\Psi(r) \bPsi{l}{m} + F_{l \, m}^\Phi(r) \bPhi{l}{m}\right\} ,
\end{align}
where the coefficients are given by
\begin{subequations}
    \label{eq:VSHexpansionCoefficients}
    \begin{align}
        F_{l \, m}^Y(r) &= \intdOmega \bY{l}{m}^* \cdot\bF, \\[4pt]
        F_{l \, m}^\Psi(r) &= \frac{1}{l(l+1)}\intdOmega \bPsi{l}{m}^* \cdot \bF, \\[4pt]
        F_{l \, m}^\Phi(r) &= \frac{1}{l(l+1)}\intdOmega \bPhi{l}{m}^* \cdot \bF.
    \end{align}
\end{subequations}
The vector spherical harmonics as vectors in three-dimensional space satisfy many useful relations. Important for us is the fact that
\begin{align}
    \label{eq:ercrossvsh}
    \er \times \bY{l}{m} = 0 \komma  \er \times \bPsi{l}{m} = \bPhi{l}{m}, \und \er \times \bPhi{l}{m} = - \bPsi{l}{m}.
\end{align}
Furthermore, the scalar products satisfy
\begin{align}
    \label{eq:VSHscalarProducts}
    \bPsi{l}{m} \cdot \bPhi{l}{m} = 0, &&
    \bY{l}{m} \cdot \bPsi{l'}{m'} = 0, \und
    \bY{l}{m} \cdot \bPhi{l'}{m'} = 0 .
\end{align}
As we consider the multipole expansion of electromagnetic fields, it is useful to know how the nabla operator acts on a vector field of the form \eqref{eq:VSHexpansion}. Using
\begin{subequations}
    \label{eq:divVSH}
    \begin{align}
        \bnabla \cdot \left( F_{l \, m}^Y(r) \, \bY{l}{m}\right)
        &= \hphantom{-} \left[\frac{1}{r^2} \frac{\mathrm{d}}{\mathrm{d}r} \left(r^2 F_{l \, m}^Y(r) \right)\right] \sY{l}{m}, \\[4pt]
        \bnabla \cdot \left( F_{l \, m}^\Psi(r) \, \bPsi{l}{m}\right)
        &= - \left[l(l+1) \frac{F_{l \, m}^\Psi(r)}{r}\right] \sY{l}{m}, \\[4pt]
        \bnabla \cdot \left( F_{l \, m}^\Phi(r) \, \bPhi{l}{m}\right)
        &= 0,  \vphantom{\bigg[}
    \end{align}
\end{subequations}
one can determine the divergence of a vector field by adding those three terms and summing with respect to $l$ and $m$. Likewise one can determine the rotation by using
\begin{subequations}
    \label{eq:rotVSH}
    \begin{align}
        \bnabla \times \left( F_{l \, m}^Y(r) \bY{l}{m}\right)
        &= -\left[\frac{F_{l \, m}^Y(r)}{r}\right] \bPhi{l}{m}, \\[4pt]
        \bnabla \times \left( F_{l \, m}^\Psi(r) \bPsi{l}{m}\right)
        &= \hphantom{-} \left[ \frac{1}{r} \frac{\mathrm{d}}{\mathrm{d}r} \left(r F_{l \, m}^\Psi(r)\right) \right] \bPhi{l}{m}, \\[4pt]
        \bnabla \times \left( F_{l \, m}^\Phi(r) \bPhi{l}{m}\right)
        &= - \left[ l(l+1) \frac{F_{l \, m}^\Phi(r)}{r} \right] \bY{l}{m}
        - \left[ \frac{1}{r} \frac{\mathrm{d}}{\mathrm{d}r} \left( r F_{l \, m}^\Phi(r) \right) \right] \bPsi{l}{m}.
    \end{align}
\end{subequations}
\section{Bessel's Equation and Spherical Bessel Functions}
\label{app:bessel}
The spherical Bessel differential equation (cf., e.g., \cite{kristensson16}, App. B.3) occurs in the classical as well as the quantum theory of scattering when considering a spherical coordinate system. It is given by
\begin{align}
    z^2 \frac{\partial^2}{\partial z^2} u_{l \, m}
    + 2z \frac{\partial}{\partial z} u_{l \, m}
    + \left[z^2 - l(l+1) \right] u_{l \, m} = 0.
\end{align}
Its fundamental solutions are the spherical Bessel function (of first kind) $j_l(z)$ and the spherical Bessel function of second kind $y_l(z)$, often denoted as spherical Neumann functions. For real arguments $z$, these functions are real. Especially important is the observation that close to the origin, they satisfy
\begin{align}
    \begin{aligned}
        j_l(z) &= \frac{2^l \, l!}{(2l+1)!}\, z^l + \order{z^{l+1}}, \\[4pt]
        y_l(z) &= -\frac{(2l-1)!}{2^{l-1} \, (l-1)!}\, z^{-l-1}+ \order{z^{-l+1}},
    \end{aligned}
    &&
    z \to 0,
\end{align}
meaning that $j_l(z)$ is a regular solution and $y_l(z)$ diverges at the origin. \par
For scattering processes, one often introduces the spherical Hankel functions, also denoted as spherical Bessel functions of third kind, as the complex linear combinations
\begin{subequations}
    \label{eq:sphHankelDef}
    \begin{align}
        h_l^\one(z) &= j_l(z) + i \, y_l(z),\\[4pt]
        h_l^\two(z) &= j_l(z) - i \, y_l(z).    
    \end{align}
\end{subequations}
In the far field, they satisfy
\begin{align}
    \begin{aligned}
    h_l^\one(z) &= \frac{e^{+iz-i(l+1)\pi/2}}{z} \left( 1+\order{z^{-1}}\right), \\[4pt]
    h_l^\two(z) &= \frac{e^{-iz+i(l+1)\pi/2}}{z} \left( 1+\order{z^{-1}}\right),    
    \end{aligned}
    &&
    z \to \infty,
\end{align}
thus they describe, in the case of $h_l^\one(z)$, out-going and, in the case of $h_l^\two(z)$, incoming spherical waves. \par
Important properties of the spherical Bessel functions include their parity
\begin{align}
    \begin{split}
        \begin{aligned}
            j_l(-z)&=(-1)^l\,j_l(z), & y_l(-z)&=(-1)^{l+1}\,y_l(z),\\[4pt]
            h_l^\one(-z)&=(-1)^l\,h_l^\two(z), & h_l^\two(-z)&=(-1)^l\,h_l^\one(z),
        \end{aligned}
    \end{split}
\end{align}
and their behavior under complex conjugation (cf., e.g., \cite{kristensson16}, Sec. 8.2.3)
\begin{align}
    \label{eq:besselAnalyticContinuation}
    j_l(z^*)&= [j_l(z)]^*, & y_l(z^*)&=[y_l(z)]^*, \\[4pt]
    h_l^\one(z^*)&= [h_l^\two(z)]^*, & h_l^\two(z^*)&= [h_l^\one(z)]^*.
\end{align}
Finally, the spherical Bessel functions satisfy many recurrence relations. Important for the main text are
\begin{align}
    f_l'(z) &= \frac{l}{z} f_l(z) - f_{l+1}(z), \\
    f_{l-1}(z) &= \frac{2l+1}{z} f_l(z) - f_{l+1}(z),
\end{align}
where $f_l(z)$ is any combination of $j_l(z)$, $y_l(z)$, $h_l^\one(z)$ and $h_l^\two(z)$.
\section{Matrix Elements and Properties}
\label{app:mat}
In Section \ref{sec:applicationBCPT} we formally expanded the perturbation operator $\hmM$ in powers of $\epsilon$ and stated properties of the $\hmM^{(n)}$ in \ref{sec:properties}. 
This appendix is dedicated to proof the properties used in the main text. \par
To begin with, we recall that we defined the perturbation operator in terms of the perturbation matrix in \eqref{eq:Mdef}, and the perturbation matrix $\bM_{l \, m}^{l'\, m'}$ is defined in \eqref{eq:bMdef}. Thus, to determine an expansion of the perturbation matrix
\begin{align}
    \bM_{l \, m}^{l'\, m'} = \bM_{l \, m}^{l'\, m' \, \z} + \epsilon \, \bM_{l \, m}^{l'\, m' \, \one} + \epsilon^2 \, \bM_{l \, m}^{l'\, m' \, \two} + \dots \, ,
\end{align}
we need to determine the expansions
\begin{subequations}
    \label{eq:ABmatricesExpansion}
    \begin{align}
        [A_\alpha^{V}]_{l \, m}^{l' \, m'} &= [A_\alpha^{V \, \z}]_{l \, m}^{l' \, m'} + \epsilon \, [A_\alpha^{V \, \one}]_{l \, m}^{l' \, m'} + \epsilon^2 \, [A_\alpha^{V \, \two}]_{l \, m}^{l' \, m'} + \dots \, , \\[4pt]
        [B_\alpha^{V}]_{l \, m}^{l' \, m'} &= [B_\alpha^{V \, \z}]_{l \, m}^{l' \, m'} + \epsilon \, [B_\alpha^{V \, \one}]_{l \, m}^{l' \, m'} + \epsilon^2 \, [B_\alpha^{V \, \two}]_{l \, m}^{l' \, m'} + \dots\,.
    \end{align}    
\end{subequations}
Recalling the definition of these matrix elements in \eqref{eq:ABmatrices}, we see that they include the quantities $A_{\alpha \, l}^X(k_\alpha\,R(\theta,\phi))$ defined in \eqref{eq:Ar}. Thus by finding an expansion of these quantities, we can go up the chain of equations to prove the properties of the perturbation operator.
\subsection{Expansion of the Radial Functions}
As a first step, we want to expand the radial functions $A_{\alpha \, l}^X(k_\alpha\, R(\theta,\phi))$. To do so, we recall that we parametrized the surface of the deformed sphere in \eqref{eq:Rf} as $R(\theta,\phi) = r_0(1+\epsilon \, f(\theta,\phi))$. Furthermore, we defined $k_\alpha \, r_0 = n_\alpha \, x$ and expanded the eigenvalue $x$ of the perturbed system in \eqref{eq:xExpansion} as $x(\epsilon) = x^\z + \epsilon \, x^\one + \epsilon^2 x^\two + \dots$\,. With this we find for example
\begin{align}
    A_{1 \, l}^\Phi (k_1 \,R(\theta,\phi)) &= \frac{j_l(n_1 (x^\z+ \epsilon \, x^\one + \epsilon^2 \, x^\two \+ \dots) (1+\epsilon f(\theta,\phi))}{j_l(n_1 (x^\z+ \epsilon \, x^\one + \epsilon^2 \, x^\two \+ \dots))} \\[4pt]
    &= 1 + \epsilon \left[ (n_1x^\z) \frac{j_l'(n_1x^\z)}{j_l(n_1x^\z)} \right] \, f(\theta,\phi) + \dots \, .
\end{align}
To write down these series expansions in a more symbolic manner, we introduce
\begin{align}
    R_{\alpha \, l}^\Phi(\theta,\phi) &\equiv A_{\alpha \, l}^\Phi (k_\alpha \,R(\theta,\phi)) \\[4pt]
    &\equiv R_{\alpha \, l}^{\Phi \, \z} + \epsilon \, R_{\alpha \, l}^{\Phi \, \one}(\theta,\phi) + \epsilon^2 \, R_{\alpha \, l}^{\Phi \, \two}(\theta,\phi) + \dots \, ,
\end{align}
as well as
\begin{align}
    R_{\alpha \, l}^\Psi(\theta,\phi) &\equiv  A_{\alpha \, l}^\Psi (k_\alpha \,R(\theta,\phi)) \\[4pt]
    &\equiv R_{\alpha \, l}^{\Psi \, \z} + \epsilon \, R_{\alpha \, l}^{\Psi \, \one}(\theta,\phi) + \epsilon^2 \, R_{\alpha \, l}^{\Psi \, \two}(\theta,\phi) + \dots \,.
\end{align}
For the last radial function we slightly adapt the definition. By recalling \eqref{eq:bB} we notice that $A_{\alpha \, l}^Y$ is always multiplied by $\bn_\parallel = \epsilon/(1+\epsilon f(\theta,\phi)) \, \be_\parallel$. Hence we define
\begin{align}
    R_{\alpha \, l}^Y(\theta,\phi) &\equiv  \frac{\epsilon}{1 + \epsilon f(\theta,\phi)} A_{\alpha \, l}^Y (k_\alpha \,R(\theta,\phi)) \label{eq:RY} \\[4pt]
    &\equiv R_{\alpha \, l}^{Y \, \z} + \epsilon \, R_{\alpha \, l}^{Y \, \one}(\theta,\phi) + \epsilon^2 \, R_{\alpha \, l}^{Y \, \two}(\theta,\phi) + \dots \, ,
\end{align}
in order to shift the entire $\epsilon$-dependence into this function. \par
Let us first mention that the zeroth-order coefficients are given by
\begin{align}
    \label{eq:R0}
    R_{1 \, l}^{\Phi \, \z} &= 1, & R_{1 \, l}^{\Psi \, \z} &= -i \, \frac{[(n_1x^\z)j_l(n_1x^\z)]'}{(n_1x^\z)j_l(n_1x^\z)}, & R_{1 \, l}^{Y\, \z} &= 0, \\[6pt]
    R_{2 \, l}^{\Phi \, \z} &= 1, & R_{2 \, l}^{\Psi \, \z} &= -i \, \frac{[(n_2x^\z)h_l(n_2x^\z)]'}{(n_2x^\z)h_l(n_2x^\z)}, & R_{2 \, l}^{Y\, \z} &= 0,
\end{align}
which are just the quantities at $\epsilon=0$. We already discussed in Section \ref{sec:zeroth} that this implies that the zeroth-order quantities correspond to the unperturbed ones. \par
In order to treat the higher-order terms $(\nu = 1,2)$ in a convenient way, we introduce
\begin{align}
    R_{\alpha \, l}^{X \, (\nu)}(\theta,\phi) &= a_{\alpha \, l}^{X \, (\nu)} x^{(\nu)} + b_{\alpha \, l}^{X \, (\nu)} + c_{\alpha \, l}^{X \, (\nu)} f(\theta,\phi) + d_{\alpha \, l}^{X \, (\nu)} f^2(\theta,\phi),
\end{align}
which is a power-series in $f$ with the tweak that it separates the $f^0$ term. \par
Now doing the explicit series expansion, we find in first order for $\alpha = 1$ the non-vanishing coefficients
\begin{subequations}
    \label{eq:R1coeff}
    \begin{align}
        c_{1 \, l}^{\Phi \, \one} &=  J_l', \\[4pt]
        a_{1 \, l}^{\Psi \, \one} &= \frac{i}{n_1(x^\z)^2} \left[ 1 + (J_l')^2 - J_l''\right], &
        c_{1 \, l}^{\Psi \, \one} &= \frac{i}{n_1x^\z} \left[ 1 - J_l' - J_l''\right], \\[4pt]
        b_{1 \, l}^{Y \, \one} &= - l(l+1) \frac{i}{n_1 x^\z},
    \end{align}    
\end{subequations}
where we introduced
\begin{subequations}
    \label{eq:J}
    \begin{align}
        J_l'    &= (n_1 x^\z) \, \frac{j_l'(n_1 x^\z)}{j_l(n_1 x^\z)} , \\[4pt]
        J_l''   &= (n_1 x^\z)^2 \, \frac{j_l''(n_1 x^\z)}{j_l(n_1 x^\z)} , \\[4pt]
        J_l'''  &= (n_1 x^\z)^3 \, \frac{j_l'''(n_1 x^\z)}{j_l(n_1 x^\z)},
    \end{align}
\end{subequations}
which should not be confused with the Bessel function of the first kind. \par
We want to mention that the quantities defined in \eqref{eq:J}, and therefore also the coefficients in \eqref{eq:R1coeff}, could be rewritten using Bessel's equation and other properties discussed in Section \ref{app:bessel}. For our intends and purposes this form is adequate. Furthermore, we get the $\alpha=2$ terms using the substitutions
\begin{align}
    \label{eq:subs2}
    n_1 \rightarrow n_2, \und J_l \rightarrow H_l,
\end{align}
where $H_l$ is defined as in $\eqref{eq:J}$ with $j_l \rightarrow h_l^{(1)}$ and should not be confused with the Hankel functions. \par
With these definitions the non-vanishing second-order coefficients for $R_{1 \, l}^{\Phi \, \two}(\theta,\phi)$ are
\begin{align}
    c_{1 \, l}^{\Phi \, \two} = \frac{x^\one}{x^\z} \left[ J_l' - (J_l')^2 + J_l'' \right], \und
    d_{1 \, l}^{\Phi \, \two} = \frac{1}{2} J_l'',
\end{align}  
for $R_{1 \, l}^{\Psi \, \two}(\theta,\phi)$ we get
\begin{align}
    \begin{split}
        a_{1 \, l}^{\Psi \, \two} &= \frac{i}{n_1 (x^\z)^2} \Big[ 1 + (J_l')^2 - J_l'' \Big], \\[4pt]
        b_{1 \, l}^{\Psi \, \two} &= \frac{i}{n_1 x^\z} \left(\frac{x^\one}{x^\z}\right)^2 \left[ -1 -(J_l')^3 + \frac{3}{2} J_l'J_l'' -\frac{1}{2} J_l''' \right], \\[4pt]
        c_{1 \, l}^{\Psi \, \two} &= \frac{i}{n_1 x^\z} \frac{x^\one}{x^\z} \Big[ -1 +(J_l')^2 -2J_l'' +J_l'J_l'' -J_l''' \Big], \\[4pt]
        d_{1 \, l}^{\Psi \, \two} &= \frac{i}{n_1 x^\z} \left[ -1 +J_l' -\frac{1}{2}J_l'' -\frac{1}{2} J_l''' \right],
    \end{split}
\end{align}
and finally for $R_{1 \, l}^{Y \, \two}(\theta,\phi)$ we find
\begin{align}
    b_{1 \, l}^{Y \, \two} = \frac{i}{n_1 x^\z} l(l+1) \frac{x^\one}{x^\z}, \und
    c_{1 \, l}^{Y \, \two} = \frac{i}{n_1 x^\z} l(l+1) \left[ 2 - J_l' \right],
\end{align}
and the coefficients for $\alpha=2$ can be determined using \eqref{eq:subs2}.
\subsection{\texorpdfstring{Expansion of the $A$- and $B$-Matrices}{Expansion of the A- and B-Matrices}}
Now we want to determine the expansion of matrix elements of the $A$- and $B$-matrices defined in \eqref{eq:ABmatricesExpansion}. First of all we notice that due to \eqref{eq:R0}, the zeroth-order matrix elements correspond to \eqref{eq:ABmatricesUnperturbed}, i.e.
\begin{subequations}
    \begin{align}
        [A_\alpha^{V \, \z}]_{l \, m}^{l' \, m'} &= \delta_{l' \, l}\, \delta_{m' \, m}\, \delta_{V \, \Phi}, \\[4pt]
        [B_\alpha^{V \, \z}]_{l \, m}^{l' \, m'} &= \delta_{l' \, l}\, \delta_{m' \, m} \, \delta_{V \, \Psi} \, R_{\alpha \, l}^{\Psi \, \z}.
    \end{align}
\end{subequations}
\par Let us first consider the higher order matrix elements $[A_\alpha^{V \, (\nu)}]_{l \, m}^{l' \, m'}$. Using \eqref{eq:Amatrices} and $R_{\alpha \, l}^{\Phi \, \one} = c_{\alpha \, l}^{\Phi \, \one} f(\theta,\phi)$ from our previous considerations, we find
\begin{align}
    \label{eq:Asomething1}
    [A_\alpha^{V \, \one}]_{l \, m}^{l' \, m'} = c_{\alpha \, l}^{\Phi \, \one} \intdOmega \mathbf{V}_{l' \, m'}^*(\theta,\phi) \cdot \bPhi{l}{m}(\theta,\phi) f(\theta,\phi).
\end{align}
For later convenience, we introduce the matrix elements
\begin{align}
    [V' f^k V]_{l \,m}^{l'\,m'} = \frac{1}{l'(l'+1)} \intdOmega \mathbf{V'}_{l'\,m'}^*(\theta,\phi) \cdot \bV_{l\,m}(\theta,\phi)f^k(\theta,\phi),
\end{align}
where we notice that for $k=0$, the matrix elements are given by $[V' \, V]_{l \,m}^{l'\,m'} = \delta_{l' \, l} \delta_{m' \, m} \delta_{V' \, V}$ due to the orthogonality of the vector spherical harmonics \eqref{eq:VSHorthogonality}. Using this definition, we can write \eqref{eq:Asomething1} compactly as
\begin{align}
    [A_\alpha^{V \, \one}]_{l \, m}^{l' \, m'} = c_{\alpha \, l}^{\Phi \, \one} \, [V f \Phi]_{l \,m}^{l'\,m'}.
\end{align}
Likewise we find the second-order matrix elements
\begin{align}
    \label{eq:AMatrix1}
    [A_\alpha^{V \, \two}]_{l \, m}^{l' \, m'} = c_{\alpha \, l}^{\Phi \, \two} [V f \Phi]_{l \,m}^{l'\,m'} + d_{\alpha \, l}^{\Phi \, \two} \, [V f^2 \Phi]_{l \,m}^{l'\,m'}.
\end{align}
\par Secondly we want to determine the higher-order matrix elements $[B_\alpha^{V \, (\nu)}]_{l \, m}^{l' \, m'}$. Therefore we notice that with the definition \eqref{eq:RY}, we can rewrite \eqref{eq:bB} as
\begin{align}
    \bB_{\alpha , l \, m}(\theta,\phi) &= R_{\alpha \, l}^\Psi(\theta,\phi) \bPsi{l}{m}(\theta,\phi) + R_{\alpha \, l}^Y(\theta,\phi) \sY{l}{m}(\theta,\phi) \be_\parallel (\theta,\phi).
\end{align}
Now we can use \eqref{eq:Bmatrices} to find
\begin{align}
    [B_\alpha^{V \, \one}]_{l \, m}^{l' \, m'} &=
    a_{\alpha \, l}^{\Psi \, \one} x^\one \delta_{l' \, l} \, \delta_{m' \, m} \, \delta_{V \, \Psi}
    + c_{\alpha \, l}^{\Psi \, \one} [V f \Psi]_{l \,m}^{l'\,m'}
    + b_{\alpha \, l}^{Y \, \one} [V e_\parallel]_{l\,m}^{l'\,m'} ,
\end{align}
where we introduced
\begin{align}
    [V f^k e_\parallel]_{l\,m}^{l'\,m'} = \frac{1}{l'(l'+1)} \intdOmega \bV_{l' \, m'}^*(\theta,\phi) \cdot \be_\parallel(\theta,\phi) \Y_{l\,m}(\theta,\phi) f^k(\theta,\phi) .
\end{align}
Similarly, we find the second order matrix elements
\begin{multline}
    [B_\alpha^{V \, \two}]_{l \, m}^{l' \, m'} = a_{\alpha \, l}^{\Psi \, \two} x^\two \delta_{l' \, l} \, \delta_{m' \, m} \, \delta_{V \, \Psi} + b_{\alpha \, l}^{\Psi \, \two} \delta_{l' \, l} \, \delta_{m' \, m} \, \delta_{V \, \Psi} \\[4pt]
    + c_{\alpha \, l}^{\Psi \, \two} [V f \Psi]_{l \,m}^{l'\,m'} + d_{\alpha \, l}^{\Psi \, \two} [V f^2 \Psi]_{l \,m}^{l'\,m'}
    +b_{\alpha \, l}^{Y \, \two} [V e_\parallel] + c_{\alpha \, l}^{Y \, \two} [V f e_\parallel].
\end{multline}
\subsection{Properties of the Perturbation Matrix}
Collecting this results, we can proof all properties we used in Section \ref{sec:properties}. First of all we notice that for $\nu=1,2$, the only quantity containing $x^{(\nu)}$ is $[B_\alpha^{\Psi \, (\nu)}]_{l \, m}^{l' \, m'}$. Thus let us write this dependence explicitly as
\begin{subequations}
    \label{eq:ABMatrixSplit}
    \begin{align}
        [A_\alpha^{\Psi \, (\nu)}]_{l \, m}^{l' \, m'} &= [\mathcal{A}_\alpha^{\Psi \, (\nu)}]_{l \, m}^{l' \, m'}, &
        [A_\alpha^{\Phi \, (\nu)}]_{l \, m}^{l' \, m'} &= [\mathcal{A}_\alpha^{\Phi \, (\nu)}]_{l \, m}^{l' \, m'}, \\
        [B_\alpha^{\Psi \, (\nu)}]_{l \, m}^{l' \, m'} &= [\mathcal{B}_\alpha^{\Psi \, (\nu)}]_{l \, m}^{l' \, m'} + x^{(\nu)} \delta_{l' \, l} \, \delta_{m' \, m} a_{\alpha \, l}^{\Psi \, (\nu)}, &
        [B_\alpha^{\Phi \, (\nu)}]_{l \, m}^{l' \, m'} &= [\mathcal{B}_\alpha^{\Phi \, (\nu)}]_{l \, m}^{l' \, m'} ,
    \end{align}
\end{subequations}
where all matrix elements denoted with calligraphic letters are independent of $x^{(\nu)}$, i.e.
\begin{align}
    \label{eq:ABindependent}
    \frac{\mathrm{d}}{\mathrm{d}x^{(\nu)}} \, [\mathcal{A}_\alpha^{V \, (\nu)}]_{l \, m}^{l' \, m'} = 0,
    \und
    \frac{\mathrm{d}}{\mathrm{d}x^{(\nu)}} \, [\mathcal{B}_\alpha^{V \, (\nu)}]_{l \, m}^{l' \, m'} = 0,
\end{align}
as well as
\begin{align}
    \label{eq:aindependent}
    \frac{\mathrm{d}}{\mathrm{d}x^{(\nu)}} \, a_{\alpha \, l}^{\Psi \, (\nu)} = 0 .
\end{align}
Furthermore $a_{\alpha \, l}^{\Psi \, (\nu)}$ is independent of the geometry of the deformation. \par
With this, we immediately find the needed properties discussed in Section \ref{sec:properties}: The zeroth-order property \eqref{eq:M0Decomposition} was already found in Section \ref{sec:zeroth}, and inserting \eqref{eq:ABMatrixSplit} into \eqref{eq:bMdef} results in \eqref{eq:MnuDecomposition}. We also find \eqref{eq:VDindependent} using \eqref{eq:ABindependent} together with \eqref{eq:aindependent}. This finishes our proof.
\newpage
\printbibliography[heading=bibintoc]
\newpage
\section*{Erklärung}
Ich versichere, dass ich diese Masterarbeit ohne Hilfe Dritter und ohne Benutzung anderer als der angegebenen Quellen und Hilfsmittel angefertigt habe und die aus benutzten Quellen wörtlich oder inhaltlich entnommenen Stellen als solche kenntlich gemacht habe. Diese Arbeit hat in gleicher oder ähnlicher Form noch keiner Prüfungsbehörde vorgelegen.\\
\ \\
\ \\
\ \\
\ \\
\rule{0.4\textwidth}{0.4pt}\\
\noindent Erlangen, den 1.9.2020

\end{document}